\documentclass[usenatbib]{mn2e}
\usepackage{apjfonts,amsfonts,amsmath,amssymb,bm,ctable,verbatim}

\newcommand{\Alf}{{Alfv\'en}}

\newcommand{\bhat}{\hat{\bf b}}

\newcommand{\gizmourl}{\href{http://www.tapir.caltech.edu/~phopkins/Site/GIZMO.html}{\url{http://www.tapir.caltech.edu/~phopkins/Site/GIZMO.html}}}

\newcommand{\paperone}{Paper {\small I}}

\newcommand{\etal}{et al.}

\newcommand{\acknowledgments}[1]{\begin{small}\section*{Acknowledgments}\end{small}{\noindent #1}\vspace{5pt}}
\newcommand{\datastatement}[1]{\begin{small}\section*{Data Availability Statement}\end{small}{\noindent #1}\vspace{5pt}}

\newcommand{\figsuffixspecial}{_centered_r79n1r27}

\newcommand{\figrowlimv}[2]{
\includegraphics[width=0.495\textwidth]{figures/#1/spectrum_E_plot\figsuffixspecial}
\includegraphics[width=0.495\textwidth]{figures/#2/spectrum_E_plot\figsuffixspecial}\\
\hspace{0.1cm}
\includegraphics[width=0.488\textwidth]{figures/#1/spectrum_BC_plot\figsuffixspecial}
\hspace{0.03cm}
\includegraphics[width=0.488\textwidth]{figures/#2/spectrum_BC_plot\figsuffixspecial}\\
\includegraphics[width=0.495\textwidth]{figures/#1/spectrum_Pos_plot\figsuffixspecial}
\includegraphics[width=0.495\textwidth]{figures/#2/spectrum_Pos_plot\figsuffixspecial}\\
}
\newcommand{\figsix}[2]{
\includegraphics[width=0.48\textwidth]{figures/#1/spectrum_J_plot#2}
\includegraphics[width=0.48\textwidth]{figures/#1/spectrum_E_plot#2} \\
\hspace{0.2cm}\includegraphics[width=0.46\textwidth]{figures/#1/spectrum_Be_plot#2}
\hspace{0.3cm}\includegraphics[width=0.46\textwidth]{figures/#1/spectrum_BC_plot#2} \\
\includegraphics[width=0.47\textwidth]{figures/#1/spectrum_PBar_plot#2}
\hspace{0.05cm}\includegraphics[width=0.47\textwidth]{figures/#1/spectrum_Pos_plot#2} \\
}

\title[Failure of SC \&\ ET Models]{Standard Self-Confinement and Extrinsic Turbulence Models for Cosmic Ray Transport are Fundamentally Incompatible with Observations}

\author[Hopkins \etal]{
\parbox[t]{\textwidth}{
Philip F.~Hopkins$^1$, Jonathan Squire$^{2}$, Iryna S.\ Butsky$^1$, Suoqing Ji$^1$ 
}\vspace*{4pt} \\
$^1$ TAPIR, Mailcode 350-17, California Institute of Technology, Pasadena, CA 91125, USA. E-mail:phopkins@caltech.edu \\
$^2$ Physics Department, University of Otago, 730 Cumberland St., Dunedin 9016, New Zealand 
}

\date{}
\begin{document}
\maketitle

\begin{abstract}
Models for cosmic ray (CR) dynamics fundamentally depend on the rate of CR scattering from magnetic fluctuations. In the ISM, for CRs with energies $\sim$MeV-TeV, these fluctuations are usually attributed either to `extrinsic turbulence' (ET) -- a cascade from larger scales -- or `self-confinement' (SC) -- self-generated fluctuations from CR streaming. Using simple analytic arguments and detailed `live' numerical CR transport calculations in galaxy simulations, we show that both of these, in standard form, cannot explain even basic qualitative features of observed CR spectra. For ET, any spectrum that obeys critical balance or features realistic anisotropy, or any spectrum that accounts for finite damping below the dissipation scale, predicts qualitatively incorrect spectral shapes and scalings of B/C and other species. Even if somehow one ignored both anisotropy and damping, observationally-required scattering rates disagree with ET predictions by orders-of-magnitude. For SC, the dependence of driving on CR energy density means that it is nearly impossible to recover observed CR spectral shapes and scalings, and again there is an orders-of-magnitude normalization problem. But more severely, SC solutions with super-\Alf{ic} streaming are unstable. In live simulations, they revert to either arbitrarily-rapid CR escape with zero secondary production, or to bottleneck solutions with far-too-strong CR confinement and secondary production. Resolving these fundamental issues without discarding basic plasma processes requires invoking different drivers for scattering fluctuations. These must act on a broad range of scales with a power spectrum obeying several specific (but plausible) constraints.
\end{abstract}


\section{Introduction}
\label{sec:intro}

Understanding how cosmic rays (CRs) propagate and interact as they travel through the inter-stellar medium (ISM) and circum/inter-galactic medium (CGM/IGM) is a problem with crucial implications for a wide variety of questions in astrophysics, including star and planet and galaxy formation and evolution, astro-chemistry and chemo-dynamics, and space plasma physics \citep[for reviews, see][]{Zwei13,zweibel:cr.feedback.review,2018AdSpR..62.2731A,2019PrPNP.10903710K}. Most of the energy density in CRs (which determines their ability to ionize, heat, and interact with the gas) resides around $\sim$\,GeV energies, and in the range $\sim$\,MeV-TeV. At these energies, CR gyro radii ($r_{g,{\rm cr}} \sim 10^{9}-10^{15}\,{\rm cm}$) are vastly smaller than the characteristic scale-lengths of the galactic disk and the driving scales of ISM turbulence. As such, CRs cannot simply ``free stream'' out of the galaxy at speeds $\sim c$, but rather scatter in pitch angle from magnetic-field fluctuations, giving rise to some effective scattering rate $\nu_{\rm s}$. This in turn leads to bulk CR transport which can be parameterized by some effective diffusivity $\kappa_{\rm eff} \sim c^{2}/\nu_{\rm s}$ or streaming speed $v_{\rm st} \sim \kappa_{\rm eff} |\nabla P_{\rm cr}|/P_{\rm cr} \ll c$. 

These scattering rates have major implications for all of the astrophysical and space plasma fields above, and are probably the most important uncertainty in our understanding of the role of CRs. In understanding star and galaxy formation and the effect of CRs on the ISM/CGM/IGM, for example, a multitude of studies have shown that if the effective diffusivity or streaming speed is ``too small,'' CRs will be trapped in dense gas, and rapidly lose their energy to a variety of processes (e.g.\ pionic, catastrophic, synchrotron/inverse Compton losses) before they can have a significant effect on the gas properties. In the opposite limit, if the diffusivity is ``too large,'' CRs will free-stream rapidly out of the CGM and either effectively decouple from the gas or build up so little pressure that they will again have no effect. But, at diffusivities ``in between'' these values, $\sim$\,GeV CRs can have energy densities that are comparable to magnetic or thermal energy densities and have important effects on the gas \citep{Giri16,wiener:2017.cr.streaming.winds,Buts18,butsky:2020.cr.fx.thermal.instab.cgm,farber:decoupled.crs.in.neutral.gas,su:turb.crs.quench,hopkins:cr.mhd.fire2,hopkins:2020.cr.transport.model.fx.galform}. Likewise, the effect of CRs on astro-chemistry, planet formation, and dense gas systems depends sensitively on how effectively very low-energy ($\lesssim 100\,$MeV) CRs are trapped and their penetration depth into dense clouds and protostellar disks \citep{wolfire:1995.neutral.ism.phases,scalo:2004.turb.fx.review,indriolo:2009.high.cr.ionization.rate.clouds.alt.source.models,padovani:2009.cr.ionization.gmc.rates.model.w.alt.sources,thompson:2013.planet.form.dense.env.crs,lee:2020.hopkins.stars.planets.born.intense.rad.fields,bustard:2020.crs.multiphase.ism.accel.confinement,2020RSOS....701271P}. And details of CR plasma physics, in particular their micro-scale interactions with the multi-phase ISM/CGM/IGM gas, both shape and are determined by the CR scattering rates \citep[see e.g.][and references therein]{zank:2014.book,bai:2015.mhd.pic,bai:2019.cr.pic.streaming,lazarian:2016.cr.wave.damping,holcolmb.spitkovsky:saturation.gri.sims,zweibel:cr.feedback.review,thomas.pfrommer.18:alfven.reg.cr.transport,2019MNRAS.490.1156V}. 

The overwhelming majority of work studying and attempting to constrain CR transport, either in the Milky Way (MW) galaxy from observations in and around the Solar system (from e.g.\ terrestrial and space-based CR experiments) or from $\gamma$-ray observations, has focused on simple, phenomenological constraints. This generally involves parameterizing CR transport in terms of  effective diffusion coefficients or streaming speeds or other transport parameters \citep[see e.g.][and references therein]{blasi:cr.propagation.constraints,strong:2001.galprop,vladimirov:cr.highegy.diff,gaggero:2015.cr.diffusion.coefficient,2016ApJ...819...54G,2016ApJ...824...16J,cummings:2016.voyager.1.cr.spectra,2016PhRvD..94l3019K,evoli:dragon2.cr.prop}. These ``effective'' coefficients represent, by definition, some weighted average in the ISM between CR sources (e.g.\ SNe remnants, in the MW) and the Solar system, and are often parameterized as e.g.\ a power-law function of the CR rigidity $R_{\rm cr}$, such as $\kappa_{\rm eff} = \kappa_{0}\,\beta_{\rm cr}\,(R_{\rm cr}/R_{\rm cr,\,0})^{\delta_{\rm s}}$.

\begin{footnotesize}
\ctable[caption={{\normalsize Commonly-Used Variables in This Paper}\label{tbl:vars}},center,star
]{cl}{
}{
\hline\hline
$f_{\rm cr}$, $\mu$, $p_{\rm cr}$, $E_{\rm cr}$ & CR distribution function (DF) $f_{\rm cr} \equiv f_{\rm cr}({\bf x},\,{\bf p}_{\rm cr},\,t,\,s,\,...)$, pitch angle $\mu \equiv \hat{\bf p}_{\rm cr} \cdot \bhat$, momentum $p_{\rm cr}=|{\bf p}_{\rm cr}|$, energy $E_{\rm cr}$ \\ 
$\Omega_{\rm cr}$, $v_{\rm cr}$, $r_{g,{\rm cr}}$, $R_{\rm cr}$  & CR gyro-frequency $\Omega_{\rm cr}$, velocity $v_{\rm cr}=\beta_{\rm cr}\,c$, gyro-radius $r_{g,{\rm cr}} \equiv v_{\rm cr}/\Omega_{\rm cr}$, rigidity $R_{\rm cr}$ \\
$j_{\rm cr}(R_{\rm cr})$ & CR injection rate/spectrum as a function of rigidity $R_{\rm cr}$ (from SNe and other sources) \\
$e_{\pm}$, $e_{A}$ & Energy of forward(+) or backward(-) CR-scattering waves $e_{\pm} \equiv k_{\|} \mathcal{E}({\bf k}\cdot\bhat = \pm k_{\|})$ at wavenumber $\pm k_{\|}$, with $e_{A} \equiv e_{+} + e_{-}$ \\
$\nu_{\rm s,\,\pm}$, $\bar{\nu}_{\rm s,\,\pm}$ & CR scattering rate from forward/backward propagating waves $\nu_{\rm s,\,\pm}$, and pitch-angle averaged $\bar{\nu}_{\rm s,\,\pm}$ \\
$\delta_{\rm s}$ & Average dependence of CR scattering rate on rigidity, e.g.\ $\bar{\nu}_{\rm s} \propto \beta_{\rm cr}\,R_{\rm cr}^{-\delta_{\rm s}}$ (observationally-required $0.4 \lesssim \delta_{\rm s}\lesssim 0.7$) \\
${\bf k}$, $k_{\|}$, $k_{\bot}$ & Wavenumber ${\bf k}$ of CR-scattering modes, with parallel ($k_{\|}\equiv {\bf k}\cdot\bhat$) and perpendicular ($k_{\bot}$) components \\
$e_{\rm cr}^{\prime}$, $\epsilon_{\rm cr}$, $P_{\rm cr}^{\prime}$ & Differential CR energy density/pressure at a given momentum $p_{\rm cr}$, $e_{\rm cr}^{\prime} \equiv d e_{\rm cr} / d\ln{p_{\rm cr}}$, $\epsilon_{\rm cr}=(\gamma_{\rm cr}-1)\,e_{\rm cr}^{\prime}$, $P_{\rm cr}^{\prime} = \beta_{\rm cr}^{2}\,e_{\rm cr}^{\prime}/3$ \\ 
 ${\bf B}$, $\bhat$, $e_{\rm B}$, $\delta{\bf B}$ & Magnetic field ${\bf B}$, direction $\bhat\equiv {\bf B}/|{\bf B}|$, energy $e_{\rm B}\equiv |{\bf B}|^{2}/8\pi$, fluctuations $\delta{\bf B}$ on scale $\sim k$ \\
$v_{A,\,{\rm ideal}}$, $v_{A,\,{\rm eff}}$ & Ideal-MHD \Alf\ speed $v_{A,\,{\rm ideal}} \equiv (|{\bf B}|^{2}/4\pi\rho)^{1/2}$, speed of gyro-resonant \Alf\ waves $v_{A,\,{\rm eff}}$ (Eq.~\ref{eqn:va.eff}) \\
$\ell_{A}$ & \Alf\ scale of large-scale turbulence (scale where $\langle |\delta{\bf v}_{\rm turb}(k \sim 1/\ell_{A})|^{2}\rangle^{1/2} \sim v_{A,\,{\rm ideal}}$) \\
\hline
$S_{\pm}$ & Source terms for CR scattering modes ($D_{t} e_{\pm} + ... = S_{\pm}$) \\
$Q_{\pm}$, $\Gamma_{\pm}$ & Damping terms for CR scattering modes ($D_{t} e_{\pm} + ... = -Q_{\pm} \equiv - \Gamma_{\pm}\,e_{\pm}$) \\
\hline\hline
}
\end{footnotesize}

\begin{footnotesize}
\ctable[caption={{\normalsize Parameterization of Generalized Damping/Driving Rates for CR-Scattering Modes}\label{tbl:coeffs}},center,star
]{clccc}{
}{
\hline\hline
$\xi_{k}$, $\xi_{A}$, $\xi_{\rm cr}$ & Coefficients for damping rates: $Q_{\pm} \equiv \Gamma_{\pm} \,e_{\pm}$ 
with $\Gamma_{\pm} \propto k_{\|}^{\xi_{k}}\,e_{\pm}^{\xi_{A}}\,\epsilon_{\rm cr}^{\xi_{\rm cr}}$ & $\xi_{k}$ & $\xi_{A}$ & $\xi_{\rm cr}$ \\
\hline
\, &  Values explored in our simulation survey & $0 \le \xi_{k} \le 2$ & $0 \le \xi_{A} \le 1$ & $0 \le \xi_{\rm cr} \le 1$ \\
\hline
$X_{\rm in}$ & Quantities (e.g.\ $Q$, $\Gamma$, $\xi$) for ion-neutral damping & 0 & 0 & 0 \\
$X_{\rm dust}$ & Quantities for dust damping & $0.5 \rightarrow 0.75$ & 0 & 0 \\
$X_{\rm nll}$ & Quantities for non-linear Landau damping & 1 & 1 & 0 \\
$X_{\rm turb/LL}$ & Quantities for linear Landau or ``turbulent'' damping & $0.4 \rightarrow 0.5$ & 0 & 0 \\
$X_{\rm new,\, damp}$ & Quantities for proposed novel damping that could fit observations & $\sim 0.1 \rightarrow 0.4$ & $\sim 0$ & $\sim 1$ \\
\hline\hline\hline
$\zeta_{k}$, $\zeta_{A}$, $\zeta_{\rm cr}$ & Coefficients for driving/source rates: $S_{\pm} \propto k_{\|}^{\zeta_{k}}\,e_{\pm}^{\zeta_{A}}\,\epsilon_{\rm cr}^{\zeta_{\rm cr}}$ & $\zeta_{k}$ & $\zeta_{A}$ & $\zeta_{\rm cr}$ \\
\hline
\, & Values explored in our simulation survey & $-2 \lesssim \zeta_{k} \lesssim 2$ &  $0 \le \zeta_{A} \le 1$ & $0 \le \zeta_{\rm cr} \le 1$ \\
\hline
$X_{\rm sc}$ & Quantities (e.g.\ $S$, $\zeta$) for self-confinement (SC) driving (non-steady-state $\zeta$ in [])  & 0 [1] & 0 [1] & 1 [1] \\
$X_{\rm et}$ & Quantities for extrinsic turbulence (ET) driving (with anisotropy/damping) & $\lesssim -1+\xi_{k}^{\rm turb}$ &  $\sim0$ & 0 \\
$X_{\rm new,\, lin}$ & Quantities for proposed novel linear source terms which could fit observations 
& $\sim 0.6 \rightarrow 0.9$ & $\sim 1$ & $\sim 0$ \\
$X_{\rm new,\, ext}$ & Quantities for proposed novel extrinsic source terms which could fit observations 
& $\sim -0.25 \rightarrow -0.1$ & $\sim 0$ & $\sim 0$ \\
\hline\hline
}
\end{footnotesize}

But Solar system constraints only measure CR transport in an average sense at one point in space and time, while ISM properties -- both along the CR ``path'' and in different galaxies and cosmic epochs -- vary tremendously in both space and time (by many orders of magnitude for quantities of interest like magnetic energy density). Further,  phenomenological models do not explain how such coefficients arise in the first place. What is therefore  required is  a {\em physical} model of CR transport that can reproduce these effective constraints and be tested in other regimes. However, this is particularly challenging at the MeV-TeV CR energies of greatest interest, because (1) the observational constraints are limited, (2) the extremely small gyro radii are much smaller than spatially-resolvable scales in most astrophysical ISM studies, (3) the ``back reaction'' of magnetic fields and  gas from CRs, e.g.\ via gyro-resonant instabilities and macroscopic CR ``pressure'' effects, is maximized around this energy scale, and (4) the ISM, CGM, and IGM phase structure and turbulence  itself remains uncertain.

Broadly speaking, historical models that  attempt to {\em predict} CR scattering rates and transport parameters  at these energies   fall into one of two broad categories: ``extrinsic turbulence'' (ET) and ``self-confinement'' (SC) models. In the simplest ET models, going back to e.g.\ \citet{jokipii:1966.cr.propagation.random.bfield,wentzel:1968.mhd.wave.cr.coupling,1975MNRAS.172..557S,1975RvGSP..13..547V}, CRs scatter from gyro-resonant fluctuations in ${\bf B}$, i.e.\ those with wavenumbers $k_{\|} \equiv {\bf k}\cdot \bhat \sim 1/r_{g,{\rm cr}}$. Those early models assumed $\delta{\bf B}(k_{\|})$ was sourced by an isotropic, undamped, inertial-range \citet{kolmogorov:turbulence}-type (K41) cascade from larger ISM scales. This gives rise to a scattering rate $\nu_{\rm s} \sim \Omega_{\rm cr}\,|\delta{\bf B}(k_{\|})|^{2}/|{\bf B}|^{2} \propto |{\bf B}|^{1/3}\,\ell_{A}^{2/3}\,R_{\rm cr}^{-1/3}$ where $\ell_{A}$ is the \Alf\ scale of the cascade.\footnote{We define the ``\Alf\ scale'' $\ell_{A}$ of any large-scale turbulent cascade as the scale where, extrapolating the inertial range, $\langle |\delta{\bf v}_{\rm turb}(k \sim 1/\ell_{A})| \rangle \approx v_{A,\,{\rm ideal}}$ (the ideal \Alf\ speed).}
In SC models, going back to \citet{wentzel.1969.streaming.instability,skilling:1971.cr.diffusion,holman:1979.cr.streaming.speed}, CRs themselves source the scattering modes, which they excite  via various instabilities as they stream down magnetic field lines \citep{wentzel:1968.mhd.wave.cr.coupling,kulsrud.1969:streaming.instability}. The instabilities grow until reaching some saturation amplitude that is determined by a wave damping rate $\Gamma$, thus giving rise to scattering rates that scale as $\nu_{\rm s} \sim \Omega_{\rm cr}\,v_{A}\,|\nabla P_{\rm cr}|/(\Gamma\,|{\bf B}|^{2})$ \citep{1975MNRAS.173..255S}. 

Until recently, it has not been possible to directly test and compare these models with local CR observations, for a variety of reasons.  Perhaps most importantly,  even in the simplest ET and SC models, scattering rates are not constant but depend strongly on ISM properties. These in turn vary dramatically across the ISM by as much as $\sim 10$ orders of magnitude, in a manner that cannot be captured by simplified  models discussed above that assume some steady-state CR distribution and solve e.g.\ a ``leaky-box'' or ``flat-halo diffusion'' model with a simple analytic galaxy model \citep[see the review in][]{hopkins:cr.transport.constraints.from.galaxies}. Moreover, only recently has the fluid theory of CRs been developed to the point where self-confinement  theories can be ``coarse-grained'' self consistently into fluid-like MHD-CR transport and scattering models \citep{Zwei13,zweibel:cr.feedback.review,thomas.pfrommer.18:alfven.reg.cr.transport,hopkins:m1.cr.closure}, while modern versions of SC and ET models that account for important effects such as damping and anisotropy have only been developed in the last two decades \citep{chandran00,yan.lazarian.02,yan.lazarian.04:cr.scattering.fast.modes,yan.lazarian.2008:cr.propagation.with.streaming,farmer.goldreich.04,zweibel:cr.feedback.review,squire:2021.dust.cr.confinement.damping}. Finally, only recently has CR data become available from outside of the heliopause, which is crucial for the CRs of greatest interest ($\lesssim 100\,$GeV energies) because these are strongly modulated by the Sun \citep{cummings:2016.voyager.1.cr.spectra,2017AdSpR..60..865B,bisschoff:2019.lism.cr.spectra}. These new observations help to remove the order-of-magnitude degeneracies that plagued previous attempts to test CR transport/scattering theories. 

In this paper, we therefore revisit the question of whether or not state-of-the-art ET or SC models can possibly explain the state-of-the-art CR observations. We first consider the problem in a purely analytic fashion, synthesizing CR transport theories (beginning from  general considerations before considering approximations such as steady-state behavior) and reviewing the state of the art in both SC and ET theories in order  to treat all potentially important damping terms. We then test these models in even greater detail with fully non-equilibrium, non-linear, non-steady state  CR transport  in high-resolution galaxy simulations, which explicitly resolve the plasma properties that determine CR scattering. While a first attempt at such comparisons was presented in \citet{hopkins:cr.transport.constraints.from.galaxies}, which already argued that present ET and SC models failed to reproduce the observations, that paper simplified by considering a ``single-bin'' CR approximation, essentially modeling only CR protons in a narrow range of  energies at $\sim 1\,$GeV. Here we expand this to a full spectrum of CRs with a wide range of secondary species. This dramatically expands the range of observational constraints and will allow us to show that the scope of the discrepancy between SC and ET models and observations is much larger than previously believed. In particular, some of the  possible resolutions to the discrepancies noted in \citet{hopkins:cr.transport.constraints.from.galaxies} -- e.g.\ changing the normalization of SC-induced scattering rates by accounting for certain pitch-angle effects --  cannot possibly provide the full solution. We use these constraints to propose that a new class of sources for gyro-resonant scattering waves is required, which obeys a well-constrained (but plausible) set of requirements.

In \S~\ref{sec:analytic} we set up the analytic background, including review of some key definitions (\S~\ref{sec:definitions}) and description of the CR dynamics equations (\S~\ref{sec:eqns}), relevant \Alf\ wave properties (\S~\ref{sec:alfven}), and expressions for scattering rates (\S~\ref{sec:scattering}). We then review standard damping mechanisms (\S~\ref{sec:damping}) and drivers of scattering fluctuations in both SC (\S~\ref{sec:sc}) and ET (\S~\ref{sec:et}) limits, and the resulting steady-state behaviors (\S~\ref{sec:steady.state}). In \S~\ref{sec:problems} we discuss the problems that follow: first we review what empirical CR transport models require (\S~\ref{sec:problems:obs}) then describe how both ET (\S~\ref{sec:problems:et}) and SC (\S~\ref{sec:problems:sc}) models cannot satisfy these constraints, then propose phenomenological solutions (\S~\ref{sec:rescue}) involving either modified damping (\S~\ref{sec:rescue:damp}) or driving (\S~\ref{sec:rescue:drive}) terms. We then proceed to explore these in detailed simulations. \S~\ref{sec:methods} describes the numerical methods, outlining the non-CR (\S~\ref{sec:methods:overview}) and CR (\S~\ref{sec:methods:crs}) physics simulated, a theoretically motivated ``reference'' model (\S~\ref{sec:reference.model}), and extensive variations to that model that we have considered (\S~\ref{sec:methods:variations}). \S~\ref{sec:results} describes the results of these simulations, first (\S~\ref{sec:results:confirmation}) confirming the analytically-predicted ``failure modes'' of SC (\S~\ref{sec:results:confirmation:sc}) and ET (\S~\ref{sec:results:confirmation:et}) models, then testing the proposed alternative damping (\S~\ref{sec:results:damping}) or driving (\S~\ref{sec:results:sources}) scalings to see if these can reproduce observations. We summarize and conclude in \S~\ref{sec:discussion}. Appendices~\ref{sec:sc.equilibrium.models}-\ref{sec:turb.review} contain more detailed analytic derivations of steady-state CR behaviors and turbulent scalings.

\section{Analytic Background}
\label{sec:analytic}

\subsection{Key Scales and Definitions}
\label{sec:definitions}

To begin, we review some important concepts. Table~\ref{tbl:vars} collects definitions of some of the most-commonly used variables in this paper. Per \S~\ref{sec:intro}, CRs with some rigidity $R_{\rm cr}$ and corresponding gyro radius $r_{g,{\rm cr}}=R_{\rm cr}/|{\bf B}|$ ($\sim 10^{-6}\,{\rm pc}$ in the diffuse ISM, for CRs with $R_{\rm cr} \sim 1\,$GV) are scattered in pitch angle $\mu$ by fluctuations in the magnetic field $\delta{\bf B}$ with some effective scattering rate $\nu_{\rm s}$. In most models (though not all, as we discuss below), the CR scattering rate is strongly dominated by gyro-resonant scattering of CRs from \Alf\ waves with parallel wavenumbers $k_{\|}\equiv{\bf k}\cdot \bhat \sim 1/r_{g,{\rm cr}}$. The power in these modes ($e_{A} \sim \langle |\delta{\bf B}(k_{\|}\sim 1/r_{g,{\rm cr}})|^{2} \rangle/8\pi$), which determines $\nu_{\rm s}$, is set by competition between some source/driving terms $S$ and damping or dissipation rates $Q = \Gamma\,e_{A}$. 

We stress that this encompasses both SC and ET models: the difference comes down to which dominates $S$. In SC models, $S$ is sourced by parallel \Alf\ waves excited directly by CRs (via e.g.\ gyro-resonant and streaming instabilities), which we denote $S_{\rm sc}$. In ET models, the dominant contribution to $S$ comes from a turbulent cascade $S_{\rm et}$ operating over a large dynamic range in scale. Also note that by definition $\Gamma$ includes any terms which remove power from the scattering modes, e.g.\ both traditional collisional damping but also processes which transfer energy to other modes with different wavenumbers or weaker scattering effects.

We will show that it is useful to parameterize $S$ and $\Gamma$ in terms of their approximate scaling with parallel wavenumber ($k_{\|}$), total kinetic energy density of CRs around a given rigidity ($\epsilon_{\rm cr}$), and energy in scattering modes at some $k_{\|}$ ($e_{A}$ or $e_{\pm}$), as $S \propto k_{\|}^{\zeta_{k}}\,e_{A}^{\zeta_{A}}\,\epsilon_{\rm cr}^{\zeta_{\rm cr}}$ and $\Gamma \propto k_{\|}^{\xi_{k}}\,e_{A}^{\xi_{A}}\,\epsilon_{\rm cr}^{\xi_{\rm cr}}$ shown in Table~\ref{tbl:coeffs}.\footnote{Note that $k_{\|}$, $\epsilon_{\rm cr}$, and $e_{A}$ do not need to be strictly independent variables for this parameterization.} The key qualitative problems and failure modes of SC and ET theories can be encapsulated entirely in these coefficients ($\zeta_{k}$, $\zeta_{A}$, $\zeta_{\rm cr}$, $\xi_{k}$, $\xi_{A}$, $\xi_{\rm cr}$). Essentially, we will show that whether or not a theory of CR scattering can potentially reproduce CR observations (independent of normalization parameters) depends on these few numbers.

It is also helpful to recall some key scales in turbulence. Most of the power in ISM/CGM turbulence is on the driving scale, typically $\gtrsim 0.1-1\,$kpc (on which scale the turbulence is often trans or super-\Alf{ic}, $\langle |\delta {\bf v}_{\rm turb}(k)|^{2} \rangle^{1/2} \gtrsim v_{A,\,{\rm ideal}}$). Below the \Alf\ scale $\ell_{A}$ (typically $\sim 10-100\,$pc in the ISM), the turbulent fluctuations are sub-\Alf{ic} ($\langle |\delta {\bf v}_{\rm turb}(k\gtrsim1/\ell_{A})|^{2} \rangle^{1/2} \lesssim v_{A,\,{\rm ideal}}$). At the vastly-smaller gyro scale $r_{g,{\rm cr}} \ll \ell_{A}$, the scattering fluctuations are fractionally small (quasi-linear; $|\delta {\bf B}(k \sim 1/r_{g,{\rm cr}})|/|{\bf B}| \ll 1$), and we can treat fluctuations (approximately) as a superposition of \Alf, slow and fast magnetosonic modes. 

\Alf\ modes are only weakly-damped down to scales much smaller than CR gyro-resonant scales (at least down to ion gyro radii). When we refer below to ``damping'' terms acting directly on the CR scattering modes ($Q_{\pm}$ and $\Gamma_{\pm}$), we generally are referring to this ``weak'' damping. Specifically, \Alf-mode-damping times ($\sim \Gamma^{-1}$) at some $k$ are much longer than the mode-crossing times $\sim 1/(k\,v_{A,\,{\rm eff}})$ (by typical factors $\sim 10^{4}-10^{8}$). However, as we discuss below (and in more detail in Appendix~\ref{sec:turb.review}), it is well-established that an \Alf{ic} (or slow magnetosonic) cascade must be highly-anisotropic on scales below the \Alf\ scale $\ell_{A}$ ($k \gg 1/\ell_{A}$): an isotropic \citet{iroshnikov:1963.ik.aniso.turb,kraichnan:1965.ik.aniso.turb}-type (IK) cascade, for example, simply cannot exist (it is not mathematically self-consistent) on scales $r_{g,{\rm cr}} \ll \ell_{A}$. This means it is crucial to distinguish between parallel $k_{\|}$ and perpendicular components of ${\bf k}$. 

Fast magnetosonic modes, on the other hand, are orders-of-magnitude more strongly damped on small scales by both collisional and collisionless/Landau damping (see Appendix~\ref{sec:turb.review}). On scales below the dissipation scale $k_{\rm diss} \sim 1/\ell_{\rm diss}$ (with typical $\ell_{\rm diss} \gtrsim 0.001$\,pc in the ISM), the magnetosonic mode damping time $\Gamma_{\rm magnetosonic}^{-1}$ becomes shorter than turbulent ``cascade'' or decoherence or energy-transfer timescale $\tau_{\rm cas}$ (with e.g.\ $\tau_{\rm cas}^{-1}(k) \sim k\,\langle |\delta{\bf v}_{\rm turb}^{2}(k)|\rangle^{1/2}$ in the classical K41 picture), so the cascade must be truncated or strongly modified by the energy losses. For essentially all plausible ISM/CGM conditions, the gyro-resonant scales are much smaller than the dissipation scale ($r_{g,{\rm cr}} \ll \ell_{\rm diss}$) at rigidities $\lesssim 100-1000\,$GV, so one cannot simply extrapolate an un-damped inertial-range magnetosonic cascade of any form (let alone K41) down to gyro-resonant scales.

One additional clarification is important. To be consistent with the previous literature, when we refer to the ``turbulent damping'' of  gyro-resonant modes, $\Gamma_{\rm turb}$ (see \S~\ref{sec:damping}), we refer specifically to a process by which interactions between gyro-resonant scattering modes and other turbulent modes transfer energy from the weakly-damped gyro-resonant scattering modes (with $k_{\|} \sim 1/r_{g,{\rm cr}}$) to either higher-$k$ or more strongly-damped modes. This is different from ``damping or dissipation of turbulence,'' which we will use to refer to the phenomena described above for e.g.\ fast magnetosonic modes, in which a turbulent cascade is strongly modified by sufficiently-strong damping on some dissipation scale $\ell_{\rm diss}$ larger than the gyro-resonant scales.

\subsection{Cosmic Ray Dynamics Equations}
\label{sec:eqns}

Consider an arbitrary CR distribution function (DF) $f_{\rm cr}=f_{\rm cr}({\bf x},\,{\bf p}_{\rm cr},\,t,\,s_{\rm cr},\,...)$ as a function of position ${\bf x}$, CR momentum ${\bf p}_{\rm cr}$, time $t$, and CR species $s_{\rm cr}$, on macroscopic scales much larger than CR gyro-radii. Assuming the DF is approximately gyrotropic, and the background gas velocities ${\bf u}_{\rm gas}$ are non-relativistic, the general Vlasov equation for $f_{\rm cr}$ can be written to leading order in $\mathcal{O}(|{\bf u}_{\rm gas}|/c)$ as the usual focused transport equation \citep{skilling:1971.cr.diffusion,1975MNRAS.172..557S,1997JGR...102.4719I,2001GeoRL..28.3831L,2005ApJ...626.1116L,zank:2014.book,2015ApJ...801..112L}, with the standard quasi-linear theory slab scalings for the scattering terms from \citet{schlickeiser:89.cr.transport.scattering.eqns}. 
As shown in \citet{hopkins:m1.cr.closure}, taking the zeroth and first pitch-angle ($\mu$) moments of that equation (retaining all terms to leading order in $\mathcal{O}(|{\bf u}_{\rm gas}|/c)$) gives the evolution equations for the isotropic part of the DF $\bar{f}_{{\rm cr},0}$ (i.e.\ the CR number density at a given differential $p_{\rm cr}=|{\bf p}_{\rm cr}|$) and its flux:\footnote{As shown in \S~\ref{sec:steady.state} below, Eq.~\ref{eqn:f0} reduces to the somewhat more familiar anisotropic ``streaming+diffusion'' equation for $\bar{f}_{{\rm cr},0}$ if one assumes $D_{t} \bar{f}_{{\rm cr},1}$ is small (i.e.\ the flux is in ``local steady-state'').}
\begin{align}
\label{eqn:f0} 
D_{t} \bar{f}_{\rm cr,\,0}& +  \nabla \cdot  (v_{\rm cr}\,\bhat\,\bar{f}_{\rm cr,\,1}) =  {j_{\rm cr,\,0}} +  \\
\nonumber &\ \  \frac{1}{p_{\rm cr}^{2}}\frac{\partial }{\partial p_{\rm cr}}
\bigg [ p_{\rm cr}^{2}\,
\bigg \{ \mathcal{R}_{\rm loss}\,\bar{f}_{\rm cr,\,0}  
+ \left( \mathbb{D}_{\rm cr}:\nabla{\bf u}_{\rm gas} \right)\,p_{\rm cr}\,\bar{f}_{\rm cr,\,0} + \\
\nonumber & \ \ \  \ \ \ \ \ \ \ \  \ \ \ \ \ \ \ \ \   \ \ \ \ \ \ \ \ \ \tilde{D}_{p \mu}\,\bar{f}_{\rm cr,\,1}
+ \tilde{D}_{p p}\, \frac{\partial \bar{f}_{\rm cr,\,0}}{\partial p_{\rm cr}} 
\bigg \} \bigg ]  \\
\label{eqn:f1} 
D_{t} \bar{f}_{\rm cr,\,1} &+  
\bhat\cdot \left[ \nabla\cdot \left( v_{\rm cr}\,\mathbb{D}_{\rm cr}\,\bar{f}_{\rm cr,\,0} \right) \right] = - \left[ \tilde{D}_{\mu\mu}\,\bar{f}_{\rm cr,\,1} + \tilde{D}_{\mu p}\,\frac{\partial \bar{f}_{\rm cr,\,0}}{\partial p_{\rm cr}} \right] \\
\nonumber
\tilde{D}_{p p}  = &\chi\,\frac{p_{\rm cr}^{2}\,v_{A}^{2}}{v_{\rm cr}^{2}}\,\bar{\nu}_{\rm s} 
\  , \ 
 \tilde{D}_{p \mu} = \frac{p_{\rm cr}\,\bar{v}_{A}}{v_{\rm cr}}\,\bar{\nu}_{\rm s} 
\  , \
\tilde{D}_{\mu\mu} = \bar{\nu}_{\rm s} 
\  , \ 
\tilde{D}_{\mu p} = \chi\,\frac{p_{\rm cr}\,\bar{v}_{A}}{v_{\rm cr}}\,\bar{\nu}_{\rm s}
\end{align}
where $\bar{f}_{\rm cr,\,n} \equiv \langle \mu^{n}\, f_{\rm cr} \rangle_{\mu}$ is the $n$'th pitch-angle moment (so e.g.\ $\bar{f}_{\rm cr,\,0}$ is the isotropic part of the DF, and $\bar{f}_{\rm cr,\,1} = \langle \mu \rangle\,\bar{f}_{\rm cr,\,0}$). In Eq.~\ref{eqn:f0}-\ref{eqn:f1}, $D_{t} X \equiv \partial_{t}  X + \nabla \cdot({\bf u}_{\rm gas}\,X) \equiv \rho\,{\rm d}_{t}(X/\rho)$ is the conservative co-moving derivative (with $\rho$ the gas mass density) , $v_{\rm cr}=\beta_{\rm cr}\,c$ is the CR velocity, $p=\gamma_{\rm cr}\,\beta_{\rm cr}\,m_{\rm cr}\,c$ the CR momentum, $\bhat \equiv {\bf B}/|{\bf B}|$ the unit magnetic field vector, $j_{\rm cr}$ represents injection \&\ catastrophic losses, $\mathcal{R}_{\rm loss}$ represents continuous loss processes, $v_{A}$ is the \Alf\ speed, the coefficients $\tilde{D}$ are defined in terms of the scattering rate $\bar{\nu}_{\rm s} \equiv \bar{\nu}_{\rm s,+} + \bar{\nu}_{\rm s,-}$ (the scattering contributed by forward-and-backward propagating modes with respect to $\bhat$), the signed $\bar{v}_{A} \equiv v_{A}\,(\bar{\nu}_{\rm s,+} - \bar{\nu}_{\rm s,-})/(\bar{\nu}_{\rm s,+} + \bar{\nu}_{\rm s,-})$, and the Eddington tensor $\mathbb{D}_{\rm cr} \equiv \chi\,\mathbb{I} + ( 1-3\,\chi )\,\bhat\bhat$ and scattering terms are defined in terms of $\chi \equiv (1-\langle \mu^{2} \rangle)/2 = (1-\bar{f}_{\rm cr,\,2}/\bar{f}_{\rm cr,\,0})/2$.

Integrating over an infinitesimal range in momentum for a CR group or ``packet,'' this can be further transformed into the differential CR energy equation, which will be useful below:
\begin{align}
\label{eqn:specific.cr.energy} D_{t} e_{\rm cr}^{\prime}  + \nabla \cdot \left( {F_{e,\,{\rm cr}}^{\prime}}\,\bhat \right) &= \tilde{\mathcal{S}}^{\prime}_{\rm sc} - \mathbb{P}_{\rm cr}^{\prime}:\nabla{\bf u}_{\rm gas} + \mathcal{S}_{\rm other,\,{\rm cr}}^{\prime}  \\ 
\nonumber D_{t} {F_{e,\,{\rm cr}}^{\prime}}  + c^{2}\,\bhat\cdot \left( \nabla \cdot \mathbb{P}_{\rm cr}^{\prime} \right) &= 
-{\bar{\nu}_{\rm s}}\,\left[ F_{e,\,{\rm cr}}^{\prime} - 3\,\chi\,\bar{v}_{A}\,(e_{\rm cr}^{\prime}+P_{\rm cr}^{\prime}) \right] 
\end{align}
where $e_{\rm cr}^{\prime} \equiv d e_{\rm cr}/d\ln{p_{\rm cr}} = p_{\rm cr}^{3}\,\int d\mu\,\int d\phi\,E(p_{\rm cr})\,f$ is the total CR energy in a differential range of momentum $p_{\rm cr}$, $F_{e,\,{\rm cr}}^{\prime} \equiv d F_{e,\,{\rm cr}}/d\ln{p_{\rm cr}} = p_{\rm cr}^{3}\,\int d\mu\,\int d\phi\,E(p_{\rm cr})\,v\,\mu\,f$ is its flux, $\mathcal{S}_{\rm other,\,{\rm cr}}^{\prime}$ collects any other arbitrary sources/sinks (e.g.\ catastrophic losses, injection at shocks, etc.), $\mathbb{P}_{\rm cr}^{\prime} \equiv 3\,P_{\rm cr}^{\prime}\,\mathbb{D}_{\rm cr}$ (with $P_{\rm cr}^{\prime} \equiv \beta_{\rm cr}^{2}\,e_{\rm cr}^{\prime}/3$), and scattering gives rise to the energy loss/gain term\footnote{As shown in \citet{hopkins:m1.cr.closure} the $\tilde{D}_{\mu p}$ and $\tilde{D}_{pp}$ terms, and  corresponding dependence on $\partial/\partial p_{\rm cr}$, do appear implicitly to leading-order in $\mathcal{O}(u/c)$ in Eq.~\ref{eqn:specific.cr.energy} in the $\tilde{\mathcal{S}}_{\rm sc}^{\prime}$ term, contributing to the ``streaming loss'' and ``diffusive reacceleration'' portions of $\tilde{\mathcal{S}}_{\rm sc}^{\prime}$, respectively.} $\tilde{\mathcal{S}}_{\rm sc}^{\prime} \equiv -(\bar{\nu}_{\rm s}/c^{2})\,\left[ \bar{v}_{A}\,F^{\prime}_{e,\,{\rm cr}} - 3\,\chi\,{v}^{2}_{A}\,\left( e_{\rm cr}^{\prime} + P_{\rm cr}^{\prime} \right)  \right]$.

\subsection{Which \Alf\ Speed?}
\label{sec:alfven}

In partially-ionized gas, the \Alf\ speed $v_{A}$ is not wavelength-independent. The correct \Alf\ speed in the CR dynamics equations should be \Alf\ speed of gyro-resonant \Alf\ waves, as the original derivation of the relevant scattering terms takes $v_{A} = v_{A,\,{\rm eff}} \equiv {\rm Real}(\omega_{A}/k_{\|})$ with $k_{\|}=1/\mu\,r_{g,{\rm cr}}$ (for gyro-radius $r_{g,{\rm cr}}$; see e.g.\  \citealt{1975MNRAS.172..557S}). In the MHD limit (assuming e.g.\ non-relativistic electron gyro-radii are vanishingly small), the relevant  dispersion relation  is:
\begin{align}
\label{eqn:va.dispersion} \omega_{A}^{3} + i\,\omega_{A}^{2}\,\nu_{\rm in}\,(1+\psi_{\rm ion}) - k_{\|}^{2}\,v_{A,\,{\rm ion}}^{2}\,\omega_{A} - i\,k_{\|}^{2}\,v_{A,\,{\rm ion}}^{2}\,\psi_{\rm ion}\,\nu_{\rm in} = 0,
\end{align}
where $\nu_{\rm in}$ is the ion-neutral collision frequency (this is distinct from $\nu_{\rm ni} \equiv \psi_{\rm ion}\,\nu_{\rm in}$), $v_{A,\,{\rm ion}} \equiv (B^{2}/4\pi\rho_{\rm ion})^{1/2}$, and $\psi_{\rm ion} \equiv \rho_{\rm ion}/\rho_{\rm neutral} \equiv f_{\rm ion} / (1-f_{\rm ion})$, in terms of the ion and neutral mass densities $\rho_{\rm ion}$, $\rho_{\rm neutral}$ ($\rho = \rho_{\rm ion}+\rho_{\rm neutral}$). 
The exact solutions to this are quite cumbersome, but the real part of interest can be well approximated in all relevant limits by:
\begin{align}
\label{eqn:va.eff} v_{A,\,{\rm eff}}^{2} &\approx v_{A,\,{\rm ideal}}^{2}\,\left[ 1 +  \frac{1}{\psi_{\rm ion}\,[1 + \psi_{\rm ion}\,(\psi_{\rm ion}+1/4)\,\tilde{\psi}_{\rm in}^{2}]} \right] \\
\nonumber \tilde{\psi}_{\rm in} &\equiv \frac{\nu_{\rm in}}{k_{\|}\,v_{A,\,{\rm ideal}}} \approx 0.01\,(1-f_{\rm ion})\,\rho_{-24}^{3/2}\,R_{\rm GV}\,T_{1000}^{1/2}\,B_{\rm \mu G}^{-2}\,(k_{\|}\,r_{g,{\rm cr}})^{-1}
\end{align}
where $v_{A,\,{\rm ideal}} \equiv (B^{2} / 4\pi\,\rho)^{1/2}$, $f_{\rm ion}$ is the ionized fraction,  $T_{1000} \equiv T/1000\,{\rm K}$, $B_{\rm \mu G} \equiv |{\bf B}|/{\rm \mu G}$, $\rho_{-24} \equiv \rho/10^{-24}\,{\rm g\,cm^{-3}}$, and $R_{\rm GV}\equiv R_{\rm cr}/{\rm GV}$.
This essentially interpolates between the ``ideal MHD \Alf\ speed'' $v_{A,\,{\rm ideal}}$ for long-wavelength modes with frequencies $\omega_{A}$ much lower than the ion-neutral collision frequency $\nu_{\rm in}$, and the ``ion \Alf\ speed'' $v_{A,\,{\rm ion}}$ for short-wavelength modes with $\omega_{A} \gg \nu_{\rm in}$. For most ISM conditions at the (short) gyro-resonant wavelengths of interest for CR dynamics, $\omega_{A} \gg \nu_{\rm in}$.

\subsection{Scattering Rates of CRs from a Population of Magnetic Fluctuations}
\label{sec:scattering}

Everything needed to evolve the CRs in Eq.~\ref{eqn:f0}-\ref{eqn:f1} is determined by the local plasma properties, except for the scattering rates $\bar{\nu}_{\rm s,\pm}$, which crucially determine how CRs propagate. 
Following \citet{Zwei13,zweibel:cr.feedback.review}, quasi-linear theory gives the scattering coefficients: 
\begin{align}
\label{eqn:nu} \nu_{\rm s,\pm}(p_{\rm cr},\,\mu) = \frac{\pi}{4}\,\Omega_{\rm cr}\,\frac{k_{\|}\,\mathcal{E}_{\pm}(k_{\|})}{e_{\rm B}} \rightarrow \bar{\nu}_{\rm s,\pm} \equiv \frac{\pi}{4}\,\hat{\nu}_{\rm s}\,\Omega_{\rm cr}\,\frac{e_{\pm}}{e_{\rm B}}
\end{align}
where $e_{\rm B}\equiv |{\bf B}|^{2}/8\pi$, $k_{\|}=\Omega_{\rm cr}/(\mu\,v_{\rm cr})$ from the gyro-resonant condition, and $e_{\pm} \equiv k_{\|}\,\mathcal{E}_{\pm}({\bf k} \cdot \bhat = \pm k_{\|})$ is the energy of scattering waves at parallel wavenumber $k_{\|}$ (it is important here that we distinguish $k_{\|} \equiv |{\bf k} \cdot \bhat|$ from $k=|{\bf k}|$). We parameterize our ignorance of the pitch-angle dependence with $\bar{\nu}_{\rm s,\pm} \equiv (\pi/4)\,\hat{\nu}_{\rm s}\,\Omega_{\rm cr}\,e_{\pm}/e_{\rm B}$, where $\hat{\nu}_{\rm s}\,e_{\pm}$ reflects the angle-averaged energy of wave-packets that    interact significantly with the relevant CRs (with $\hat{\nu}_{\rm s}$ a dimensionless order-unity constant coming from the integration over pitch angle), traveling in either the forward ($+$) or backward ($-$) direction along $\bhat$. As shown in \citet{Zwei13,zweibel:cr.feedback.review} and later in \citet{thomas.pfrommer.18:alfven.reg.cr.transport}, one can write a fluid equation for the wavepackets: 
\begin{align}
\label{eqn:ea} D_{t} e_{\pm} &+ \nabla \cdot \left( {v}_{A,\,\pm}\,e_{\pm}\,\bhat \right) = -\frac{e_{\pm}}{2}\,\nabla \cdot {\bf u}_{\rm gas} + S_{\pm}- Q_{\pm}
\end{align}
where one can think of $e_{\pm}/2$ as the ``pressure'' or ``PdV'' term (with $\nabla \cdot {\bf u}_{\rm gas}$ being the change of comoving volume), we define ${v}_{A,\,\pm} = \pm v_{A,\,{\rm eff}}$ corresponding to the $e_{\pm}$ sign, and $S_{\pm}$ and $Q_{\pm}$ correspond to source and damping terms. 
We can write 
\begin{align}
\label{eqn:scr.decomposition} S_{\pm} & \equiv S_{\rm sc,\,\pm} + S_{\rm et,\,\pm} + S_{\rm new,\,\pm} \\ 
\nonumber Q_{\pm} &\equiv \Gamma_{\pm}\,e_{\pm} = \left[ \Gamma_{\rm in} + \Gamma_{\rm dust} + \Gamma_{\rm turb/LL} + \Gamma_{\rm nll,\,\pm} + \Gamma_{\rm new,\,damp,\,\pm} \right]\,e_{\pm}
\end{align}
Here $S_{\rm sc,\,\pm}$ corresponds to energy transfer from the CRs themselves as they scatter off the waves, $S_{\rm et,\,\pm}$ corresponds to ``extrinsic turbulence'' driving (defined below), and $S_{\rm new,\,\pm}$ corresponds to some other, new source(s) of driving that we will consider later. 
Likewise, $\Gamma_{\pm}$ is an effective damping rate, which we will take in general to be the sum of ion-neutral ($\Gamma_{\rm in}$), dust ($\Gamma_{\rm dust}$), linear Landau or ``turbulent'' ($\Gamma_{\rm turb/LL}$), non-linear Landau ($\Gamma_{\rm nll,\,\pm}$), and some other arbitrary additional ($\Gamma_{\rm new,\,damp,\,\pm}$) damping rates (all defined below). 

From Eq.~\ref{eqn:specific.cr.energy}, note following \citet{hopkins:m1.cr.closure} that the $\tilde{\mathcal{S}}_{\rm sc}^{\prime}$ term arises directly from taking the moments of the quasi-linear theory scattering rate equations for any scattering rate expressions $\nu_{\rm s}(\mu,\,{\bf p}_{\rm cr},\,...)$: it is the total energy exchange between CRs and the ``scatterers.'' If we assume gyro-resonance,  so that CRs of a given $R_{\rm cr}$ interact only with the gyro-resonant  wavepacket,\footnote{More generally as noted above, at a given momentum CRs can resonate with short-wavelength modes $k_{\|} = \Omega_{\rm cr}/(\mu\,v_{\rm cr}$ which depends on particle pitch angle, so this should be taken to be some effective pitch-angle-average over different wavenumbers. For a close-to-isotropic CR DF (as required by observations), it is straightforward (albeit tedious) to show this does not change any of our conclusions. A more detailed calculation (e.g.\ Kempsi et al., private communication) shows the same even for anisotropic DFs, if the dependence of scattering rate on energy predicted by SC theory $\delta_{\rm s} \approx 0$ (as we show below).} then without making any specific assumptions about the mechanism for this exchange, energy conservation {\em imposes} the form of $S_{\rm sc,\,\pm}$: 
\begin{align}
\label{eqn:scr} S_{\rm sc,\,\pm} &= \sum_{\rm species} 
\bar{\nu}_{\rm s,\pm}\,\frac{v_{A,\,\pm}}{c^{2}}\,\left[ F_{e,\,\rm cr}^{\prime} - v_{A,\,\pm}\,3\,\chi\,(e_{\rm cr}^{\prime}+P_{\rm cr}^{\prime}) \right]
\end{align}
Here $\sum_{\rm species}$ represents the sum over all CR species with a given gyro-resonant wavelength/rigidity $k_{\|} \sim 1/r_{g,{\rm cr}} \propto 1/R_{\rm cr}$.

\subsection{Damping of Parallel \Alf\ Waves: Standard Mechanisms}
\label{sec:damping}

We stress that there are many known damping processes contributing to $Q_{\pm}=\Gamma_{\pm}\,e_{\pm}$ for the relevant high-frequency scattering modes. Here we briefly review a few which are commonly invoked, focusing on the terms which apply to weakly-damped \Alf\ modes (as compared to fast magnetosonic modes, which are vastly more strongly-damped on gyro-resonant scales, a case we discuss in \S~\ref{sec:et} and Appendix~\ref{sec:turb.review} below). 

\begin{enumerate}

\item{Ion-Neutral Damping:} 
Ion-neutral collisions generically lead to a damping rate in Eq.~\ref{eqn:va.dispersion} which is rather complicated but for all limits where it is relevant can be accurately approximated as $Q_{\rm in,\,\pm} \equiv \Gamma_{\rm in}\,e_{\pm}$ with $\Gamma_{\rm in} \approx (\alpha_{\rm iH} + \alpha_{\rm iHe})/2\,\rho_{\rm i} \approx 10^{-9}\,{\rm s^{-1}}\,f_{\rm neutral}\,(T/1000\,{\rm K})^{1/2}\,(\rho/10^{-24}\,{\rm g\,cm^{-3}})$.

\item{Dust Damping:} 
From \citet{squire:2021.dust.cr.confinement.damping}, charged dust will have gyro motion excited by \Alf\ waves on the wavelengths of interest, removing some of the scattering-wave energy (and dissipating it with dust collsions with ions+neutrals), giving a damping term: $Q_{\rm dust,\,\pm} \equiv \Gamma_{\rm dust}\,e_{\pm}$ with $\Gamma_{\rm dust} \approx 0.02\,k\,v_{A,\,\rm eff}\,f_{\rm dg}\,(k/k_{d})^{-\xi_{d}}$ where $f_{\rm dg} \approx 0.01\,Z/Z_{\odot}$ is the dust-to-gas ratio (normalized to the LISM value), and $k_{d} \sim 7.4\times10^{-11}\,{\rm cm^{-1}}\,(n/{\rm cm^{-3}})^{1/2}\,\psi_{d}$ with $\xi_{d} = 1/4$ at $k\ll k_{d}$ and $\xi_{d}=1/2$ at $k\gg k_{d}$, and $\psi_{d} \equiv \bar{U}_{0}\,(\bar{\rho}_{d}/{\rm g\,cm^{-3}})\,(T/10^{4}\,{\rm K})^{\xi_{d,\,T}} \sim 1$ depending on $\bar{U}_{0}$ and $\bar{\rho}_{d}$ which parameterize the grain charge and internal grain density, $\xi_{d,\,T}=0-1$ depending on the wavelength and grain charge regime (see Eq.~18 therein).

\item{Non-Linear Landau Damping:} 
On gyro-resonant scales, oblique magnetosonic waves are rapidly damped by resonant ion interactions, pressure anisotropy, and other effects 
\citep{lee.volk:1973.damping.alfven.waves.smalscale,foote.kulsrud:1979.damping,cesarsky.kulsrud:1981.cr.confinement.damping.hot.gas,volk.cesarsky:1982.nonlinear.landau.damping,squire:2016.alfvenic.perturbations.limits.anisotropy,squire:2017.max.braginskii.scalings}. 
This strongly suppresses isotropic magnetosonic modes at the scales we follow, as noted above. But even for the weakly-damped waves of interest (e.g.\ parallel \Alf\ modes), wave-wave interactions and field-line wandering transfer energy from the weakly-damped modes to strongly-damped modes, giving rise to the usual non-linear Landau damping expression $Q_{\rm nll,\,\pm} \equiv \Gamma_{\rm nll,\,\pm}\,e_{\pm} \equiv \Gamma_{\rm nll}^{0}\,(e_{\pm}/e_{\rm B})\,e_{\pm}$ with $\Gamma_{\rm nll}^{0} \approx (\sqrt{\pi}/8)\,c_{s}\,k$ \citep{kulsrud.1969:streaming.instability,volk.mckenzie.1981}.

\item{``Turbulent,'' Linear Landau, and Collisionless Damping:} In addition to non-linear Landau damping, if there are other modes present as part of an extrinsic turbulent cascade (in addition to the parallel scattering modes themselves), these will also contribute to shearing apart or mixing waves such that power is transferred either to (a) a weakly-damped but higher-$k$ modes (``turbulent damping''; \citealt{yan.lazarian.02,farmer.goldreich.04}), or (b) strongly-damped magnetosonic modes (``linear Landau damping''; \citealt{zweibel:cr.feedback.review}). This gives $Q_{\rm turb/LL,\,\pm} \equiv \Gamma_{\rm turb/LL}\,e_{\pm}$ with $\Gamma_{\rm turb/LL}({\bf k}) \sim 1/\tau_{\rm cas}({\bf k})$ scaling with the cascade timescale $\tau_{\rm cas}$ for modes of the given ${\bf k}$. 
Considering a realistically anisotropic \citet{GS95.turbulence}-type (GS95) cascade (which is not strongly damped on gyro-resonant scales), interacting with primarily parallel modes, gives $\Gamma_{\rm turb/LL}=\Gamma_{\rm turb/LL,\,GS95} \sim [(v_{A,\,{\rm ideal}}+0.4\,c_{s})/\ell_{A}]\,(k_{\|}\,\ell_{A})^{1/2}$ (where the $v_{A,\,{\rm ideal}}$ and $c_{s}$ terms represent ``turbulent'' and ``linear Landau'' respectively from \citealt{farmer.goldreich.04,zweibel:cr.feedback.review}).\footnote{\citet{lazarian:2016.cr.wave.damping} note that on scales $1/k_{\|}$ approaching or larger than the driving and \Alf\ scales this could steepen to $\Gamma_{\rm turb/LL} \propto k_{\|}^{2/3}$, but that is well outside the relevant range of scales for $\lesssim$\,TeV CRs (although assuming such a scaling has no effect on our conclusions). Likewise, the \citet{farmer.goldreich.04} argument that field-line fluctuations set a minimum $|k_{\bot}/k_{\|}| \sim |\delta B_{\rm ext}(k_{\bot})|/|B_{0}|$ can in principle be generalized for any external critically-balanced shearing cascade with different intermittency effects modifying the perpendicular spectrum \citep{schekochihin:2020.mhd.turb.review}, but these generally lead to only minor modifications of $\Gamma_{\rm turb/LL} \sim [(v_{A,\,{\rm ideal}} + 0.4\,c_{s})/\ell_{A}]\,(k_{\|}\,\ell_{A})^{\xi_{k}}$ with $0.4 \lesssim \xi_{k} \lesssim 0.5$.}

\end{enumerate}

As discussed above, it is instructive to write the damping terms in the generic form $Q_{\pm} \equiv \Gamma_{\pm}\,e_{\pm}$ with $\Gamma_{\pm} =  k_{\|}^{\xi_{k}}\,e_{\pm}^{\xi_{A}}\,\epsilon_{\rm cr}^{\xi_{\rm cr}}\,f^{\Gamma}_{\rm ISM}$, where $f^{\Gamma}_{\rm ISM}\equiv f^{\Gamma}_{\rm ISM}(\rho,\,T,\,f_{\rm neutral},\,|{\bf B}|,\,f_{\rm dg},\,...)$ is a function of ``bulk'' ISM plasma properties (which do not directly depend on the CRs or ${\bf k}$ or $e_{\pm}$), and we have separated the dependence on the CR kinetic energy density $\epsilon_{\rm cr}$, scattering-wave  energy $e_{\pm}$, and wavelength $k_{\|}$. All the damping mechanisms above give $\xi_{\rm cr}=0$ with $0\le \xi_{\rm A}\le 1$ and $0 \le \xi_{k} \le 1$, as summarized for reference in Table~\ref{tbl:coeffs}.

\subsection{``Self-Confinement'' Driving}
\label{sec:sc}

The standard ``self-confinement'' limit arises if $S_{\pm} \rightarrow S_{\rm sc,\,\pm}$ (or $S_{\rm et,\,\pm}$, $S_{\rm new,\,\pm}  \rightarrow 0$),  i.e.\ the only source term for $e_{\pm}$ is the CR scattering itself. This excites parallel \Alf\ modes, which then compete against the different damping mechanisms in \S~\ref{sec:damping} to set $e_{\pm}$. As shown below, if the CR flux ($D_{t} F^{\prime}_{e,\,{\rm cr}}$) and $D_{t} e_{\pm}$ equations reach local steady-state, then one of $e_{\pm}$ is damped to negligible values while the other (opposing the direction of the CR flux) becomes large, and the salient driving term becomes $S_{\rm sc} \rightarrow -v_{A,\,{\rm eff}}\,\bhat\cdot \nabla P_{\rm cr}^{\prime}$. 

Akin to the damping terms, it is useful to parameterize this in terms of $S_{\rm eff} = k_{\|}^{\zeta_{k}}\,e_{\pm}^{\zeta_{A}}\,\epsilon_{\rm cr}^{\zeta_{\rm cr}}\,f^{S}_{\rm ISM}$, where $f^{S}_{\rm ISM}$ parameterizes the ISM structure dependence. In local flux-steady-state, the SC limit therefore gives $\zeta_{k} = 0$, $\zeta_{A}=0$, $\zeta_{\rm cr} \approx 1$, as denoted in Table~\ref{tbl:coeffs}.\footnote{When the CR flux and $e_{\pm}$ equations are far from local quasi-equilibrium, then from Eq.~\ref{eqn:scr} the more general form of $S_{\rm cr}$ ($\propto \bar{\nu}_{\rm s}\,(v_{A,\pm}/c)\,[F^{\prime}_{e,{\rm cr}}-v_{A,\pm} 3\chi\,(e_{\rm cr}^{\prime}+P_{\rm cr}^{\prime}) \propto \Omega_{\rm cr}\,(e_{A}/e_{\rm B})\,\epsilon_{\rm cr} \propto k\,e_{A}\,\epsilon_{\rm cr}$) would have coefficients closer to $\zeta_{k} \sim 1$, $\zeta_{A} \sim 1$, $\zeta_{\rm cr} \sim 1$. But because these reach equilibrium on short timescales, and both expressions have $\zeta_{\rm cr} \sim 1$ which is the important feature that drives the qualitative behavior discussed below, in our analytic models we will typically work with the local steady-state expressions.} But this is not the only possible source of $e_{\pm}$ -- other drivers can be included as we discuss below.

\subsection{``Extrinsic Turbulence'' Driving}
\label{sec:et}

In the classic extrinsic turbulence picture, the source term for $e_{\pm}$ is dominated by an external turbulent cascade from much larger scales, which we will denote $S_{\rm et,\,\pm}$ (with symmetric $S_{\rm et,\,+}\approx S_{\rm et,\,-} = S_{\rm et}$). The traditional K41 scenario, for example, is immediately recovered if we assume an isotropic, un-damped (except for cascade transfer), inertial-range cascade, so $S_{\rm et,\,\pm} \rightarrow S_{\rm et,\,K41} \sim |\delta v^{3}|\,k \sim $\,constant is just the turbulent dissipation/cascade rate, balanced by the damping (cascade transfer) term defined in \S~\ref{sec:damping} above $Q_{\pm} \rightarrow \Gamma_{\rm turb,\,K41}\,e_{\pm}$ with $\Gamma_{\rm turb,\,K41} \sim k\,|\delta v|$, so we obtain $e_{\pm} \approx e_{\rm B}\, (k\,\ell_{A})^{-2/3}$, with $k_{\|} \sim k$. Similarly one could in principle imagine models that might give different isotropic power spectra such as $e_{\pm} \sim e_{\rm B}\,(k\,\ell_{A})^{-1/2}$ (often called ``Kraichnan'' or ``IK-like'' in the CR literature).
But as noted above (\S~\ref{sec:definitions}), and reviewed in more detail in Appendix~\ref{sec:turb.review}, these scalings cannot physically apply at gyro-resonant scales for CRs with rigidities $\lesssim 0.1-1\,$TV, far smaller than the \Alf\ and magnetosonic dissipation scales of turbulence.

First, consider an \Alf{ic} (or slow-mode) cascade. It is well-established that an \Alf{ic} cascade cannot be isotropic on scales smaller than the \Alf\ scale $\ell_{A}$, and recall $r_{g,{\rm cr}} \ll \ell_{A}$ by a huge factor. In any cascade in which the anisotropy obeys some kind of critical balance-type condition (as seen in the solar wind, \citealp{Chen2016a}, and essentially all simulations of MHD turbulence cascades; see e.g.\ \citealt{1994ApJ...432..612S,GS95.turbulence,Boldyrev2006,terry:2018.critical.balance,beresnyak:2019.mhd.turbulence.review,schekochihin:2020.mhd.turb.review}, and references therein), the cascade power spectrum as a function of the {\em parallel} wavenumber $k_{\|}$ must obey $\mathcal{E}(k_{\|}) \propto k_{\|}^{-2}$ (where critical balance gives $k_{\|} \sim |\delta v(k_{\bot})/v_{A,\,{\rm ideal}}|\, k_{\bot} \ll k_{\bot}$, so $|k_{\bot}| \sim k$ and this is independent of the form of $\mathcal{E}(k_{\bot})$). In other words, regardless of the dominant structure of the cascade,  $e_{\pm} = e_{\rm B}\,\alpha_{t}(k_{\|})\,(k_{\|}\,\ell_{A})^{-1}$ and\footnote{The statement $S_{\rm et} \propto \Gamma_{\rm turb}$ here is just a rephrasing of the usual relation between the turbulent power spectrum and the cascade rate $\tau_{\rm cas} \sim 1/\Gamma_{\rm turb}$, i.e.\ $e_{A} \sim k_{\|}\,\mathcal{E}(k_{\|}) \propto S(k_{\|})/\Gamma_{\rm turb}(k_{\|}) \sim S(k_{\|})\,\tau_{\rm cas}(k_{\|})$.} $S_{\rm et,\,\pm} = \alpha_{t}(k_{\|})\,e_{\rm B}\,\Gamma_{\rm turb}\,(k_{\|}\,\ell_{A})^{-1}$ where more careful calculation gives the dimensionless pre-factor $\alpha_{t}(k_{\|}) \approx 7\,(v_{A,\,{\rm eff}}/v)\,\ln{(k_{\|}\,\ell_{A})} \ll 1$ as a geometric factor that accounts for gyro-averaging over the modes with $k_{\bot} \gg k_{\|}$ \citep{chandran00}. We further show in Appendix~\ref{sec:turb.review} that any mathematically-consistent \Alf{ic} cascade, even one which does not follow critical balance, must obey a similar constraint on $\mathcal{E}(k_{\|})$.

Alternatively, consider a fast-magnetosonic-mode cascade, which at least in principle could be isotropic. But recall, $r_{g,{\rm cr}}$ is well below the relevant dissipation/Kolmogorov scale ($k_{\rm diss}\sim 1/\ell_{\rm diss}$) of the magnetosonic cascade, so we cannot extrapolate an un-damped isotropic magnetosonic cascade from large scales. For strictly gyro-resonant interactions, this generally leads to a strong suppression of the power $S_{\rm et,\,\pm}$ on scales $r_{g,{\rm cr}} \ll \ell_{\rm diss}$. \citet{yan.lazarian.04:cr.scattering.fast.modes} argue that if the ``resonance function'' is strongly broadened by super-\Alf{ic} turbulence on large scales (of order the CR mean-free-path), then under the right conditions (plasma $\beta_{\rm plasma} < 1$ or $v_{A,\,{\rm ideal}} > c_{s}$, and negligible ion-neutral damping or $f_{\rm neutral} \lesssim 0.001\,(v_{A,\,{\rm ideal}}^{2}/n_{1}\,c_{s}^{2})^{3/4} (c_{s}/10\,{\rm km\,s^{-1}})^{1/4} ({\rm kpc}/R_{\rm GV}\,\ell_{A})^{1/2}$), a substantial contribution to the CR scattering rate can come from the transit-time-damping terms owing to magnetosonic modes with $k \sim k_{\rm diss} \ll 1/r_{g,{\rm cr}}$. In that limit, the resulting scattering rates from the \citet{yan.lazarian.04:cr.scattering.fast.modes} model can be written as $\bar{\nu}_{\rm s} \sim \Omega_{\rm cr}\,e_{\rm B}^{-1}\,(\delta B^{2}[k_{\rm diss}])\,(k_{\rm diss}/k_{\|})$ \citep{hopkins:cr.transport.constraints.from.galaxies}. Even though by definition in this scenario some of the scattering modes come from larger scales, in our mathematical formalism this is identical to assuming an equivalent gyro-resonant mode or cascade power  with $e_{\pm} \sim (\delta B^{2}[k_{\rm diss}])\,(k_{\rm diss}/k_{\|})$ or (since $\delta B[k_{\rm diss}]$ comes, by assumption in this model, from some isotropic cascade on larger scales with $|\delta B^{2}[k_{\rm diss}]| \sim B^{2}\,(k_{\rm diss}\,\ell_{A})^{-\psi_{\rm turb}}$) therefore $e_{\pm} \sim e_{\rm B}\,(k_{\rm diss}\,\ell_{A})^{-(\psi_{\rm turb}+1)}\,(k_{\|}\,\ell_{A})^{-1}$. Again, we show in Appendix~\ref{sec:turb.review} that a similar constraint is generic to any magnetosonic cascade with $\ell_{\rm diss} \gtrsim r_{g,{\rm cr}}$ (independent of its detailed form or which terms dominate the CR scattering). 

Parameterizing again as $S_{\rm eff} = k_{\|}^{\zeta_{k}}\,e_{\pm}^{\zeta_{A}}\,\epsilon_{\rm cr}^{\zeta_{\rm cr}}\,f^{S}_{\rm ISM}$, we see that in  the extrinsic turbulence limit generically $\zeta_{\rm cr}=\zeta_{A}=0$ and accounting for anisotropy and/or damping we must have $\zeta_{k} \lesssim -1 + \xi_{k}^{\rm turb}$ (where $\xi_{k}^{\rm turb}$ is the dependence of $\Gamma_{\rm turb} \propto k_{\|}^{\xi_{k}^{\rm turb}}$), which is equivalent to $\delta_{\rm s} \lesssim 0$. For a magnetosonic ET cascade, if dissipation is non-negligible outside the special limits above (e.g.\ in gas with plasma $\beta_{\rm plasma} >1$, or partially-neutral gas), this gives a super-exponential cutoff to the ET power spectrum on small scales which is equivalent to $\zeta_{k} \ll -1 + \xi_{k}^{\rm turb}$. 

In Appendix~\ref{sec:turb.review}, we present a much more detailed review and discussion of the anisotropy and damping constraints above. There we show that the key conclusion that any cascade model must predict  $\zeta_{k} < -1 + \xi_{k}^{\rm turb}$, i.e.\ $e_{\pm}\propto k^{-1}$ or steeper if $r_{g,{\rm cr}}$ is smaller than the dissipation and \Alf\ scale, is robust to any specific assumptions about the turbulent cascade, dissipation mechanism, or mode structure.

\subsection{Steady-State Solutions}
\label{sec:steady.state}

Eq.~\ref{eqn:ea} and Eq.~\ref{eqn:f1} converge to their ``local steady state'' or quasi-equilibrium values with $D_{t}e_{\pm} \rightarrow0$ and $D_{t} \bar{f}_{\rm cr,\,1} \rightarrow 0$ (or more formally, $|D_{t}e_{\pm}| \ll |\bar{\nu}_{s}\,e_{\pm}|$, $|D_{t} \bar{f}_{\rm cr,\,1}| \ll |\bar{\nu}_{s} \bar{f}_{\rm cr,\,1}|$), in approximately the scattering time $\bar{\nu}_{\rm s}^{-1}$. This is $\bar{\nu}_{\rm s}^{-1}\sim 30\,{\rm yr}\,R_{\rm GV}^{0.5}$ from empirically-fitted models, much faster than other timescales over which e.g.\ bulk ISM properties evolve (see Appendix~\ref{sec:local.steady.state.detailed.solns}). In this limit, assuming strong scattering, the DF becomes nearly isotropic ($\chi \approx 1/3$) and Eqs.~\ref{eqn:f0}-\ref{eqn:f1} can be combined into a single anisotropic diffusion equation:
\begin{align}
D_{t} \bar{f}_{\rm cr,\,0} &\approx \nabla \cdot \left( \kappa_{\|}\,\bhat\bhat \cdot \nabla \bar{f}_{\rm cr,\,0} + \frac{\bar{v}_{A}}{3}\,\bhat\,p_{\rm cr}\,\frac{\partial \bar{f}_{\rm cr,\,0}}{\partial p_{\rm cr}} \right) + j_{\rm cr,\,0}  \\
\nonumber &+ \frac{1}{p_{\rm cr}^{2}}\frac{\partial }{\partial p_{\rm cr}} \bigg[ p_{\rm cr}^{3}\, \bigg\{ \frac{\mathcal{R}_{\rm loss}}{p}\,\bar{f}_{\rm cr,\,0}  
+ \frac{\nabla \cdot {\bf u}_{\rm gas}}{3}\,\bar{f}_{\rm cr,\,0}\\
\nonumber & \ \ \ \ \ \ \ \ \ \ \ \ \ \ \ \ \ \ \ \ \  \ \ \ \ \ \ \ \  \ - \frac{\bar{v}_{A}}{3}\,\bhat\cdot\nabla \bar{f}_{\rm cr,\,0} + 
  \frac{(v_{A,\,{\rm eff}}^{2}-\bar{v}_{A}^{2})}{9\,\kappa_{\|}}\,p\, \frac{\partial \bar{f}_{\rm cr,\,0}}{\partial p} 
\bigg\}\bigg ]  
\end{align}
with $\kappa_{\|} \equiv v_{\rm cr}^{2}/3\,\bar{\nu}_{\rm s}$. 
The relevant behaviors here are more obvious if we again take the CR energy equation, Eq.~\ref{eqn:specific.cr.energy}:
\begin{align}
\label{eqn:dtecr.eqm} D_{t} e_{\rm cr}^{\prime}  \approx & \ 
\nabla \cdot \left[ \kappa_{\|} \,\bhat\bhat\cdot \nabla e_{\rm cr}^{\prime} 
- \bar{v}_{A}\,\bhat\,e_{\rm cr}^{\prime} \right] \\ 
\nonumber & - P_{\rm cr}^{\prime}\,\nabla \cdot ({\bf u}_{\rm gas} + \bar{v}_{A}\,\bhat)  
+ \frac{v_{A,\,{\rm eff}}^{2}-\bar{v}_{A}^{2}}{c^{2}}\,\bar{\nu}_{\rm s}\,\left(e_{\rm cr}^{\prime} + P_{\rm cr}^{\prime} \right)
+ \mathcal{S}^{\prime}_{\rm other,\,cr},
\end{align}
illustrating that we have anisotropic ``diffusion'' with $\kappa_{\|} \equiv v_{\rm cr}^{2}/3\,\bar{\nu}_{\rm s} = v_{\rm cr}^{2}/[3\,(\bar{\nu}_{\rm s,+}+\bar{\nu}_{\rm s,-})]$, and ``streaming'' along $\bhat$ with speed $\bar{v}_{A} \equiv v_{A,\,{\rm eff}}\,(\bar{\nu}_{\rm s,+} - \bar{\nu}_{\rm s,-})/(\bar{\nu}_{\rm s,+} + \bar{\nu}_{\rm s,-})$.

Meanwhile, Eq.~\ref{eqn:ea} becomes $\nabla \cdot ({v}_{A,\,\pm}\,e_{\pm}\,\bhat) + e_{\pm}\,(\nabla\cdot{\bf u}_{\rm gas})/2 + Q_{\pm} - S_{\pm} \approx 0$, with $S_{\rm sc\,\pm}$ as a function of $F^{\prime}_{\rm cr}$ given by solving the steady-state flux equation 
$F_{e,\,{\rm cr}}^{\prime} \approx \bar{v}_{A}\,(e_{\rm cr}^{\prime} + P_{\rm cr}^{\prime}) -(c^{2}/\bar{\nu}_{\rm s})\,\bhat\cdot \nabla P_{\rm cr}^{\prime}$. This gives
$S_{\rm sc,\,\pm} \approx 
 -(\nu_{\rm s,\,\pm}/\bar{\nu}_{\rm s})\,v_{A,\,\pm}\,\bhat\cdot \nabla P_{\rm cr}^{\prime} - 2\,(v_{A}/c)^{2}\,(\nu_{\rm s,\,\mp}/\bar{\nu}_{\rm s})\,(e_{\rm cr}^{\prime} + P_{\rm cr}^{\prime})$. 
Assuming we know the form of $S_{\rm et,\,\pm}$, $\Gamma_{\pm}$, etc., the pair of steady-state equations for $e_{\pm}$, can then be numerically solved exactly, given a locally fixed background (${\bf B}$, etc.; see Appendix~\ref{sec:local.steady.state.detailed.solns}). But the exact solution is given by a fifth-order polynomial in $e_{+}+e_{-}$, without closed-form solutions, which is not particularly useful or instructive (defeating the purpose of our steady-state assumption here). 

It is much more useful to consider two limiting cases (justified in Appendix~\ref{sec:local.steady.state.detailed.solns}). First, if the extrinsic term dominates the ``source'' with $|S_{\rm et,\pm}| \gg |S_{\rm sc,\,\pm}|$, then assuming the extrinsic driver is symmetric with respect to waves in the $\pm\bhat$ direction, $S_{\rm et,\,+} \approx S_{\rm et,\,-}$, we have $\bar{\nu}_{\rm s,+} \approx \bar{\nu}_{\rm s,-}$, and $|\bar{v}_{A}| \ll v_{A}$. Alternatively, if the self-confinement term dominates, then either $\bar{\nu}_{\rm s,+}\gg \bar{\nu}_{\rm s,-}$ or $\bar{\nu}_{\rm s,+} \ll \bar{\nu}_{\rm s,-}$ (the larger corresponding to the opposite direction of $\bhat\cdot \nabla P_{\rm cr}^{\prime}$), so $|\bar{v}_{A}| \approx v_{A,\,{\rm eff}}$. In either case, the ``streaming'' speed $\bar{v}_{A}$ is not especially important for CR transport, because it is sub-dominant to the ``diffusive'' term at most energies of interest for empirically-allowed models (shown below). This can be seen both by considering its normalization, which is far smaller than allowed by empirical considerations (with a halo size $\sim 10\,$kpc, the CR ``escape time'' due to this term would be $\sim 10\,$Gyr, compared to the observed $\sim 5\,R_{\rm GV}^{-0.5}\,$Myr),
or by noting that this would produce a residence/escape time that is completely independent of rigidity, again in contradiction to CR observations. So in either the SC or ET limit we can reasonably reduce the key transport physics to understanding $\kappa_{\|} \propto 1/\bar{\nu}_{\rm s}$, and can approximate Eq.~\ref{eqn:ea} with a single equation for $\bar{\nu}_{\rm s}$ to leading order. Moreover, for the types of models discussed above, the second and third terms in Eq.~\ref{eqn:ea} (${\sim} \nabla ({v}_{A,\,{\rm eff}}\,e_{\pm})$ and ${\sim} e_{\pm}\,\nabla {\bf u}_{\rm gas}$) are relatively small ($\ll S \sim Q$), so if we write $Q_{\pm} \equiv \Gamma_{\pm}\,e_{\pm}$, then we can very generically approximate $S\approx\Gamma_{\pm}\,e_{\pm}$ in steady-state, giving
\begin{align}
\label{eqn:kappa.steady.state} \kappa_{\|} &\equiv \frac{v_{\rm cr}^{2}}{3\,\bar{\nu}_{\rm s}} = \frac{4\,\beta_{\rm cr}}{3\pi\,\hat{\nu}_{\rm s}}\,c\,r_{g,{\rm cr}}\,\left( \frac{e_{\rm B}}{e_{+}+e_{-}} \right) \sim v_{\rm cr}\,r_{g,{\rm cr}}\,\left( \frac{\Gamma_{\pm}\,e_{\rm B}}{S_{\pm}} \right).
\end{align}
We will return to this approximate scaling below.

\begin{figure*}
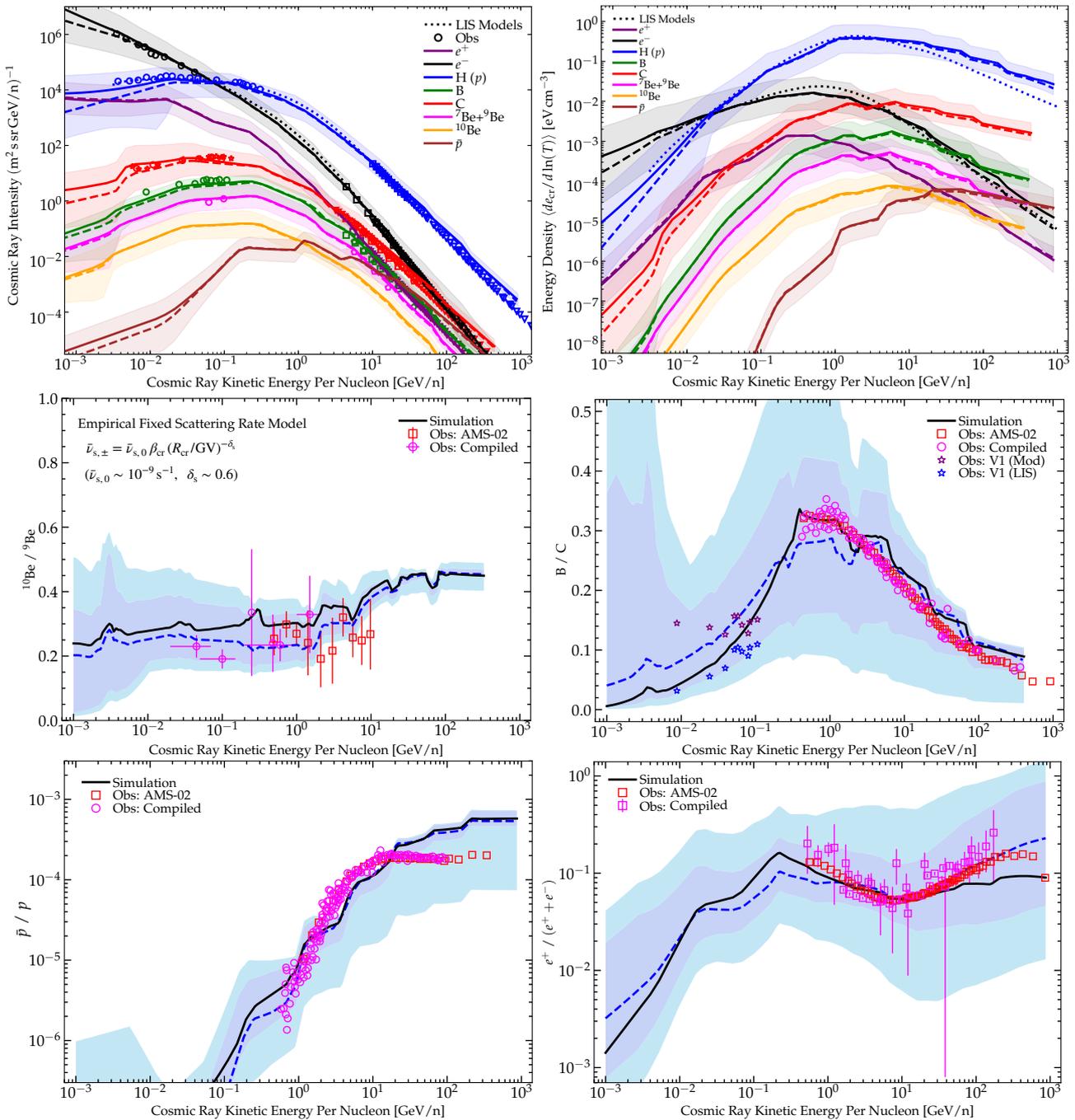

	\figsix{output_12_16_20_newrsol_m10k_coeff700_anisoclosure_Dcoeff0pt6_loinit}{}
	\vspace{-0.1cm}
	\caption{Example of an ``empirical model'' from \paperone\ (see \S~\ref{sec:problems:obs}) where CR scattering rates are assumed to be a simple constant power-law function of rigidity $\bar{\nu}_{\rm s} = 10^{-9}\,{\rm s^{-1}}\,\beta_{\rm cr}\,R_{\rm GV}^{-0.6}$. CR spectra are calculated by integrating CR dynamics and losses in a full live galaxy-formation simulation at redshift $z=0$ (\S~\ref{sec:methods}). 
{\em Top:} CR intensity/kinetic energy density spectra for different species (labeled). Lines show median ({\em dashed}) and mean ({\em solid}) values in the simulation for local ISM (LISM) gas in the Solar circle ($r=7-9\,$kpc) with density $n=0.3-3\,{\rm cm^{-3}}$. Shaded dark (light) range corresponds to $\pm1\sigma$ ($\pm2\sigma$) range, allowing for a broader range of galacto-centric radii ($4-10\,$kpc) and LISM densities ($n=0.1-10\,{\rm cm^{-3}}$). Points show compiled observations (see text; \S~\ref{sec:results}). 
	{\em Middle Left:} $^{10}$Be/$^{9}$Be; dark purple (light cyan) range shows the $\pm1\,\sigma$ ($\pm2\,\sigma$) range.	
	{\em Middle Right:} B/C ratio. 
	{\em Bottom Left:} $\bar{p}/p$ ratio. Note the value at the highest-energies is significantly affected by our upper boundary (we do not evolve $p$ or heavier ions with rigidity $\gtrsim1000\,$GV). 
	{\em Bottom Right:} $e^{+}/(e^{+}+e^{-})$ ratio.
	All of these properties can be reasonably well-fit with a simple empirical power-law scaling. The spectra are nearly independent of the initial CR spectra after $\sim 100$\,Myr of evolution.
	\label{fig:demo.cr.spectra.fiducial}\vspace{-0.5cm}}
\end{figure*}

\section{Problems of Both Self-Confinement \&\ Extrinsic Turbulence Models}
\label{sec:problems}

\subsection{Empirical Models}
\label{sec:problems:obs}

It is well-known that one can reproduce almost all of the observed local Solar neighborhood CR data and Galactic $\gamma$-ray constraints by assuming an empirically {\em parameterized} $\kappa_{\|} \sim \beta_{\rm cr}\,c\,r_{0}(R_{\rm cr})$.\footnote{In models which assume an isotropic diffusion Fokker-Planck equation for CRs, it is common to quote $D_{x x}$. For isotropically tangled magnetic fields, $D_{x x} \approx \kappa_{\|}/3$.}. Most modern studies favor a scaling close to $r_{0} \sim 10^{19}\,{\rm cm}\,R_{\rm GV}^{1/2}$, or equivalently $e_{A}/e_{\rm B} = (e_{+}+e_{-})/e_{\rm B} \sim 3\times 10^{-7}\,B_{\mu {\rm G}}^{-1}\,R_{\rm GV}^{1/2}$, or $e_{A} \sim (k_{\|}\,\ell_{0})^{-1/2}\,e_{\rm B}$ with $\ell_{0} \sim 3\times 10^{25}\,B_{\mu {\rm G}}^{2}\,{\rm cm}$ \citep{blasi:cr.propagation.constraints,vladimirov:cr.highegy.diff,gaggero:2015.cr.diffusion.coefficient,2016ApJ...819...54G,2016ApJ...824...16J,cummings:2016.voyager.1.cr.spectra,2016PhRvD..94l3019K,evoli:dragon2.cr.prop,2018AdSpR..62.2731A}. For example, \citealt{delaTorre:2021.dragon2.methods.new.model.comparison} show that, even allowing for a wide range of model variations with different systematic uncertainties and assumptions, the largest plausible deviations in the empirical models lie within the range $r_{0} \sim 10^{18.5 - 19.5}\,R_{\rm GV}^{0.4 - 0.7}\,{\rm cm}$, i.e.\ $e_{A} \sim (10^{-7}-10^{-6})\,\langle B_{\mu {\rm G}}^{-1} \rangle \,R_{\rm GV}^{0.3-0.6}\,e_{\rm B}$, or equivalently $| \delta B(k_{\|} \sim 1/r_{g,{\rm cr}})| / |{\bf B}| \sim (0.3-1)\,0.001\,R_{\rm GV}^{(0.15 - 0.3)}$. This required scaling seems, at first, remarkably simple and plausible. Yet in practice, it proves remarkably difficult to actually produce even {\em qualitatively} similar scaling from either ET or SC models at energies $\lesssim$\,TeV.

For reference below, in the CR propagation literature the slope of the dependence of diffusivity on rigidity is usually parameterized as $\kappa_{\|} \sim (\beta_{\rm cr}\,c^{2}/\bar{\nu}_{\rm s,\,0})\,R_{\rm GV}^{\delta_{\rm s}}$, so $\bar{\nu}_{\rm s} = \bar{\nu}_{\rm s,\,0}\,\beta_{\rm cr}\,R_{\rm GV}^{-\delta_{\rm s}}$, with  $\bar{\nu}_{\rm s,\,0}$ a constant and $0.4 \lesssim \delta_{\rm s} \lesssim 0.7$ allowed by observations (above). To reproduce this, we require $e_{A} \propto R_{\rm cr}^{1-\delta_{\rm s}} \propto r_{g,{\rm cr}}^{1-\delta_{\rm s}} \propto k_{\|}^{\delta_{\rm s}-1}$.

\subsection{The Problems With Extrinsic Turbulence}
\label{sec:problems:et}

First let us  consider the standard ET models introduced in \S~\ref{sec:et}. Naively, the empirically inferred slope $\delta_{\rm s}$ appears quite consistent with the expectation from an isotropic, undamped, inertial-range cascade with $\mathcal{E}(k) \propto k^{-3/2}$, which gives $e_{A} \sim k_{\|}\,\mathcal{E}_{\rm turb}(k_{\|}) \sim k_{\|}^{-1/2}$ and thus $\delta_{\rm s} = 1/2$). 
It is also  marginally consistent with an isotropic undamped inertial range K41 cascade ($\delta_{\rm s} = 1/3$). But there are three major  problems: (1) anisotropy and (2) damping and (3) normalization. Once again, recall that for all CR energies of interest $r_{g,{\rm cr}} \sim 10^{12}\,{\rm cm}\,R_{\rm GV}/B_{\rm \mu G}$ is much smaller than both of the \Alf\ scale -- below which the turbulence is sub-\Alf{ic} ($\ell_A\gtrsim$\,pc) --  and the magnetosonic dissipation/Kolmogorov scales ($\ell_{\rm diss}\gtrsim 10^{15}\,{\rm cm}$). 

\begin{itemize}

\item{\bf Anisotropy:} First (1) below the \Alf\ scale, theory  and simulations robustly predict something akin to critical balance must apply to the  \Alf{ic} cascade. But as noted in \S~\ref{sec:et} (see \citealt{schekochihin:2020.mhd.turb.review}), {\em any} energy-conserving cascade that obeys critical balance automatically predicts $\mathcal{E}(k) \propto k_{\|}^{-2}$ ($e_{A} \propto k_{\|}^{-1}$). This implies  $\delta_{\rm s} = 0$, regardless of how the cascade scales with the perpendicular components of ${\bf k}$, because it is $k_{\|}$ that plays the key role for CR scattering.\footnote{In Appendix~\ref{sec:turb.review}, we show that $\delta_{\rm s} \le 0$ is generic to any \Alf{ic} cascade, independent of its form, and mathematically-allowed violations of critical balance or other conditions within a cascade generically lead to $\delta_{\rm s} < 0$, making the problem worse.}
But this further implies scattering rates are {\em independent} of CR energy, which is {\em strongly} ruled-out. 

\item{\bf Damping/Dissipation:} Second (2) strong dissipation, particularly of magnetosonic modes, causes two problems. First, it damps fast magnetosonic fluctuations, which makes the normalization of $\bar{\nu}_{\rm s}$ much too low  for scattering from a magnetosonic cascade (generally by $\sim 3-6$ orders of magnitude for realistic damping rates, as shown in \citealt{hopkins:cr.transport.constraints.from.galaxies}). But, equally important, dissipation can only make the spectrum steeper ($\mathcal{E}(k)$ decreases more-rapidly at high $k$). This in turn implies that $\delta_{\rm s}$ generically becomes {\em negative}. One might  argue that at some point, one should ignore the strongly-damped smaller-scale gyro-resonant magnetosonic modes and only integrate the contribution from larger magnetosonic modes above the dissipation/Kolmogorov scale -- this is the argument in \citealt{yan.lazarian.04:cr.scattering.fast.modes} (YL04). But, as noted in \S~\ref{sec:et}, this {\em also} gives $\delta_{\rm s} \le 0$, always. In fact $\delta_{\rm s} = 0$, i.e.\ $e_{A} \propto k_{\|}^{-1}$,  corresponds to the ``most efficient'' possible case of the YL04 model so long as $\ell_{\rm diss}\gg r_{g,{\rm cr}}$, and in many other limits $\delta_{\rm s} \ll -1$, e.g.\ when there is an exponential-like cutoff due to non-zero ion-neutral damping, or plasma $\beta_{\rm plasma} \equiv c^{2}_{s}/v^{2}_{A,\,{\rm ideal}} > 1$. Again, as shown rigorously in Appendix~\ref{sec:turb.review}, this applies to {\em any} magnetosonic cascade regardless of details, if the gyro scale is smaller than the (fast-mode) dissipation scale. So the only way to ``salvage'' even the qualitative scaling of $\delta_{\rm s}$ in ET models, with turbulence that is either Alfv\'enic or magnetosonic in character,  is to ignore {\em both} anisotropy and damping/dissipation effects. This would require discarding almost everything that is known about the structure of MHD turbulence. 

\item{\bf Normalization:} Third (3), even if we did invoke an isotropic, undamped $\mathcal{E}(k) \propto k^{-3/2}$ cascade, which leads to the observationally favored  slope, then we would obtain $e_{A} \sim e_{\rm B}\,(k\,\ell_{A})^{-1/2}$. This is systematically larger than the empirically required value of $e_{A}$ by a factor $\sim  1000\,B_{\rm \mu G}\,(\ell_{A}/10\,{\rm pc})^{-1/2}$ -- i.e.\ this would typically over-predict observed CR scattering rates by factors of several thousand. Thus, some anisotropy and/or damping {\em must} be present to prevent extrinsic turbulence from over-confining CRs, but as soon as those are invoked, the predicted shape ($\delta_{\rm s}$)  is incorrect. 

\end{itemize}

As reviewed in \citet{hopkins:cr.transport.constraints.from.galaxies} and Appendix~\ref{sec:turb.review}, almost all proposed more-detailed corrections and modifications to traditional ET models in the literature make the problems above worse, not better (i.e.\ they make the scattering rates even more different from those observationally required).

\subsection{Generic Alternatives to Extrinsic Turbulence}
\label{sec:et.alternatives}

A generic alternative to ET is to  have scattering modes that are directly excited ``at each scale'' by some process, rather than arising through a cascade from large scales. In this scenario, the driver must operate over a wide range of scales -- scattering CRs in the range $\sim\,$MeV-TeV implies a factor $\sim 10^{6}$ in $k$ -- and excite the relevant $k_{\|}$ modes. 

Of course self-confinement (SC), where the excitation comes from the CRs themselves, does exactly this, and is very natural --  indeed, it should occur to some extent. And a simple order-of-magnitude calculation shows that the SC source term $S_{\rm sc}$ should almost always be dominant over the ``standard'' ET source terms for CR energies $\lesssim$\,TeV, if we account for either  anisotropy or damping of ET (let alone both). For this reason, SC has been the most popular model to explain the scaling of $e_{A}$ (hence CR scattering rates) at these energies.

\subsection{The Problems With Self-Confinement}
\label{sec:problems:sc}

However,  there are also three major qualitative problems with SC models: (1) normalization, (2) spectral shape/scaling of scattering rates, and (3) instability or ``solution collapse.'' 

\begin{itemize}

\item{\bf Normalization:} The normalization problem (1) is discussed in detail in both analytic models and full dynamical simulations in \citet{hopkins:cr.transport.constraints.from.galaxies}. In brief, for $\sim1-10\,$GeV CRs, which contain most of the energy, while damping is large in neutral gas, the CR energy density in a multi-phase ISM is determined by the volume-filling phases with the {\em lowest} diffusivity (the ``boundary condition'') i.e.\ the WIM and inner CGM. In these regions, non-linear Landau and ion-neutral damping are both inefficient, so  standard models suggest that turbulent or linear-Landau damping dominates \citep[see][]{hopkins:cr.transport.constraints.from.galaxies,hopkins:2020.cr.transport.model.fx.galform,buck:2020.cosmic.ray.low.coeff.high.Egamma}. Considering CRs near the peak of the CR energy spectrum, and assuming local steady state and dominant turbulent damping, the effective diffusivity from Eq.~\ref{eqn:kappa.steady.state} would then be (see \citealt{hopkins:cr.transport.constraints.from.galaxies}) $\kappa_{\|} \sim v_{\rm cr}\,r_{g,{\rm cr}}\,(\Gamma_{\pm}\,e_{\rm B}/S_{\rm sc,\,\pm}) \approx  10^{28}\,\delta v_{10}^{3/2}\,\ell_{\nabla,{\rm cr},{\rm kpc}}\,\ell_{A,\,{\rm 10}}^{-1/2}\,e_{\rm cr,\,eV}^{-1}\,n_{1}^{3/4}$ (with $\delta v_{10}\equiv \delta v_{\rm turb}/10\,{\rm km\,s^{-1}}$, $\ell_{\nabla,{\rm cr},{\rm kpc}}\equiv \ell_{\nabla,{\rm cr}}/{\rm kpc}$ with $\ell_{\nabla,{\rm cr}} \equiv P_{\rm cr}^{\prime} / | \nabla P_{\rm cr}^{\prime}|$, $\ell_{A,\,10}\equiv \ell_{A}/10\,{\rm pc}$, $e_{\rm cr,\,eV}\equiv e_{\rm cr}^{\prime}/{\rm eV\,cm^{-3}}$, $n_{1} \equiv n/{\rm cm^{-3}}$), or 
$e_{A}/e_{\rm B} \sim 10^{-5}\,e_{\rm cr,\,eV}\,\ell_{A,\,10}^{1/2}\,B_{\rm \mu G}^{-1}\,\delta v_{10}^{-3/2}\,\ell_{\nabla,{\rm cr},{\rm kpc}}^{-1}\,n_{1}^{-3/4}$. This is  a factor $\sim 30-100$ smaller than the empirically favored value of $\kappa_{\|}$ for  $\sim1-10\,$GeV CRs. 

This issue is potentially important, but perhaps the least-serious of the three problems: it can be ameliorated by (a) including additional damping mechanisms such as the recently proposed dust damping, which can be larger than turbulent damping invoked above (raising $\Gamma_{\pm}$) by factors $\sim 10-100$ at $\sim 1\,$GV under MW-like conditions \citep{squire:2021.dust.cr.confinement.damping}; (b) invoking somewhat slower growth rates $S$, for the relevant modes as excited by CRs, which can occur accounting more accurately for e.g.\ the full spectrum of modes which contribute to scattering (instead of assuming strict gyro-resonance plus the gray approximation), as argued in \citet{bai:2019.cr.pic.streaming}; (c) accounting more accurately for geometric and other non-gray effects, which can lower the effective scattering rate for gyro-resonant modes for a given $e_{A}$ (i.e.\ our ``$\hat{\nu}_{\rm s}$'' parameter). These corrections arise from accounting more accurately for the full shape of the CR spectrum (Kempski et al., in prep), accounting for the $\mu=0$ pitch-angle scattering barrier, which has been shown to  be significant in some PIC simulations \citep{bai:2019.cr.pic.streaming}, and accounting for anisotropy that arises from the fact that the CRs only source one helicity of modes in some regimes \citep{holcolmb.spitkovsky:saturation.gri.sims}. It is noteworthy that all of these corrections go in the ``favored'' direction. 

\item{\bf Spectral Shapes:} However problem (2), regarding the spectral-shape-dependence of scattering rates, is not resolved by the possible solutions above. The issue is that the growth term $S_{\rm sc,\,\pm}$ scales with the CR pressure or energy itself, as $S_{\rm sc,\,\pm} \rightarrow \pm v_{A,\,{\rm eff}}\,\bhat \cdot \nabla P_{\rm cr}^{\prime} \sim (v_{A,\,{\rm eff}}/\ell_{\nabla,{\rm cr}})\,\epsilon_{\rm cr}$, where $\ell_{\nabla,{\rm cr}}$ is the CR gradient scale length above and $\epsilon_{\rm cr}$ is the CR kinetic energy density in a logarithmic interval in rigidity $R_{\rm cr}$ (or equivalently, gyro-resonant $k_{\|}$). But, for realistic CR spectra, $\epsilon_{\rm cr} \propto R_{\rm cr}^{-\alpha_{\rm cr}} \propto k_{\|}^{\alpha_{\rm cr}}$, with  $-1.4 \lesssim \alpha_{\rm cr} \lesssim -0.7$ at energies $\ll 1\,$GV, and $\alpha_{\rm cr} \approx 0.7$ at energies from $\sim 1-300\,$GV \citep[e.g.][and references therein]{cummings:2016.voyager.1.cr.spectra,bisschoff:2019.lism.cr.spectra}. So, in our parameterization above, we would have $S_{\rm sc,\,\pm} \propto k_{\|}^{\alpha_{\rm cr}}$, with $\Gamma_{\pm} \propto k_{\|}^{\xi_{k}}\,e_{A}^{\xi_{A}}$ (for all the known damping mechanisms discussed in \S~\ref{sec:damping}).   In steady state,  i.e.\ solving $S_{\pm} \sim \Gamma_{\pm}\,e_{A}$, this leads to $\delta_{\rm s} \approx 1 - (\xi_{k} - \alpha_{\rm cr})/(1+\xi_{A})$. But for $\alpha_{\rm cr} \sim 0.7$ (i.e.\ $R_{\rm cr} \gtrsim $\,GV), this gives $0.8 \lesssim \delta_{\rm s} \lesssim  1.7$ for all known damping mechanisms ($0 \le \xi_{k},\,\xi_{A} \le 1$), while for $\alpha_{\rm cr} \sim -1$ (i.e.\ $R_{\rm cr} \lesssim $\,GV), this gives $-1 \lesssim \delta_{\rm s} \lesssim 0.5$. In other words, $\delta_{\rm s}$ is generically much too low at $R_{\rm cr} \ll 1\,$GV, owing to the fact that the spectrum is rising, and much too high at $R_{\rm cr} \gg 1\,$GV, owing to the
fact that the spectrum is falling. This produces a  strong minimum in $\kappa_{\|}$, a much-too-sharply-peaked CR spectrum, and the incorrect dependence in both limits of e.g.\ secondary-to-primary ratios on $R_{\rm cr}$. In principle, this could be solved by imposing a new dominant damping mechanisms with a dependence on $k$ or $e_{A}$ that is different to the mechanisms discussed in \S~\ref{sec:damping}. But, this would require rather unusual values of $\xi_{k}$, in particular, $\xi_k \approx \alpha_{\rm cr} + (1-\delta_{\rm s})\,(1+\xi_{A})$ implies $1.2 \lesssim \xi_{k} \lesssim 1.7$ for $R_{\rm cr} \gtrsim 1\,$GV, and $-0.5 \lesssim \xi_{k} \lesssim 0$ for $R_{\rm cr} \lesssim 1\,$GV. This in turn requires a dramatic difference in the dominant damping mechanism between waves that resonate with low and high CR energies -- longer wavelength waves require larger $\xi_{k}$, while shorter wavelength waves require much smaller (negative) $\xi_{k}$ -- compared to the known mechanisms described in \S~\ref{sec:damping}. In other words, it is not possible to resolve this problem by simply ``tweaking'' coefficients or power-law scalings of standard damping or growth-rate terms.

\item{\bf Instability:} Even if (1) and (2) were solved with something like the possibilities mentioned above, a potentially more fundamental problem with self confinement is that the commonly adopted ``local steady state'' SC solutions for the CR scattering rates and fluxes  (given some CR energy density at a given CR momentum/rigidity) are not stable equilibria of the CR energy equation. This is derived in detail in Appendix~\ref{sec:sc.equilibrium.models}. Briefly, the problem arises because the SC driving/growth rate $S_{\rm sc,\,\pm}$ is proportional to the CR flux $F_{e,\,\rm cr}^{\prime}$ or (in local steady-state) to $\propto \nabla P_{\rm cr}^{\prime}$ (i.e.\ $\zeta_{\rm cr} \sim 1$), while all standard damping processes are independent of $e_{\rm cr}^{\prime}$ (i.e.\ $\xi_{\rm cr} = 0$). This implies $e_{A} \sim S_{\rm sc,\,\pm}/\Gamma_{\pm} \propto \nabla e_{\rm cr}^{\prime}$, where we assumed linear damping dominates for simplicity (this assumption is relaxed in Appendix~\ref{sec:sc.equilibrium.models} and our simulation tests). But this in turn means that the steady-state CR flux $F_{e\,{\rm cr}}^{\prime} \propto \kappa_{\|}\,\nabla e_{\rm cr}^{\prime} \propto e_{A}^{-1}\,\nabla e_{\rm cr}^{\prime} \propto $\,constant is {\em independent} of $e_{\rm cr}^{\prime}$; or equivalently that the CR ``escape time'' (the streaming/diffusion time to some distance $\ell$) is $t_{\rm esc} \sim \ell^{2}/\kappa_{\|} \sim \ell/v_{\rm stream,\,eff} \propto \ell^{2}\,e_{A}/(v_{\rm cr}\,r_{g,{\rm cr}}\,e_{\rm B}) \propto e_{\rm cr}^{\prime}$. This is a fundamental feature of SC models, as it is a re-statement of the fact that the confining wave energy ultimately comes from the CRs themselves. Indeed, the effect  clearly manifests in the early kinetic quasi-linear theory and solutions of \citet{skilling:1971.cr.diffusion}, as the absence of the CR distribution in the diffusion term (the final term in their Eq.~(9)). The consequence of this feature of the SC model is that if the CR flux is lower than the exact value needed to maintain global steady-state in the CR energy density equation (or even if there is a perturbation about the exact value),
the local $e_{\rm cr}^{\prime}$ will increase due to the CR overconfinement, then causing $e_{A}$ and the CR escape time to also increase, which in turn further bottlenecks\footnote{In the literature, CR ``bottlenecks'' can occur under a variety of different conditions and the term is often used to refer to distinct physical processes \citep[see e.g.][]{zweibel:cr.feedback.review,bustard:2020.crs.multiphase.ism.accel.confinement,quataert:2021.cr.outflows.diffusion.staircase,huang.davis:2021.cr.staircase.in.outflows}. Unless otherwise specified, in this paper we will generally use the term in connection to this SC ``solution collapse'' problem, specifically to the over-confined solution branch.} $e_{\rm cr}^{\prime}$. Since the steady-state CR energy density in some region $e_{\rm cr}^{\prime} \sim j_{\rm inj} \, t_{\rm esc}$, and both $j_{\rm inj}$ and $t_{\rm esc}$ have just increased, $e_{\rm cr}^{\prime}$ increases as well. This will grow on a timescale of the injection time ($e_{\rm cr}^{\prime} / j_{\rm inj}$), until $t_{\rm esc} \propto e_{\rm cr}^{\prime} $ becomes so large that the CRs can only move at the local \Alf\ speed (effectively $\bar{\nu}_{\rm s} \rightarrow \infty$), which is much slower than relevant loss times. They thus lose all their energy to catastrophic or radiative losses before they can escape. As a result, in steady-state, the energy density is no longer set by the escape time but by the calorimetric loss timescale $e_{\rm cr}^{\prime} \sim j_{\rm inj}\,t_{\rm loss}$. The outcome is that CRs at all $R_{\rm cr}$ lose all their energy near their injection sites, in gross contradiction to observations. Even if we neglect losses, the CRs would all stream at the same speed $v_{A}$, implying $\delta_{\rm s}=0$, which is also in direct contradiction to observations. 

This is the ``SC runaway'' problem described and seen in the simulations of \citet{hopkins:cr.transport.constraints.from.galaxies}, who noted it for just $\sim1\,$GV CRs. Here we note that it applies at all energy scales, becoming more severe due to the additional constraints on $\delta_{\rm s}$. Conversely, if one deviates from a balanced initial condition by lowering $e_{\rm cr}^{\prime}$, then $e_{A}$ drops, the escape time decreases, so $e_{\rm cr}^{\prime}$ drops more rapidly, etc.\ until CRs essentially ``free stream'' and produce negligible secondaries. 

We will show that in simulations, this instability or ``solution collapse'' problem is quite severe. In controlled restarts from otherwise identical initial conditions (ICs), if we start from a ``high'' CR density $e_{\rm cr}^{\prime}$ (with only SC-motivated $S$), then $e_{\rm cr}^{\prime}$ and scattering rates grow until they reach calorimetric losses across a wide range of  energies; conversely, if we start from a ``low''  $e_{\rm cr}^{\prime}$, then $e_{\rm cr}^{\prime}$ and scattering rates decrease to an extremely low value, giving negligible secondary production (much too-low B/C, $\bar{p}/p$, $e^{+}/e^{-}$) at all energies. It is inevitable that real galaxies will almost constantly undergo events that push them towards one or the other limit.

\end{itemize}

\begin{figure*}
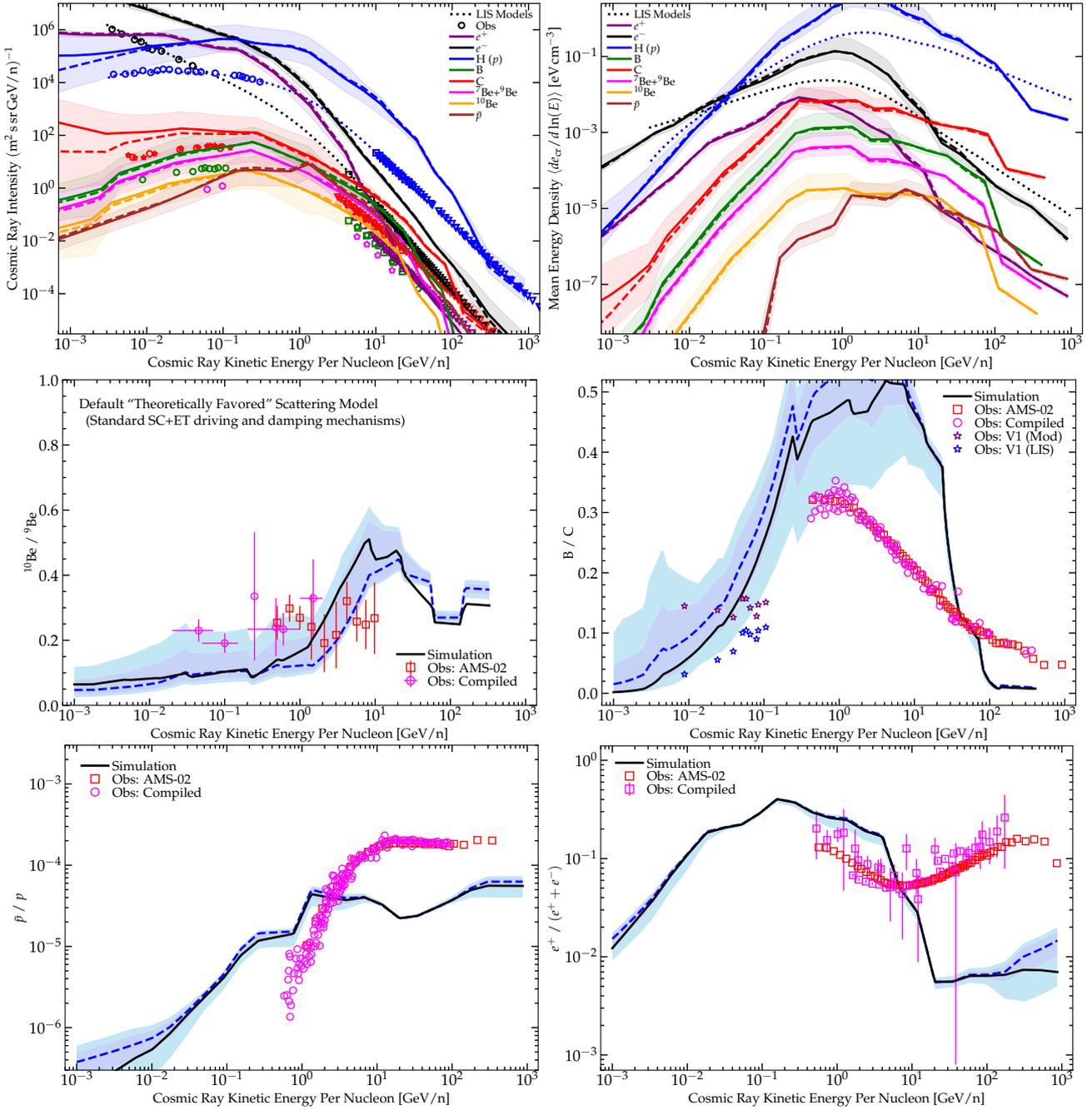

	\figsix{output_01_22_21_newrsol_m10k_SC_model0_coeff1_anisoclosure_hiinit_fire2}{\figsuffixspecial}
	\vspace{-0.1cm}
	\caption{Predicted CR spectra as Fig.~\ref{fig:demo.cr.spectra.fiducial}, but for our default, ``theoretically favored'' model. Here the driving term $S_{\pm}$ for scattering includes both self-confinement (SC; $S_{\rm sc,\,\pm}$) and extrinsic turbulence (ET; $S_{\rm et,\,\pm}$) terms, accounting self-consistently for anisotropy and damping in both, but the ET terms contribute negligibly at the energies plotted (so our results are similar to a ``pure SC'' model with $S_{\rm et}=0$). The initial conditions (ICs) have CR  spectra set to observed values (the ``normal start'' in \S~\ref{sec:important.variations}). We include all standard mode-damping mechanisms (\S~\ref{sec:damping}). From this IC, the initially super-\Alf{ic} streaming at intermediate CR energies (where the CR energy density $e_{\rm cr}^{\prime}$ is relatively high) quickly collapses to the ``bottleneck'' or ``infinite scattering'' solution which gives very slow CR transport (limited by the \Alf\ speed, and independent of rigidity), over-predicting B/C and $e^{+}/e^{-}$ and $e_{\rm cr}^{\prime}$ by an order of magnitude at $\sim1-30\,$GV. At higher CR energies ($\gtrsim 30-100\,$GV) where initial $e_{\rm cr}^{\prime}$ is lower the solutions collapse to the ``unconfined'' or ``free-streaming-at-$c$'' solution with negligible scattering (giving too-low B/C and $e^{+}/e^{-}$). 
	 This is the ``instability'' or ``solution collapse'' problem (\S~\ref{sec:problems:sc}): regardless of IC or re-normalizing the SC driving or damping rates, no stable intermediate solutions between these extremes exist in the context of standard SC models. 
	\label{fig:SC.defaults}\vspace{-0.3cm}}
\end{figure*}

\begin{figure*}
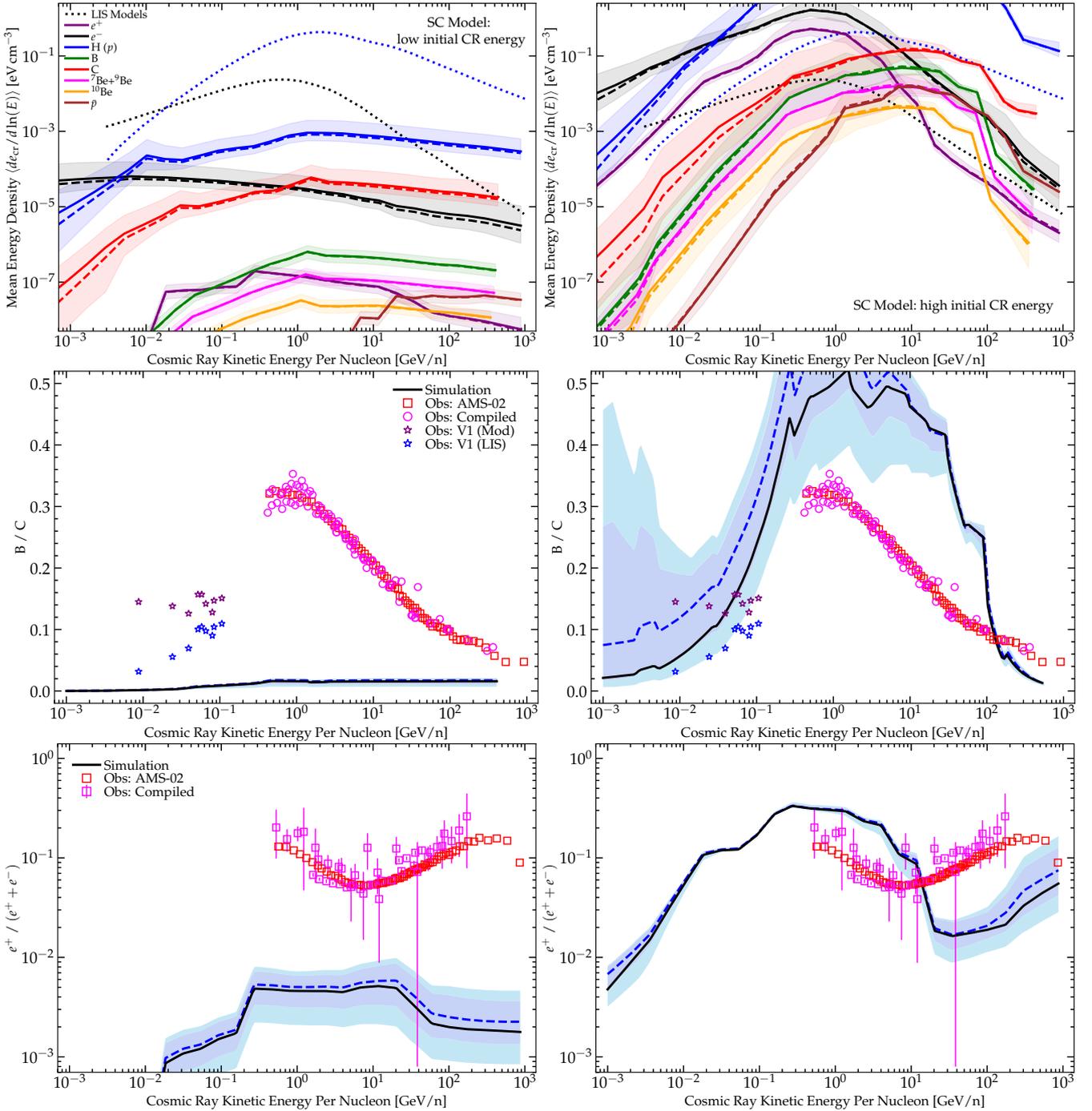

	\figrowlimv{output_01_01_21_newrsol_m10k_SC_model2_coeff1_anisoclosure}{output_01_22_21_newrsol_m10k_SC_model0_coeff1_anisoclosure_hiinit2_fire2}
	\vspace{-0.1cm}
	\caption{As Fig.~\ref{fig:SC.defaults}, but for SC models which adopt a lower or higher initial CR energy (see \S~\ref{sec:important.variations}). 
	{\em Left:} A ``low-start'' IC where we multiply the initial $e_{\rm cr}^{\prime}$ by $\sim 0.001$ relative to observed values. Now CRs at all energies collapse to the unconfined solution (the small residual scattering is from the non-zero ET terms).
	{\em Right:} A ``high-start'' IC where we impose initially flat CR spectra with total CR energy multiplied by $\sim10$ relative to observed. Now a broader range of CR energies  collapse to the ``bottleneck'' solution, near the calorimetric limit, except for the highest-energy CRs which collapse to the unconfined solutions (note the high-energy leptonic spectra are strongly modified by losses here). 
	For simplicity we omit the plots of $^{10}{\rm Be}/^{9}{\rm Be}$, $\bar{p}/p$, and the CR intensity: these disagree with observations in the same manner as expected from B/C, $e^{+}/e^{-}$, and CR spectra shown, so the information is redundant. 
	\label{fig:SC.hilow}\vspace{-0.3cm}}
\end{figure*}

\subsection{How to Rescue Things?}
\label{sec:rescue}

We now consider what would be qualitatively needed in either the damping ($\Gamma_{\pm}$) or driving ($S_{\pm}$) terms for $e_{A}$, in order to obtain some plausible consistency between observations and either SC or ET  CR transport models.

\subsubsection{Alternative Damping}
\label{sec:rescue:damp}

First, let us consider the consequences of modifying or adding only damping terms, while retaining standard SC \&\ ET driving. 

In the ET-dominated limit -- or any model where the dominant contribution to the driving term $S_{\pm}$ derives from an MHD cascade -- no addition or modification of the form of the damping terms can ``rescue'' the models. This is because, as shown above, the cascade assumption itself implies $\delta_{\rm s} <  0$ (i.e.\ the slope of the ``cascade'' is cut off to some non-zero degree) for {\em any} finite damping term if the damping is significant. Even for negligible damping in ET, anisotropy requires $\delta_{\rm s} \le 0$, independent of the form of the damping terms. However, for standard scattering expressions motivated by physical turbulence models (i.e.\ including anisotropy and damping), our the naive expectation is that the driving term $S_{\pm}$ is dominated by SC at all energies of interest (i.e.\ $S_{\rm et,\,\pm} \ll S_{\rm sc,\,\pm}$). Thus, it is most relevant to consider what (if any) damping terms could resolve the problems above for self-confined CRs.

In the SC-dominated limit, one could in principle solve both the ``normalization'' (1) and ``spectral shape'' (2) problems by invoking some arbitrary damping rate $\Gamma_{\rm new,\,damp,\,\pm}$ which has the appropriate normalization and desired scaling as a function of $k$ and $e_{A}$ (i.e.\ varying $\xi_{k}$ as a function of $R_{\rm cr}$, to new values outside the range of known damping mechanisms, as explained in \S~\ref{sec:problems:sc}). This in itself would not solve the ``instability'' problem (3). However, if the dominant damping mechanism {\em also} scaled with $e_{\rm cr}^{\prime}$, then in equilibrium ($S_{\pm} \sim Q_{\pm}$) this would cancel the $e_{\rm cr}^{\prime}$ dependence of $S_{\rm sc,\,\pm}$, resolving issue (3) as well. Specifically, following the parameterization of \S~\ref{sec:damping}, if we had damping with $\xi_{k} \approx 1/2$, $\xi_{A} \approx 0$, $\xi_{\rm cr} \approx 1$, i.e.\ $\Gamma_{\rm new,\,damp,\,\pm} \propto k_{\|}^{1/2}\,e_{\rm cr}^{\prime}$, then since the SC driving term $S_{\rm sc,\,\pm} \propto v_{A,\,{\rm eff}}\,\bhat\cdot \nabla P_{\rm cr}^{\prime}$, this would give approximately the desired $e_{A} \propto k_{\|}^{-1/2}$ independent of $e_{\rm cr}^{\prime}$, thus curing the ``spectral shape'' and ``instability'' problems. An example which gives roughly the correct normalization as well would be something like $\Gamma_{\rm new,\,damp,\,\pm} \sim v_{\rm eff}\,(k_{\|}\,\ell_{A})^{1/2}\,(e_{\rm cr}^{\prime}/e_{\rm B})$, with $v_{\rm eff} \sim c_{s} \sim v_{A,\,{\rm ideal}}$ -- i.e.\ something akin to standard turbulent or linear Landau damping rates, but  multiplied by $e_{\rm cr}^{\prime}/e_{\rm B}$. 

This is perhaps not wildly implausible, if it arose from some non-linear process involving CR back-reaction on the gas (perhaps, for example, shocks induced on small-scales by CR-gas coupling as in \citealt{tsung:2021.cr.outflows.staircase,huang.davis:2021.cr.staircase.in.outflows,2021arXiv210608404Q}). Our simulations below will test both ``microscopic'' (i.e.\ sub-grid, added manually) and ``macroscopic'' (spatially-resolved, on $\gtrsim$\,pc scales, which should emerge self-consistently from the simulation physics) versions of this scenario. However, it is not a perfect solution. Not only would one need to think of an effective damping mechanism that produced the desired scaling above (for which there is no obvious candidate, and our simulations do not appear to produce this ``macroscopically''), but one would also need to ensure that this is the {\em dominant} damping process compared to the other known damping terms in \S~\ref{sec:damping}, which  do not simply disappear. This would need to be true, at least on average in the ISM, for $0.01\,{\rm GV}\lesssim R_{\rm cr} \lesssim 1000\,$GV. But at both low and high $R_{\rm cr}$, $e_{\rm cr}^{\prime}$ is low, so it is quite unclear how a damping rate similar to that proposed, which is suppressed by $e_{\rm cr}^{\prime}$, could dominate over the entire range of interest.

Briefly, we note that an alternative damping mechanism with $\xi_{\rm cr}\approx 0$ which scales with $e_{A}$ more steeply than non-linear Landau damping ($\xi_{A} > 1$) would formally admit steady-state  solutions (solving problem (3), per Appendix~\ref{sec:sc.equilibrium.models}). But since this would give $e_{A} \propto (e_{\rm cr}^{\prime})^{1/(1+\xi_{A})}$, it would still suffer from the spectral shape problem (2), unless one takes $\xi_{A} \gg 1$, with $\xi_{k} \sim (1+\xi_{A})/2 \gg 1$. This is a rather unusual scaling. Moreover, in this limit the required normalization of this added $\Gamma_{\rm new,\,damp\,\pm} \propto e_{A}^{\xi_{A}}\,k_{\|}^{(1+\xi_{A})/2}$ term at the wavenumbers and $e_{A}$ observationally required at $\sim 1\,$GV would be problematically large: $\sim 10^{5}$ times larger than the standard non-linear Landau damping term, despite its being nominally much higher-order in $e_{A}/e_{\rm B}$. As such, we will not consider this particular class of alternative damping model further.

\subsubsection{Alternative Driving}
\label{sec:rescue:drive}

Alternatively, we could invoke a different or additional driving term $S_{\rm new,\,\pm}$, while keeping the known damping mechanisms. Consider an alternative source term as parameterized in \S~\ref{sec:sc}: $S_{\rm new,\,\pm} = d E/d\ln{k_{\|}}\,dt\,d{\rm Volume} \propto k_{\|}^{\zeta_{k}}\,e_{A}^{\zeta_{A}}$ (i.e.\ $\zeta_{\rm cr}=0$, so we avoid the problems of SC models above). Given some damping rates parameterized in similar fashion as in \S~\ref{sec:damping}, then a desired $\delta_{\rm s}$ is obtained for $\zeta_{k} = \xi_{k}\, - (1-\delta_{\rm s})\,(1 + \xi_{A} - \zeta_{A})$. 

{\bf (1) Extrinsic/External Sources:} First consider the case with $\zeta_{A} = 0$, i.e.\ $S_{\rm new,\,\pm} = S_{\rm new,\,ext}$ independent  of $e_{A}$, as appropriate for a truly ``extrinsic'' or ``external'' energy driving/pumping term (akin to ET in this limited sense). Then if we consider $\xi_{A}$, $\xi_{k}$ for all possible damping mechanisms in \S~\ref{sec:damping}, and allow $0.3 \lesssim \delta \lesssim 0.7$, then the range of possible $\zeta_{k}$ that produces the desired $\delta_{\rm s}$ is bounded by $-0.7 \lesssim \zeta_{k} \lesssim 0.5$. But more realistically if we restrict to $\delta_{\rm s} \approx 1/2$, and ignore ion-neutral damping (which has a rather different scaling $\xi_{k}$ from other rates, and is rarely relevant in the volume-filling ISM which dominates the statistics as seen in the LISM), then allowing for all other damping processes requires a much narrower range of $-0.1 \lesssim \zeta_{k} \lesssim 0.25$. In other words, a model with $S_{\rm new,\,ext} \sim $\,constant, or weakly dependent on $k_{\|}$ (hence $r_{g,{\rm cr}}$), is potentially viable. One example, which has approximately the correct normalization if we assume turbulent, linear-Landau, or dust damping dominates $\Gamma_{\pm}$, would be $S_{\rm new,\,ext} \sim 0.01\,(v_{A,\,{\rm ideal}}/\ell_{A})\,e_{\rm B}$ (this could be multiplied by a weak power of $k_{\|}$ or $r_{g,{\rm cr}}$, e.g.\ $(k_{\|}\,r_{g,{\rm cr}}[1\,{\rm GV}])^{\zeta_{k}}$ with $-0.1\lesssim \zeta_{k} \lesssim 0.25$). 

{\bf (2) Linear Sources:} Second, consider the case with $\zeta_{A} = 1$, i.e.\ $S_{\rm new,\,\pm} = S_{\rm new,\,lin} \propto e_{A}$. This would be appropriate for e.g.\ any linear instability which amplifies $e_{A}$. The  gyro-resonant CR instabilities (or non-resonant CR instabilities; \citealt{bell.2004.cosmic.rays}) invoked in SC models are one such instability, but suffer from the problems described in \S~\ref{sec:problems:sc}. For any $\zeta_{A}\approx 1$, again allowing for a broad range of $\delta_{\rm s}$ and all possible damping mechanisms bounds $0 \lesssim \zeta_{k} \lesssim 0.85$, but restricting to $\delta_{\rm s} \approx 1/2$ and ignoring ion-neutral damping gives the much narrower range $0.4 \lesssim \zeta_{k} \lesssim 0.75$. Again, an example here that has approximately the correct normalization would be e.g.\ $S_{\rm new,\,lin} \sim 0.001\,(v_{A,\,{\rm ideal}}/\ell_{A})\,(k_{\|}\,\ell_{A})^{1/2}\,e_{A}$. 

Either of these novel source terms seems to be at least plausible. For the first ``extrinsic source'' case ($\zeta_{A}=0$), the scaling $S_{\rm new,\,ext} \sim $\,constant is just that assumed in isotropic, undamped turbulent cascade models. But, the more general condition is simply  that the driving in energy space -- i.e.\ $d |\delta{\bf B}^{2}(k_{\|})| / d\ln{k_{\|}}\,dt\,d{\rm Volume}$  -- is only  weakly dependent of $k_{\|}$ (i.e.\ the driving rate is comparable across scales from $\sim $\,MeV-TeV gyro-radii). We emphasize that from the constraints in \S~\ref{sec:damping} and assumed structure of the competition with scale-dependent damping, this cannot apply to any of the standard physically motivated models for a traditional MHD turbulent cascade from  larger scales, which would introduce the damping and anisotropy problems. Rather, in order to satisfy the requirements, it is more natural to consider modes as driven and damped effectively ``independently'' on each scale, in a manner where the energy driving/injection rate is comparable per logarithmic interval in parallel $k_{\|}$, but allowing for damping and/or anisotropy and/or transfer so long as this condition is met. 
Note that the required normalization/total energy driving rate in this scenario is  quite small -- only  $\sim 1\%$ of the driving/dissipation rate of ISM turbulence on larger scales. 

For the second ``linear source'' case ($\zeta_{A}=1$), the linear scaling $S_{\rm new,\,lin} \sim \Psi_{\rm new,\,lin}\,e_{A}$ with $\Psi_{\rm new,\,lin} \propto k_{\|}^{0.4-0.75}$ is physically easy to imagine. Most obviously, a huge variety of multi-fluid instabilities present in the ISM exhibit  behavior that could lead to similar scalings. For example at a two-fluid interface, the Rayleigh-Taylor instability with $\Psi_{\rm new,\,lin} \sim \sqrt{g\,k}$ where $g$ is some acceleration, would require only very weak $g \sim 10^{-6}\,v_{A}^{2}/\ell_{A}$ to produce roughly the correct behavior. More generally any ``co-spatial fluids'' -- i.e.\ any two fluids that both share the same volume, such as dust and gas, ions and neutrals, radiation and gas, etc. -- can be unstable to the  family of resonant-drag instabilities (RDIs; \citealt{squire.hopkins:RDI}), many of which drive modes that could scatter CRs with the desired scalings.\footnote{More technically, if a second fluid (e.g.\ dust, radiation, or neutrals) can ``resonate'' with  \Alf\ waves with the desired $k_{\|}$ (i.e., if they have a natural mode with a frequency that matches that of the \Alf\ wave), and any coupling that depends on their relative streaming velocities (e.g.\ collisions, drag, Lorentz forces), then it can produce an RDI. These generally  drive  modes in $k_{\|}$ that could scatter CRs and have growth rates $\gamma \sim k_{\|}^{\alpha}$, where $1/3\lesssim \alpha\lesssim 2/3$ depending on the scale and type of  mode (see \citealt{squire.hopkins:RDI}). 
For the ``intermediate'' $k_{\|}$ or small-$f_{\rm dg}$ case (where $f_{\rm dg}$ is the mass-density ratio of the two fluids), assuming the streaming/drift velocity is sub-\Alf{ic}, this generically gives $\Psi_{\rm new,\,lin} \sim (f_{\rm dg}\,\Delta a_{\rm dg}\,k_{\|})^{1/2}$ where  $\Delta a_{\rm dg}$ is any difference in the accelerations felt by the two fluids that gives rise to non-zero streaming.} For example, the \Alf-wave dust-gas RDI \citep{hopkins:2018.mhd.rdi} is unstable on all scales of interest here, and has a growth rate $\Psi_{\rm new,\,lin} \sim (f_{\rm dg}\,\Delta a_{\rm dg}\,k_{\|})^{1/2}$ at intermediate $k_{\|}$ and $\propto k_{\|}^{1/3}$ at high $k_{\|}$, where $f_{\rm dg}$  is the dust-to-gas ratio ($f_{\rm dg}\lesssim 0.01$ in the Milky Way, depending on gas phase),  $\Delta a_{\rm dg}$  is any external acceleration felt differently by dust and gas (e.g.\ radiation pressure), and we have assumed the dust-gas drift speed is sub-\Alf{ic}. So such a mechanism would require $f_{\rm dg}\,\Delta a_{\rm dg} \sim 10^{-6}\,v_{A}^{2}/\ell_{A}$ in order to scatter CRs sufficiently, well within plausible ranges \citep[see e.g.][]{weingertner.draine:photo.forces.on.dust.hard.ism.rad}. Alternatively, very similar RDIs can arise between the ionized and neutral gas phases in partially-ionized gas, several of which are studied in \citet{tytarenko:two.fluid.drift.intabilities} with growth rates $\Psi_{\rm new,\,lin} \propto k_{\|}^{1/3-2/3}$ depending on $k_{\|}$ (although some of these are stabilized on small scales by pressure effects in the neutrals). 

\begin{figure*}
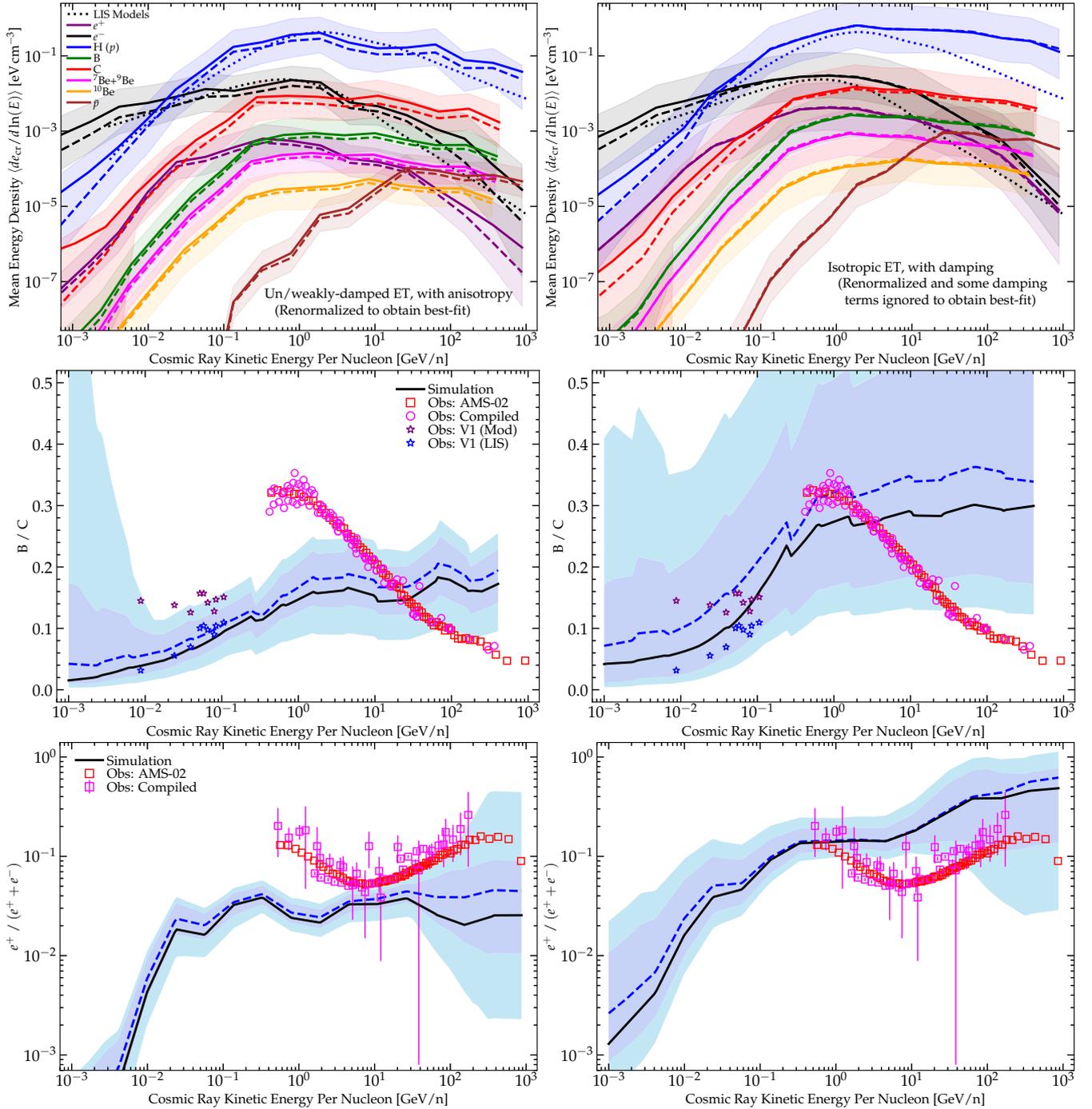

	\figrowlimv{output_05_19_21_newrsol_m10k_SC_model0_coeff1_anisoclosure_loinit_AlfMax_x10_wall}{output_05_19_21_newrsol_m10k_SC_model0_coeff1_anisoclosure_loinit_YL04_onlyturb_corrected}
	\vspace{-0.1cm}
	\caption{CR spectra (as Fig.~\ref{fig:SC.hilow}) now for pure-ET models, demonstrating the anisotropy, normalization, and damping problems (\S~\ref{sec:problems:et}). Here we disable SC driving ($S_{\rm sc,\,\pm}\rightarrow 0$), and arbitrarily re-normalize the ET driving ($S_{\rm et,\,\pm}$) or damping rates to force the models to match  CR proton spectra at $\sim 1$\,GV. 
	{\em Left:} Representative behavior of any model that accounts for anisotropy in the \Alf{ic} cascade (e.g.\ any model obeying critical balance, such as GS95). This imposes $S_{\rm et,\,\pm} = \alpha_{t}\,e_{\rm B}\,\Gamma_{\rm turb}\,(k_{\|}\,\ell_{A})^{-1}$ (\S~\ref{sec:et}), We renormalize $\alpha_{t}\sim1$ (a factor $\sim 10^{3}-10^{6}$ larger than theoretically favored) to fit spectra at $\sim 1\,$GV. But the spectral shapes are still incorrect. Accounting for damping  or deviating from critical balance makes the disagreement worse (see \S~\ref{sec:turb.review}). 
	{\em Right:} Representative behavior of any model assuming an isotropic fast-magnetosonic cascade, accounting for the fact that the spectrum is modified by damping/dissipation on a scale $\lambda_{\rm diss}$ larger than the gyro scale (here following YL04; \S~\ref{sec:et}). We renormalize to fit the $p$ spectrum by disabling ion-neutral, dust and other parallel or \Alf{ic} damping terms and adopting the YL04 scalings for plasma $\beta_{\rm plasma}<1$ everywhere (otherwise the scattering rate is reduced by $\sim 10^{6}$). 
	Even re-normalizing to force a reasonable mean scattering rate at $\sim$\,GV, these models cannot reproduce observed spectral shapes: accounting for anisotropy and/or finite dissipation scales forces ET scattering rates (therefore B/C) to be {\em independent} or even {\em increasing} functions of CR energy at $\gtrsim$\,GV (e.g.\ $\delta_{\rm s} \le 0$), contradicting observations.
	\label{fig:ET.defaults}\vspace{-0.3cm}}
\end{figure*}

These alternative source models appear more well-motivated than the alternative damping model in \S~\ref{sec:rescue:damp}. To start, there are physically-motivated, known processes  that could potentially produce the correct additional source terms. But also, they  do not  require that we ``discard''  major known damping mechanisms or other known source terms, in order to make these models ``work'' (whereas with the alternative damping model, we must invoke some other physics to explain why other other known damping processes do not dominate). In fact, we can simply ``add''  similar driving  terms on top of  the known terms in \S~\ref{sec:damping}-\ref{sec:et}. There is, however, one remaining caveat if we do so. We still need to avoid the SC bottleneck/runaway (if we still include $S_{\rm sc}$) in regions that {\em do} have high $e_{\rm cr}^{\prime}$, specifically at rigidities $\sim 1-10\,$GV where $e_{\rm cr}^{\prime}$ is maximized. In other words,  $S_{\rm sc}$ would be expected to still be large and potentially dominate over $S_{\rm new}$ by  a factor as large as $\sim 100$, given the ``normalization'' problem in \S~\ref{sec:sc} (which imposing new driving terms does not  solve). This is not a serious problem outside of the range $\sim 1-10\,$GV, because $e_{\rm cr}^{\prime}$ is smaller. But at $\sim 1-10\,$GV, all of the plausible solutions to the ``normalization'' problem discussed in \S~\ref{sec:sc} -- e.g.\ more accurately calculating gyro-resonant growth rates that account for the full CR spectrum, and more accurately calculating SC-induced scattering rates ($\hat{\nu}_{\rm s}$ arising from the SC) -- act to  reduce the SC contribution, so can potentially alleviate this problem.

\clearpage

\section{Numerical Simulation Methods}
\label{sec:methods}

We now test these analytic conclusions in detailed numerical simulations, beginning by describing our simulation methods. Briefly though, we note that there are several motivations to explore fully-dynamical simulations of global galaxy formation and structure. One is to test whether these conclusions are robust in a more realistic, turbulent, multiphase medium (in which plasma properties such as $v_{A}$, etc.\ vary on scales which are resolved but still small compared to the size of the entire Galaxy; see discussion in \paperone) as is present in these numerical simulations, but cannot be captured in even state-of-the art analytic Galactic structure models \citep[compare e.g.][]{maurin:2018.cr.sam.favored.parameters.close.to.fire,benincasa:2020.gmc.lifetimes.fire}. Another is to test whether non-equilibrium CR dynamics (i.e.\ dynamical behaviors whether either the background plasma, or CR flux or energy equations themselves are not in equilibrium), neglected in any ``steady-state'' models, could impact these conclusions \citep[see e.g.][]{bustard:2020.crs.multiphase.ism.accel.confinement,hopkins:cr.transport.constraints.from.galaxies,thomas:2022.self-confinement.non.eqm.dynamics}. Yet another motivation is to test whether ``feedback'' or back-reaction effects from CRs on the medium, necessarily ignored in any post-processing models where CRs are not evolved ``on the fly'' could somehow produce different conclusions. Examples of this include the effect of global CR pressure gradients and CR coupling to magnetic fields re-accelerating outflows to large CGM radii \citep{salem:2013.cosmic.ray.outflows,Simp16,wiener:2017.cr.streaming.winds,hopkins:2020.cr.outflows.to.mpc.scales}; or producing strong shocks, mixing via bouyancy effects, or thermally heating the CGM \citep{ensslin:2011.cr.transport.clusters,wiener:cr.supersonic.streaming.deriv,su:turb.crs.quench,su:2021.agn.jet.params.vs.quenching,wellons:2022.smbh.growth}; or altering the phase structure of the CGM via allowing gas to occupy states prohibited in strict thermal pressure equilibrium \citep{Sale16,Buts18,butsky:2020.cr.fx.thermal.instab.cgm,ji:fire.cr.cgm,ji:20.virial.shocks.suppressed.cr.dominated.halos}; or altering the vertical support, hence pressure balance, turbulent strength, or magnetic field strengths of galactic disks \citep{Wien13,hopkins:cr.mhd.fire2,chan:2021.cosmic.ray.vertical.balance,ponnada:fire.magnetic.fields.vs.obs}; or altering ionization balance in neutral gas in e.g.\ GMCs or galactic nuclei \citep{gaches:2018.protostellar.cr.acceleration,hopkins:cr.multibin.mw.comparison,armillotta:2021.cr.streaming.vs.environment.multiphase}. We emphasize that all of these effects have been studied in previous simulations with the same physics, code/numerical methods, and resolution (references given), demonstrating that they can in fact be captured -- the difference is that these previous studies have generally treated CRs in the ``single-bin'' approximation (integrating only a total CR energy, rather than the full spectrum, and neglecting differences between species) with a simple phenomenological transport/scattering rate model (e.g.\ a single universally-constant scattering rate or diffusivity or streaming speed).\footnote{Briefly, it is also worth noting that even studies using entirely different codes and numerical methods and initial conditions have largely produced similar results to the FIRE simulations referenced here, provided they include similar physics and adopt similar CR transport parameters (see e.g.\ the discussion in \citealt{armillotta:2021.cr.streaming.vs.environment.multiphase}). Moreover, while there is of course some resolution-dependence and there will necessarily be un-resolved scales (discussed in more detail below), these appear to have little effect on the mean properties predicted in the studies above (see references therein) and to manifest weakly via effects like the failure to resolve small molecular clouds leading to slightly more-clustered star formation and higher variability \citep[see][]{hopkins:fire2.methods,hopkins:fire3.methods,armillotta:2021.cr.streaming.vs.environment.multiphase}, which only strengthens our ultimate conclusions since it means the simulations sample a broader range of possibilities.}

\subsection{Non-CR Physics} 
\label{sec:methods:overview}

The simulations studied here are identical to those in \citet{hopkins:cr.multibin.mw.comparison} (hereafter \paperone), except for the expressions used for the CR scattering rates, so we only briefly summarize the methods here. The simulations are run with {\small GIZMO}\footnote{A public version of {\small GIZMO} is available at \gizmourl} \citep{hopkins:gizmo}, in meshless finite-mass (MFM) mode, with magneto-hydrodynamics (MHD) solved as in \citet{hopkins:mhd.gizmo,hopkins:cg.mhd.gizmo} with anisotropic Spitzer-Braginskii conduction/viscosity as in \citet{hopkins:gizmo.diffusion,su:2016.weak.mhd.cond.visc.turbdiff.fx}, self-gravity solved with adaptive Lagrangian force softening, and cooling, star formation, and stellar feedback following the FIRE-3 implementation of the Feedback In Realistic Environments (FIRE) physics \citep{hopkins:fire2.methods,hopkins:fire3.methods}. We explicitly follow enrichment, dynamics, and chemistry of 11 species \citep{colbrook:passive.scalar.scalings,escala:turbulent.metal.diffusion.fire}, cooling and non-equilibrium ionization/atomic/molecular chemistry from $\sim 1-10^{10}$\,K including metal-line, molecular, fine-structure, photoelectric, ionization, and other processes with local sources and a meta-galactic (self-shielded) background from \citet{cafg:2020.uv.background}. Locally self-gravitating Jeans-unstable gas in converging flows is allowed to form stars following \citet{hopkins:virial.sf,grudic:sfe.cluster.form.surface.density}, and once formed stars evolve according to explicit stellar evolution models and return metals, mass, momentum, and energy to the ISM via resolved individual SNe (both Ia \&\ core-collapse) and O/B and AGB mass-loss as in \citet{hopkins:sne.methods}, with radiative heating and momentum fluxes solved using a five-band radiation-hydrodynamic approximation from \citet{hopkins:radiation.methods}. We note resolution tests below but the default mass resolution is $\Delta m_{i} \approx 7000\,{\rm M_{\odot}}$, so the spatial/force resolution scales with density as $\Delta x_{i} \sim 10\,{\rm pc}\,(n/100\,{\rm cm^{-3}})^{-1/3}$, and the simulations naturally feature a multi-phase ISM with hot phases at $n \ll 0.01\,{\rm cm^{-3}}$ and molecular clouds (with the mass spectrum and other scalings of the most massive, resolved clouds agreeing well with observations; \citealt{guszejnov:fire.gmc.props.vs.z,benincasa:2020.gmc.lifetimes.fire,keating:co.h2.conversion.mw.sims}) up to the maximum densities where the fragmentation scale can be resolved of $n \sim 10^{3}-10^{4}\,{\rm cm^{-3}}$ \citep[see][for more details]{hopkins:fire2.methods}. The simulations here are ``controlled restarts'' where we take a fully-cosmological simulation run from $z\sim100$ to $z=0$ with a simpler CR treatment from \citet{hopkins:cr.mhd.fire2}, selected because it forms a galaxy similar in all obvious relevant properties to the Milky Way (MW), and restart it from a snapshot at $z\approx 0.05$, modifying the CR assumptions, and running for $\approx500\,$Myr to $z=0$. This is sufficient for all CR quantities in the ISM to reach their new quasi-steady-state values, but ensures (unlike running an entirely new cosmological simulation) that our comparison is ``controlled'' (bulk Galaxy properties are similar). All numerical details of the methods are described and tested extensively in \citet{hopkins:cr.multibin.mw.comparison}.

\subsection{CR Physics} 
\label{sec:methods:crs}

Following \citet{hopkins:m1.cr.closure}, we explicitly evolve the CR DF $f_{\rm cr}=f_{\rm cr}({\bf x},\,{\bf p}_{\rm cr},\,t,\,s,\,...)$ assuming a gyrotropic DF following Eqs.~\ref{eqn:f0}-\ref{eqn:f1}. By definition $\langle \mu \rangle \equiv \bar{f}_{\rm cr,\,1}/\bar{f}_{\rm cr,\,0}$ and the moments hierarchy for $\bar{f}_{\rm cr,\,2}$ is closed by the interpolated M1-like relation $\langle \mu^{2} \rangle \approx (3+4\,\langle \mu\rangle^{2}) / (5 + 2\,\sqrt{4-3\,\langle \mu\rangle^{2}})$, which captures the exact behavior in both the ``free streaming'' or weak-scattering and isotropic-DF or strong-scattering limits \citep{hopkins:m1.cr.closure}. All the variables above are functions of position and time. CRs act on the gas and radiation fields: the appropriate collisional/radiative terms are either thermalized or added to the total radiation (e.g.\ Bremstrahhlung, inverse Compton, etc.) or magnetic energy, and the CRs exert forces on the gas in the form of the Lorentz force (proportional to the perpendicular CR pressure gradient) and parallel force from scattering as detailed in \citet{hopkins:m1.cr.closure}.  

The momentum-space evolution of $\bar{f}_{\rm cr,\,0}$ is integrated independently in every resolution element using the finite-momentum-space-volume scheme in \citet{girichidis:cr.spectral.scheme}, treating $\bar{f}_{\rm cr,\,0}(p_{\rm cr})$ as a series of independent piecewise power-laws with exactly computed number and energy fluxes (so the scheme exactly conserves CR number and energy). We discretize with $\sim 11$ independent power-law intervals (each with an evolving slope and normalization) for $\bar{f}_{\rm cr,\,0}(p_{\rm cr})$ spanning $\sim$\,MeV-TeV energies, per CR species, per simulation cell. We cannot resolve first-order Fermi acceleration so we model injection by assuming $\sim10\%$ of the initial pre-shock kinetic energy goes into CRs, with $\sim2\%$ of that into leptons, at the formation of the reverse shock around each SNe and/or O/B winds, with the relative number per species set by the evolved abundances at that point (e.g.\ the test-particle limit) with a fixed injection spectrum $j_{\rm cr}(R_{\rm cr}) \propto R_{\rm cr}^{-4.2}$. We explicitly follow the CR species protons $p$ (H), CNO, $^{7}$Be, $^{9}$Be, $^{10}$Be, anti-protons $\bar{p}$, electrons $e^{-}$, and positrons $e^{+}$. In the loss terms $\mathcal{R}_{\rm loss}$ and $j_{\rm cr}$, we include Coulomb \&\ ionization, Bremstrahhlung, inverse Compton, synchrotron, pionic, fragmentation, radioactive decay, and annihilation processes, with standard cross sections compiled in \paperone. This includes secondary production of e.g.\ $e^{+}$, $e^{-}$, B, C, Be, etc.  All ISM quantities needed for these rates (e.g.\ gas densities \&\ ionization states, magnetic \&\ radiation energy densities, etc.) are taken directly from the dynamically-evolved simulation quantities in the cell. As also noted in \citet{hopkins:m1.cr.closure}, our Eq.~\ref{eqn:f0}-\ref{eqn:f1} automatically include ``adiabatic'' ($\mathbb{D}_{\rm cr}:\nabla{\bf u}_{\rm gas}$), ``streaming loss'' or ``gyro-resonant'' ($\propto \bar{v}_{A}$ or $D_{\mu p}$, $D_{p \mu}$), and ``diffusive'' ($\propto D_{p p}$) re-acceleration terms, in more general and accurate forms than usually considered. 

We calculate $\bar{\nu}_{\rm s,\pm}$ following Eq.~\ref{eqn:nu}, with $\hat{\nu}_{\rm s} \approx 3/4$ appropriate for grey scattering, and $e_{\pm}$ determined from Eq.~\ref{eqn:ea}, for a given set of sources $S_{\pm} = \sum_{i} S_{i,\,\pm}$ and damping terms $Q_{\pm} = \sum_{i} \Gamma_{i,\,\pm}\,e_{\pm}$. We have considered both the cases where we explicitly dynamically evolve the time-dependence of Eq.~\ref{eqn:nu} alongside the CR flux and energy equations, or where we simply set $e_{\pm}$ to the local-steady-state values (setting $D_{t} e_{\pm} \pm \nabla\cdot(v_{A}\,e_{\pm}\,\bhat) \rightarrow 0$); these give very similar results for our study below, so we default to the local-state-state values as it involves slightly reduced computational expense, which allows a larger parameter survey. The key physics we vary in our tests is the scaling of the sources and damping rates $S_{i,\,\pm}$ and $\Gamma_{i,\,\pm}$, or equivalently scattering rates $\bar{\nu}_{\rm s}$.

\subsection{Reference Model \&\ Quantities Measured}
\label{sec:reference.model}

We stress from the above that (1) all of the CR physics needed to resolve, in principle, any of the known relevant CR-gas interactions or feedback effects on ``macroscopic'' (simulation-resolved) scales are included; (2) all of the plasma properties (e.g.\ ${\bf B}$, $n$, $\ell_{A}$, $e_{\rm cr}$) needed to calculate the ``microscopic'' (unresolved, gyro-scale) scattering rates in the (extrapolative) models we will consider are self-consistently predicted on the resolved simulation scales; and (3) given those (assumed) scattering rates, our simulations naturally produce a self-consistent prediction for the CR spectra across the range of energies and species we consider.

The key physics of CR transport in our model therefore reduces to our expressions for the source $S_{\pm}$ and damping $Q_{\pm}$ terms in Eq.~\ref{eqn:ea}. We will explore many model variations, but it is useful to first define a ``reference model,'' which attempts to represent the best current theoretical understanding of SC+ET effects as developed in e.g.\ \citet{Zwei13,Rusz17,zweibel:cr.feedback.review,thomas.pfrommer.18:alfven.reg.cr.transport} and other references in \S~\ref{sec:damping}-\ref{sec:et}. In this ``baseline'' model, we take: $S_{\pm} = S_{\rm sc,\,\pm} + S_{\rm et,\,\pm}$ where $S_{\rm sc,\,\pm}$ follows Eq.~\ref{eqn:scr}, and $S_{\rm et,\,\pm} = \alpha_{t}(k_{\|})\,e_{\rm B}\,\Gamma_{\rm turb}\,(k_{\|}\,\ell_{A})^{-1}$ assumes an anisotropic GS95-like \Alf{ic} cascade \citep{chandran00}. We take $Q_{\pm} = (\Gamma_{\rm in} + \Gamma_{\rm dust} + \Gamma_{\rm turb/LL})\,e_{\pm} + \Gamma^{0}_{\rm nll}\,(e_{\pm}/e_{\rm B})\,e_{\pm}$ where the expressions for $\Gamma_{\rm in}$, $\Gamma_{\rm dust}$, and $\Gamma^{0}_{\rm nll}$ are in \S~\ref{sec:damping}, and for consistency with our driving terms $S_{\pm}$ (since we are assuming parallel modes being sheared out by a GS95-type cascade) we have $\Gamma_{\rm turb/LL} = [(v_{A,\,{\rm ideal}} + 0.4\,c_{s})/\ell_{A}]\,(k_{\|}\,\ell_{A})^{1/2}$ (where the $0.4\,c_{s}$ term is the ``linear Landau'' term). We use the appropriate $v_{A,\,{\rm eff}}$ in Eq.~\ref{eqn:va.eff} for the relevant $v_{A}$ terms in the CR equations, and note that these expressions self-consistently determine the relation between $\bar{v}_{A}$ and $v_{A}$.

We will focus on comparison of the models here to the Solar-neighborhood/local ISM (LISM) constraints -- the only place where the full CR spectrum of various species can be determined. In \paperone\ (where we study only phenomenological CR transport models), we consider a more extensive suite of constraints including spatially-resolved $\gamma$-ray emission and ionization constraints that span various Galactic environments, as well as comparisons of different CR species and abundances not shown here. While of course any ultimate ``successful'' model must produce agreement with {\em all} of these constraints, our focus here is ruling out a number of models which {\em cannot} reproduce the observations, for which a simpler comparison of the LISM spectral shapes and secondary-to-primary ratios is both sufficient and most useful, given that the theoretical slope $\delta_{\rm s}$ most directly manifests in the shape of the predicted secondary-to-primary ratio as a function of energy. The details of how we compare to observations are given in \paperone, but briefly we select all gas cells in a mock Solar circle (at galacto-centric radii $r = 7-9\,$kpc), in the midplane ($|z|<1\,$kpc), with gas densities similar to those observed ($n\sim 0.3-3\,{\rm cm^{-3}}$), and calculate the median CR spectrum of all gas in this ensemble. To define the ``scatter'' we allow for a wider range of both galactocentric radii ($4-10\,$kpc) -- allowing for the fact that our galaxies are not perfect Milky Way analogs -- and a wider range of densities ($n=0.1-10\,{\rm cm^{-3}}$) and compute the inter-quartile ranges of all CR spectra in all cells meeting these criteria. Of course, we expect CR spectra to vary with Galactic environment, and this is discussed extensively in \paperone. We further have examined all of our predicted CR spectra in both different Galactic annuli from $r=1-15\,$kpc and at different densities $0.001-10\,{\rm cm^{-3}}$, as well as gas selected only in different thermal phases (though this is closely related to density selection as shown in \paperone); importantly, while the normalization and detailed spectral shape of the CRs can depend on these environmental properties, none of our conclusions (particularly about the shape of B/C and $\delta_{\rm s}$, and the success or failure of different models) depends on exactly where or how we measure the CR spectra.

\subsection{Model Variations Considered} 
\label{sec:methods:variations}

We have tested a large number of model variations in our simulations (many in concert with one another), in order to systematically survey whether different changes to our default model could resolve the qualitative tensions described above. Here we outline variations considered, grouping them into those which have no appreciable effect on the {\em qualitative} behaviors of interest in this paper, and those that we find to be most significant.

\subsubsection{Variations that Do Not Alter Our Qualitative Conclusions}
\label{sec:nuisance}

The following variations -- all of which we have tested in full simulations to verify the robustness of our results -- do not alter our qualitative conclusions, even if they produce systematic or quantitative shifts in predicted quantities. We therefore will not discuss them in detail below, but list them for completeness here. 

\begin{itemize}

\item{Changing Galaxy \&\ Stellar Assumptions:} As studied in detail in \paperone\ for simpler power-law scattering rates, we have re-run adopting two different fiducial Milky Way (MW) like galaxy simulations as our initial condition ({\bf m12f} and {\bf m12m}, instead of our usual default {\bf m12i} here), all of which are similar to the real MW but differ in various details \citep{garrisonkimmel:local.group.fire.tbtf.missing.satellites,samuel:2020.plane.of.satellites.fire}. Also as in \paperone, have also arbitrarily multiplied the magnetic fields in our {\bf m12i} initial condition by $10$ and $0.1$ (even though the ``default'' values agree well with MW observations; \citealt{su:fire.feedback.alters.magnetic.amplification.morphology,guszejnov:fire.gmc.props.vs.z}), as these are both theoretically and observationally uncertain and influence the transport physics. Finally, we have also re-run using both our FIRE-3 \citep{hopkins:fire3.methods} and the older FIRE-2 \citep{hopkins:fire2.methods} implementation of the FIRE physics, the latter of which uses older stellar evolution and cooling tables leading to slightly different SNe and stellar mass-loss rates, detailed cooling physics, etc. As shown in \paperone\ these make significantly smaller differences compared to changing CR transport coefficients at the level of detail considered here.

\item{CR Injection Parameters:} We have systematically varied the injection spectrum, e.g.\ considering slopes $j_{\rm cr}\propto R_{\rm cr}^{-\psi_{\rm inj}}$ within a broad range of $\psi_{\rm inj}=3.2-5.2$, allowing for a ``broken'' power-law with a break at $\sim 1\,$GV, freeing the normalization of the injected energy fraction and normalization of different components (e.g.\ leptonic vs.\ hadronic). These variations are again discussed in detail in \paperone; they can be used to improve the agreement of a given model with observed CR spectral shapes, but do not resolve the qualitative problems that are evident in secondary-to-primary ratios.

\item{\Alf\ Speeds \&\ Streaming:} We have considered replacing the full expression $v_{A,\,{\rm eff}}$ (Eq.~\ref{eqn:va.eff}) for the gyro-resonant \Alf\ speed with either the ``ion \Alf\ speed'' ($v_{A,\,{\rm ion}}$), which is nearly identical to $v_{A,\,{\rm eff}}$, or with the ideal \Alf\ speed $v_{A,\,{\rm ideal}}$, which is much lower in dense neutral gas. While the latter has non-negligible quantitative effects (see \paperone), because the overwhelmingly-neutral gas has a relatively small volume filling-fraction (so contributes only modestly in a weighted sense, for diffusive CRs), it does not alter the qualitative conclusions here regarding the success or failure modes of different models. We have also varied the ``streaming speed'' $\bar{v}_{A} \equiv v_{A}\,(\bar{\nu}_{\rm s,+}-\bar{\nu}_{\rm s,-})/(\bar{\nu}_{\rm s,+}+\bar{\nu}_{\rm s,-})$, which we by default solve for explicitly, by replacing it with either exactly $|\bar{v}_{A}|=0$ (the expectation in e.g.\ ``pure ET models'') or $|\bar{v}_{A}|= v_{A,\,{\rm eff}}$ (the expectation in ``pure SC models''). This again has little effect, as this is generally sub-dominant to the diffusive or ``super-\Alf{ic}'' streaming speed \citep{evoli:dragon2.cr.prop,chan:2018.cosmicray.fire.gammaray,su:turb.crs.quench,2018AdSpR..62.2731A,hopkins:cr.transport.constraints.from.galaxies,delaTorre:2021.dragon2.methods.new.model.comparison}. 

\item{Renormalizing or Disabling Different ``Reference Model'' Terms:} We have re-run our reference model, multiplying each source term ($S_{\rm sc}$, $S_{\rm et}$) and damping term ($\Gamma_{\rm in}$, $\Gamma_{\rm dust}$, $\Gamma_{\rm turb/LL}$, $\Gamma_{\rm nll}$) by $100$ and by $0.01$ or $10^{-10}$ (effectively disabling it entirely). While this can have large effects in some cases (discussed below), and ameliorate some of the ``normalization'' tensions described above, none of these modifications, in and of themselves, resolves the fundamental issues of the failure of SC or ET models (whichever is dominant) -- i.e.\ there is no ``single term'' which drives the qualitative problems discussed above, and only the ``normalization'' problem is substantively addressed by these renomalization experiments \citep[see also][]{hopkins:cr.transport.constraints.from.galaxies}.

\item{Variant ET Models:} We have experimented with a number of variant ``pure ET'' models (disabling the SC source term $S_{\rm sc}$) or ET+SC models (retaining $S_{\rm sc}$) which vary $S_{\rm et,\,\pm}$ (and $\Gamma_{\rm turb}$ which must match appropriately). For each we have considered both (a) retaining all the other, usual damping $\Gamma_{\pm}$ terms (e.g.\ $\Gamma_{\rm in}$) or (b) disabling all damping terms other than the cascade transfer term $\Gamma_{\rm turb}$, so the spectrum is exactly that predicted by classical ET models. We consider each of the ET models reviewed in \citet{hopkins:cr.transport.constraints.from.galaxies} and \S~\ref{sec:et}: (1) ``standard'' \Alf{ic} turbulence (our default $S_{\rm et,\,\pm}$); (2) ``\Alf-Max,'' which assumes an anisotropic \Alf{ic} cascade but arbitrarily sets $\alpha_{t}=1$ (this ignores the gyro-averaging correction for anisotropic modes, but still retains the effect of anisotropy producing an $e_{A} \propto k_{\|}^{-1}$ spectrum); (3) ``YL04'' which follows \citet{yan.lazarian.04:cr.scattering.fast.modes} as detailed in the Appendices of \citet{hopkins:cr.transport.constraints.from.galaxies}, accounting for collisionless and viscous damping, and accounting for the much stronger effects of damping (super-exponentially suppressing $S_{\rm et,\,\pm}$) when the neutral fraction is non-zero or plasma $\beta_{\rm plasma} \equiv c^{2}_{s}/v^{2}_{A,\,{\rm ideal}} > 1$, as well as a variant where we neglect the predicted suppression from ion-neutral damping or plasma $\beta_{\rm plasma} > 1$ (the ``Fast-Max'' or ``YL04-Max'' model from \citealt{hopkins:cr.transport.constraints.from.galaxies}); (4) a model which assumes a critically-balanced \Alf{ic} cascade but with a modified cascade rate  (which might be motivated by alignment/intermittency effects, e.g.\ $\alpha=1$ in the notation of \citealt{boldyrev:2005.dynamic.alignment} or $\delta=1/8$ in \citealt{schekochihin:2020.mhd.turb.review}), which can modify the perpendicular spectrum significantly, but again (necessarily) gives an $e_{A} \propto k_{\|}^{-1}$ parallel spectrum 
with only a weakly-modified $\Gamma_{\rm turb} \rightarrow (v_{A,\,{\rm ideal}}/\ell_{A})\,(k_{\|}\,\ell_{A})^{\xi_{k}}$ (we take $\xi_{k}=0.4$ as a somewhat ad-hoc example, for the sake of comparison).

\item{Numerical Variations:} We have considered a number of numerical variations, including (1) replacing the more general second-moment closure relation from \citet{hopkins:m1.cr.closure} with the assumption that the CRs are always near-isotropic \citep[as in][]{thomas.pfrommer.18:alfven.reg.cr.transport}, or (2) in flux-steady state (reducing the CR equations to a diffusion+streaming equation; \citealt{Zwei13}), (3) testing different ``reduced speed of light'' numerical approximations from $\tilde{c} \sim 0.01-1\,c$ to ensure convergence, (4) comparing a re-simulation at 8x improved mass resolution (2x improved force resolution), or (5) directly integrating or assuming local steady-state  for the scattering modes ($D_{t} e_{\pm} \rightarrow 0$, as discussed above). As shown in \paperone, and \citet{chan:2018.cosmicray.fire.gammaray} in more detail, these have quite weak effects on our results. 

\end{itemize}

\subsubsection{Variations that Matter}
\label{sec:important.variations}

The variations we have studied that do lead to important results, discussed below, are summarized here.

\begin{itemize}

\item{``High'' or ``Low'' Initial CR Energies:} As described above, in our ``default'' initial conditions, we initialize the total CR energy density following the consistent result of a cosmological simulation, with spectral shapes matched to those observed, but these quickly converge to new equilibria. However, we have also experimented with a ``low start'' and ``high start'' IC, in which we multiply the initial CR energy (renormalizing all spectra) by factors of $0.001$ and $10$, respectively. This does not change our conclusions and neither IC resolves the SC or ET problems above; and for models where the source term $S_{\pm}$ does not depend on CR energy (non-SC-dominated) this has little effect (the simulations converge to the same equilibrium, independent of this choice). However, for SC models, where $S_{\rm sc,\,\pm} \propto e_{\rm cr}^{\prime}$, we will show that this determines which ``attractor'' solution the SC model converges towards, as described in the ``instability'' problem (\S~\ref{sec:problems:sc}) for SC. 

\item{Adding New Damping Mechanisms:} We experiment with several variant models where we add a new damping term $\Gamma_{\rm new,\,damp,\,\pm}$ (optionally disabling other damping terms in our ``reference'' model $\Gamma_{\rm in}$, $\Gamma_{\rm dust}$, $\Gamma_{\rm turb/LL}$, $\Gamma_{\rm nll}$, in turn), motivated by the discussion in \S~\ref{sec:rescue:damp}. In the most interesting of these experiments, we add a new damping term with the form $Q_{{\rm new},\,\pm} = \Gamma_{\rm new,\,damp,\,\pm}\,e_{\pm}$ with $\Gamma_{\rm new,\,damp,\,\pm} = f_{\rm ISM}^{\Gamma}\,k_{\|}^{\xi_{k}}\,(e_{\rm cr}^{\prime}/e_{\rm B})$, where $f_{\rm ISM}^{\Gamma}$ and $\xi_{k} \sim 1/2$ are varied as described below. 

\item{Adding New Sources:} We experiment with variant models where we add a new source term, considering both ``external'' and ``linear'' sources motivated by the discussion in \S~\ref{sec:rescue:drive}, with $S_{\pm} = S_{\rm new,\,\pm} = f_{\rm ISM}^{S}\,k_{\|}^{\zeta_{k}}\,e_{\pm}^{\zeta_{A}}$ with $\zeta_{A}=0$ (external) or $\zeta_{A}=1$ (linear) and $f_{\rm ISM}^{S}$, $\zeta_{k}$ varied as described below. We again consider both this added directly on top of our reference model, or disabling/renormalizing various others of the ``reference'' source or damping terms in turn.

\end{itemize}

\section{Results: Model Comparison}
\label{sec:results}

We now examine the results of the full simulations. To remind the reader, these self-consistently follow the dynamics of cosmological magnetized gas inflow into galaxies, cooling, and star formation, followed by stellar evolution, stellar mass-loss in O/B and AGB winds, radiative heating and photo-ionization and, for massive stars, supernova explosions, which inject a spectrum of cosmic rays back into the ISM alongside energy and momentum which drive galactic outflows. Phenomena such as galactic winds, turbulence, clumping, magnetic dynamo amplification, and the like are followed self-consistently. In this medium the injected cosmic rays propagate according to the full dynamics equations (e.g.\ Eqs.~\ref{eqn:f0}-\ref{eqn:f1}, incorporating diffusion, streaming, adiabatic gains/losses, diffusive re-acceleration, catastrophic losses, radiative losses, and the like), producing secondary and tertiary cosmic rays ``on the fly'' while they propagate. Importantly, the cosmic rays interact directly with the gas dynamically as they travel (via momentum exchange and scattering and heating and ionization) which allows not only for the non-linear development of coupled cosmic ray-gas instabilities, but also cosmic-ray driven winds and outflows, cosmic ray heating altering star formation or ionization coupling to thermo-chemistry, and other unique phenomenology. The CR scattering rate $\bar{\nu}_{\rm s}$ for each rigidity is calculated self-consistently at every distinct position and time, according to the different models we explore (based on the local plasma properties). We do not enforce or assume any ``steady-state'' assumptions, so non-equilbrium and non-linear phenomena can occur. We wish to understand whether this could change our key conclusions above. 

With these simulations, we specifically consider the most relevant model variations to test the analytic predictions developed above. First, for reference, Fig.~\ref{fig:demo.cr.spectra.fiducial} shows an example of an empirically-calibrated model (as \S~\ref{sec:problems:obs}) where one does {\em not} solve for the CR scattering rate $\bar{\nu}_{\rm s}$ according to any physical model at each position and time, but simply assumes or imposes a phenomenological uniform-in-time-and-space power-law scattering rate, $\bar{\nu}_{\rm s} \sim 10^{-9}\,{\rm s^{-1}}\,\beta_{\rm cr}\,(R_{\rm cr}/{\rm GV})^{-0.6}$, i.e.\ $\delta_{\rm s} = 0.6$. This was studied  in \paperone, where we show explicitly that similar quality fits could be obtained for a narrow range of $\delta_{\rm s} \sim 0.5-0.7$, independent of all the parameters listed as ``unimportant'' in \S~\ref{sec:nuisance}, as well as the normalization (``high'' or ``low'') of the CR energies in the initial conditions (i.e.\ the system rapidly converges to the same steady-state results, independent of the details of the IC).

We compare the predicted spectra of a variety of species including H ($p$), $\bar{p}$, $e^{+}$, $e^{-}$, B, $^{7}$Be, $^{9}$Be, $^{10}$Be, C, N, O, and various secondary-to-primary and radioactive species ratios. As discussed in \paperone, the most constraining combination of constraints comes from fitting the overall shape and normalization of the $p$ and $e^{-}$ spectra (which dominate $e_{\rm cr}^{\prime}$, CR ionization, and $\gamma$-ray emission), the positron-to-electron and B/C ratio (which give standard secondary-to-primary ratios but depend differently on some model parameters owing to their different sensitivity to e.g.\ leptons versus hadrons and different loss processes), and $^{10}$Be/$^9$Be (which as a diagnostic of radioactive species provides an independent ``clock,'' as compared to the secondary ratios that are more sensitive to grammage). We compare the model to the observations compiled and discussed in \paperone\ (see that paper for more detailed discussion of the comparison, along with comparisons to a range of other observables including spatially-resolved Galactic $\gamma$-ray and ionization constraints). In Fig.~\ref{fig:demo.cr.spectra.fiducial}, points show observations (colors denote species), from the local ISM (LISM) from Voyager ({\em circles}; \citealt{cummings:2016.voyager.1.cr.spectra}), AMS-02 ({\em squares}; \citealt{2018PhRvL.120b1101A,2019PhRvL.122d1102A,2019PhRvL.122j1101A}, and references therein), and compiled from other experiments including PAMELA, HEAO, BESS, TRACER, CREAM, NUCLEON, CAPRICE, Fermi-LAT, CALET, HESS, DAMPE, ISOMAX ({\em pentagons}; \citealt{engelmann:1990.heao.cosmic.rays,2007APh....28..154S,2000ApJ...532..653B,2011ApJ...742...14O,adriani:2014.pamela.experiment,2017PhRvD..95h2007A,2017ICRC...35.1091B,2017arXiv170906442H,2017ApJ...839....5Y,2017Natur.552...63D,2018PhRvL.120z1102A,2019ARep...63...66A}). For the non-Voyager data we omit observations at energies where the Solar modulation correction is estimated to be important \citep[see][and references therein]{2017AdSpR..60..865B,bisschoff:2019.lism.cr.spectra}. For the Voyager data, we show both the directly observed values and the ``modulation-corrected'' value from \citet{2007ARNPS..57..285S} who consider models where modulation could still be important for V1 data (note this would also reduce the value of B/C observed at $\sim1\,$GeV). 

More extensive comparisons to other observations, including e.g.\ observed $\gamma$-ray emissivities, and CR ionization rates as a function of Galactic position, are presented in \paperone, all of which demonstrate consistency between this particular model and the observations. In future work it will be important to compare some of the proposed alternative models below to this extended set of constraints as well.

\begin{figure*}
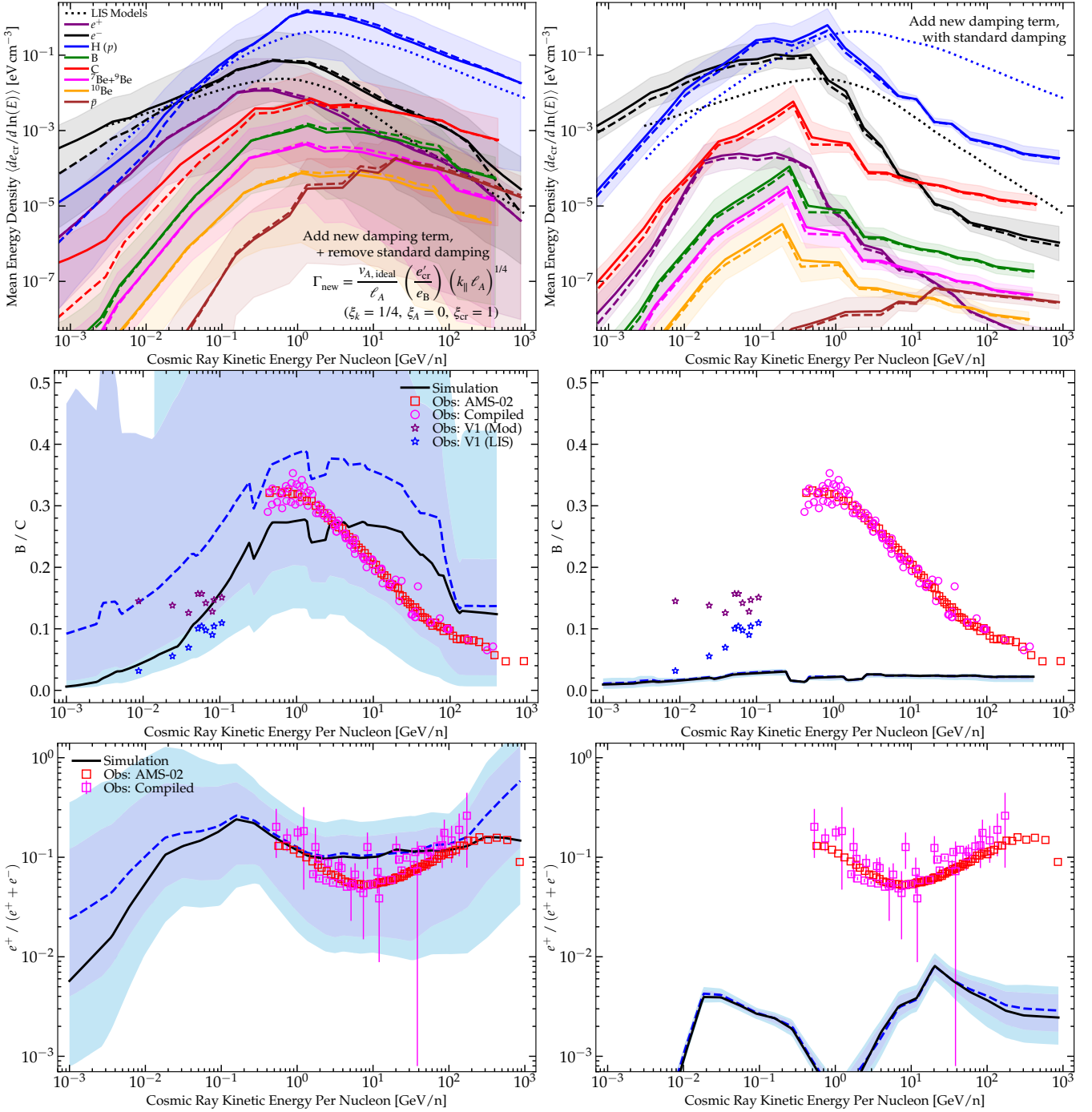

	\figrowlimv{output_01_26_21_newrsol_m10k_SC_model0_coeff1_anisoclosure_hiinit_TurbDampProptoEcr_noaltdampsource_gammadamppowermod}{output_01_26_21_newrsol_m10k_SC_model0_coeff1_anisoclosure_hiinit_TurbDampProptoEcr}
	\vspace{-0.1cm}
	\caption{As Fig.~\ref{fig:SC.hilow}, for models where we attempt to modify the damping physics to reproduce observations (\S~\ref{sec:results:damping}).
	{\em Left:} A model where we take the default SC model, remove other ET sources ($S_{\rm et,\,\pm}\rightarrow 0$), and replace all the standard known linear damping terms (\S~\ref{sec:damping}) with a damping term that scales as $\Gamma_{\rm new,\,damp} \sim (v_{A,\,{\rm ideal}}/\ell_{A})\,(k_{\|}\,\ell_{A})^{1/4}\,(e_{\rm cr}^{\prime}/e_{\rm B})$, per \S~\ref{sec:rescue:damp}. 
	This depends on $e_{\rm cr}^{\prime}$ in a way that cancels the term in SC driving which gives rise to the ``solution collapse'' problems, and allows for a reasonable and {\em stable} solution (independent of the CR energy density in the ICs).
	{\em Right:} Results if we retain this new damping term $\Gamma_{\rm new,\,damp}$, but re-introduce the ET driving and standard linear damping terms from  e.g.\ ion-neutral, linear Landau, turbulence, and dust. Any of those linear damping terms is significantly larger than $\Gamma_{\rm new,\,damp}$ (for the normalization of $\Gamma_{\rm new,\,damp}$ needed to get a reasonable scattering rate) and ``swamps'' it, producing results closer to our ``default'' SC behavior in Fig.~\ref{fig:SC.defaults}, unless we make $\Gamma_{\rm new,\,damp}$ so large that the SC models are over-damped (giving $\bar{\nu}_{\rm s}\rightarrow 0$, so CRs are unconfined). 
	\label{fig:TurbdampEcr.bestvsbad}\vspace{-0.3cm}}
\end{figure*}

\subsection{Default SC \&\ ET Models: Confirmation of Failure Modes} 
\label{sec:results:confirmation}

Having shown in Fig.~\ref{fig:demo.cr.spectra.fiducial} that it {\em is} possible to simultaneously reproduce the observations with a simple phenomenological model, we will now show that it is remarkably difficult to achieve the same in  physically motivated SC or ET models. We compare the same observations from Fig.~\ref{fig:demo.cr.spectra.fiducial} to our ``default'' model (Fig.~\ref{fig:SC.defaults}), SC-dominated models starting from lower and higher CR energy densities (Fig.~\ref{fig:SC.hilow}) and ET-dominated models (Fig.~\ref{fig:ET.defaults}), defined as in \S~\ref{sec:analytic} \&\ \ref{sec:reference.model}.

\subsubsection{SC Models}
\label{sec:results:confirmation:sc}

First, we can immediately confirm that in our ``reference model'' from \S~\ref{sec:reference.model}, the total scattering rate driving is dominated by the SC terms $S_{\rm sc,\,\pm}$ (as compared to the \Alf{ic} $S_{\rm et,\,\pm}$). This is expected since the theoretically-favored scattering rate from \Alf{ic} turbulence accounting for anisotropy (\S~\ref{sec:et}) is equivalent to a  diffusivity $\kappa \gtrsim 2\times10^{33}\,{\rm cm^{2}\,s^{-1}}\,(\ell_{A}/100\,{\rm pc})$ independent of $R_{\rm cr}$ (and larger with damping). So our qualitative conclusions and the key results in Figs.~\ref{fig:SC.defaults}-\ref{fig:SC.hilow} are identical whether we consider ``SC+(favored) ET'' or ``SC only'' ($S_{\rm et,\,\pm}\rightarrow 0$) models. 

Figs.~\ref{fig:SC.defaults} \&\ \ref{fig:SC.hilow} illustrate the fundamental ``instability'' or ``solution collapse'' problem of SC models, as discussed in \S~\ref{sec:problems:sc} and derived more rigorously in Appendix~\ref{sec:sc.equilibrium.models}. For either the regular or ``low'' or ``high'' start ICs (\S~\ref{sec:important.variations}), the system initially rapidly converges to the approximate ``local steady-state'' scattering rates (i.e.\ the steady-state scattering rates assuming the CR and plasma properties are frozen at their instantaneous values), which allow for ``super-\Alf{ic}'' streaming at some finite multiple of the \Alf\ speed (see Appendix~\ref{sec:local.steady.state.detailed.solns}). However, this is not a steady-state solution for the CR {\em energy density} equation, and the system then (on the CR transport timescale $\sim 10\,$Myr) collapses to one of the only two truly stable steady-state SC solutions for $e_{\rm cr}^{\prime}$. If the initial CR energy density at some rigidity $e_{\rm cr}^{\prime}$  is too low (and therefore also the SC-driving strength $S_{\rm sc} \propto e_{\rm cr}^{\prime}$, and the resulting scattering rates $\bar{\nu}_{\rm s}$), the CRs escape more rapidly, further lowering $e_{\rm cr}^{\prime}$, until the system collapses to the ``free streaming limit'' with no scattering (the tiny residual scattering in Fig.~\ref{fig:SC.defaults} is driven by the small ET term). This occurs at all CR rigidities in our ``low start'' (lower initial CR energy density) ICs and rigidities $\gtrsim 100\,$GV (where $e_{\rm cr}^{\prime}$ is still relatively low) in our ``normal start'' ICs. On the other hand, if the initial CR energy $e_{\rm cr}^{\prime}$ is too high, the system over-scatters ($\bar{\nu}_{\rm s}$ becomes very large) slowing transport and producing a bottleneck until the system collapses to the ``infinite scattering'' limit, where CRs can only stream at the \Alf\ speed. This gives a momentum-independent CR escape time of $\sim 1\,{\rm Gyr}\,(\ell_{\rm R,\,halo}/10\,{\rm kpc})\,(v_{A}/10\,{\rm km\,s^{-1}})^{-1}$ (where $\ell_{\rm R,\,halo}$ is the maximum of either the galacto-centric radius or height of the CR scattering halo), which is orders of magnitude larger than observationally allowed. The dependence of escape time on rigidity  is also qualitatively different from that required by observations. This produces an order-of-magnitude excess, as well as the wrong shape/rigidity-dependence, in the CR spectrum and secondary-to-primary ratios. 

We have also compared these models to observed Galactic $\gamma$-ray emissivities and ionization rates, following the identical procedure to \paperone\ (Figs.~11 \&12 therein) where we compared the phenomenological model in Fig.~\ref{fig:demo.cr.spectra.fiducial} to data from \citet{digel.2001:diffuse.gamma.ray.cr.profile.constraints,ackermann.2011:diffuse.gamma.ray.cr.profile.constraints,tibaldo.2014:diffuse.gamma.ray.cr.profile.constraints,tibaldo.2015:diffuse.gamma.ray.cr.profile.constraints,indriolo:2015.cr.ionization.rate.vs.galactic.radius,acero:2016.gamma.ray.constraints.cr.emissivity,yang.2016:diffuse.gamma.ray.cr.profile.constraints,tibaldo.2021:diffuse.gamma.ray.cr.profile.constraints}. We do not show this explicitly as the information is redundant with that in Figs.~\ref{fig:SC.defaults} \&\ \ref{fig:SC.hilow}: the ``default'' and ``high-start'' models, which lead to the over-confined limit for $\sim 0.1-10\,$GV protons that dominate the $\gamma$-ray emissivity observed, produce a $\gamma$-ray emissivity ($\propto e_{\rm cr}^{\prime}\,\rho$) about a factor $\sim 30$ larger than observed at Galacto-centric radii $\sim 1-10\,$kpc. Conversely, the ``low-start'' model produces an emissivity a factor $\sim 100$ lower than observed. Note that even if the CR proton spectrum in Figs.~\ref{fig:SC.defaults} and \ref{fig:SC.hilow} is lower in some low-density ISM gas or at larger galacto-centric radii, or even if we uniformly increased the \Alf\ speed of self-confined CR streaming by an arbitrary large factor $\sim 10-30$, it is particularly hard to avoid severely over-predicting the $\gamma$-ray flux in the ``default'' or ``high-start'' models: it only requires some dense regions where the local \Alf\ speed is low to produce excessive $\gamma$-ray emission \citep[see][]{hopkins:cr.transport.constraints.from.galaxies}. Although the CR ionization rates show the same qualitative trend, being over-predicted where the CR spectrum at low energies is high, they are less constraining. This is because low-energy CRs are well-confined (have slow diffusion) even in the phenomenological model in Fig.~\ref{fig:demo.cr.spectra.fiducial}, and losses can regulate their residence time. 

We also clearly see in Fig.~\ref{fig:SC.defaults} the ``spectral shapes'' problem predicted in \S~\ref{sec:problems:sc} for the ``normal start'' model, where the CR proton and electron spectra are much too-sharply peaked around $\sim 1\,$GV. In other words,  the shape is ``too steep'' at high energies and ``too shallow'' at low energies. This corresponds to the effective $\delta_{\rm s}$ being ``too low'' at low energies and ``too high'' at high energies. 

We have confirmed that none of the variations in \S~\ref{sec:nuisance} make any appreciable difference to these behaviors. Changing, for example, the normalization of SC source or damping terms,  removing damping terms, or changing the wavelength-dependence of the SC driving or damping terms, only shifts the value of $e_{\rm cr}^{\prime}$ that divides the two unstable ``solution collapse'' limits. In other words, if we systematically lower the normalization of $S_{\rm sc,\,\pm}$ at some wavelength $k_{\|}$ or rigidity $R_{\rm cr}$ by a factor $A$, then collapse to the ``unconfined'' solution as compared to the ``infinite scattering'' solution will occur at a factor $\sim A$ lower CR energy density $e_{\rm cr}^{\prime}$ at that $R_{\rm cr}$. We also confirm that no variant model we test is somehow able to exactly balance at the ``dividing line'' between the two regimes. This is not surprising: even if we could contrive a model that was balanced in this respect, our simulations are dynamical so the local CR energy density varies (as e.g.\ clustered SNe explode and star formation rates vary across the Galactic disk), and such a solution is unstable to variations in Galactic properties (Appendix~\ref{sec:sc.equilibrium.models}). So, the system is perturbed and immediately  collapses into either extreme.

For this reason, the results here are also insensitive to resolution or other micro-physical details of how we initialize the simulations (changing the magnetic field strengths, phase structure, resolution, etc.) -- since there are only two limits to which the simulations can collapse (each of which contradicts observations) we can only indirectly influence ``which branch'' is collapsed, or the absolute value of the \Alf\ speed in the over-confined limit (which will change the exact normalization of some predictions, but not the qualitative prediction of momentum-independent escape times far in excess of those observationally allowed).\footnote{Briefly, we note that in future work it will be particularly interesting to explore the behavior of recently-discovered instabilities which rely on the behavior of CRs in the ``collapsed'' streaming limit, such as the CR ``staircase'' \cite{quataert:2021.cr.outflows.diffusion.staircase,huang.davis:2021.cr.staircase.in.outflows,tsung:2021.cr.outflows.staircase} in these simulations, as they have thus far been studied only in idealized setups. We intentionally include all the coupling terms necessary and the resolution requirement noted in e.g.\ \citet{huang.davis:2021.cr.staircase.in.outflows} of $\Delta x \lesssim \kappa_{\|} / c_{\rm s} \sim 8\,{\rm kpc}\,(T_{\rm gas}/10^{6}\,{\rm K})^{-1/2}$ (for observationally-favored $\kappa$ at $\sim 1\,$GeV) is easily satisfied, but as noted therein the instability depends on the plasma-$\beta$ (but our experiments in described in \S~\ref{sec:nuisance} and \paperone\ vary this by factors of $\sim 10^{4}$). For now, we note that this behavior does not appear to change any of our conclusions, nor did we expect it to do so, as (1) it only appears in the \Alf-streaming (collapsed) limit; (2) in that limit if manifest in the ISM/inner CGM, it would not change the fact that the CRs have over-long residence times with $\delta_{\rm s} \le 0$; and (3) as a result the primary regime of interest for such behaviors is in the outer CGM (where more interesting observable effects for CR-driven outflows could be present), not the ISM.}

Moreover, even if we take an arbitrarily re-normalized SC model, and we choose to measure the CR spectra in low density gas in the Solar circle, such that we can reproduce roughly the correct normalization of CR spectra and B/C at $\sim 1\,$GV in the ``infinite scattering'' (\Alf{ic}-streaming or ``high-start'') limit (it is not possible to reproduce these under any circumstances in the ``free escape'' limit), we (1) still see the ``spectral shape'' problem and ``solution collapse'' at energies far from $\sim 1$\,GV, (2) see solution collapse at $\ll 1$\,GV in different environments such as molecular clouds, which would violate observational constraints on CR ionization rates \citep{indriolo:2009.high.cr.ionization.rate.clouds.alt.source.models,padovani:2009.cr.ionization.gmc.rates.model.w.alt.sources,indriolo:2012.cr.ionization.rate.vs.cloud.column,indriolo:2015.cr.ionization.rate.vs.galactic.radius}, and (3) find that for this normalization at the Solar circle, the fact that most star formation and SNe occur in the MW at radii much closer to ($<5\,$kpc from) the Galactic center, where gas densities are higher, leads to the prediction that the $\gamma$-ray emission at $\sim1-10\,$GeV  from the Galaxy would be at nearly the proton-calorimetric limit, a factor $\sim10-100$ larger than observed in the MW and other Local Group galaxies \citep[see discussion in][]{lacki:2011.cosmic.ray.sub.calorimetric,blasi:cr.propagation.constraints,fu:2017.m33.revised.cr.upper.limit,evoli:dragon2.cr.prop,2018AdSpR..62.2731A,lopez:2018.smc.below.calorimetric.crs}.

\subsubsection{ET Models}
\label{sec:results:confirmation:et}

By turning off SC driving, we  now examine ``pure ET'' models in Fig.~\ref{fig:ET.defaults}. While we have tested them to verify this, the ``most theoretically-favored'' models for ET driving from either \Alf{ic} turbulence (realistically accounting for anisotropy following e.g.\ \citealt{chandran00,Boldyrev2006,lazarian:2016.cr.wave.damping}) or fast/magnetosonic turbulence (realistically accounting for damping following e.g.\ \citealt{yan.lazarian.02,cho.lazarian:2003.mhd.turb.sims,yan.lazarian.04:cr.scattering.fast.modes,yan.lazarian.2008:cr.propagation.with.streaming}) are not interesting, as (at these CR energies $\sim$\,MeV-TeV) they predict extremely low and  approximately rigidity-independent scattering rates, which correspond  to diffusivities $\kappa \gtrsim 10^{33}\,{\rm cm^{2}\,s^{-1}}$. As a result, either ``default'' pure-ET-only (\Alf{ic}/GS95 or magnetosonic/YL04) model predicts CR scattering rates that are so low that one sees negligible secondary production at any energy, far-too-low a normalization of the CR spectra, etc. -- the results are similar to those from the ``free escape'' or ``low-start'' model shown in Fig.~\ref{fig:SC.hilow}. This is also shown explicitly around $\sim 1\,$GV in \citealt{hopkins:cr.transport.constraints.from.galaxies}.

So instead, to give ET models the best possible chance of reproducing observations, in Fig.~\ref{fig:ET.defaults} we do not show the ``most theoretically favored'' ET models with their default normalization of the ET scattering rate $S_{\rm et,\,\pm}$, but instead allow the normalization of the scattering and damping rates to be free parameters. These normalizations are then rescaled to attempt to find a ``best-fit'' to observations.

In \Alf{ic} ET models, as described in \S~\ref{sec:et} \&\ \ref{sec:problems:et}, for any type of MHD/\Alf{ic} turbulence that obeys a critical balance-type assumption, the ET driving term must have the form $S_{\rm et,\,\pm} = \alpha_{t}\,e_{\rm B}\,\Gamma_{\rm turb}/(k_{\|}\,\ell_{A})$. In the ``theoretically favored'' model, where one attempts to calculate $\alpha_{t}$ self-consistently from the same GS95-type MHD turbulence model as used for the cascade itself, one predicts $\alpha_{t} \ll 1$ \citep[as small as $\sim 10^{-6}-10^{-3}$; see][]{chandran00,lazarian:2016.cr.wave.damping}. Instead treating $\alpha_{t}$ as a free parameter, we find that, in order to approach roughly the correct order-of-magnitude normalization of the CR spectra and B/C ratios at $\sim1-10$\,GV, we require $\alpha_{t}\sim1$. But even then, if we include the standard damping terms (e.g. ion-neutral, non-linear Landau, dust), the cascade can be ``over-damped,'' and still produces poor agreement with observations. So to give the best possible chance of reproducing observations (and also to highlight the ``pure ET'' prediction), we ignore any damping other than the cascade transfer itself $\Gamma_{\rm turb}$. In other words, we have essentially assumed a pure, undamped, \Alf{ic} cascade, with arbitrary fitted normalization, so the {\em only} constraint on this ET model is the functional dependence on $k_{\|}$ that is {\em required} by critical balance.

Alternatively, we can consider a YL04-like magnetosonic model, which assumes the inertial-range cascade is isotropic, which is possible for e.g.\ fast modes on scales larger than the turbulent dissipation scales. But, this must account for the fact that, at $\lesssim$\,TeV energies,  the dissipation/Kolmogorov scale for magnetosonic modes is orders-of-magnitude larger than the gyro-resonant scale. The ``theoretically-favored'' version of this model is again over-damped (giving much-too-low $\bar{\nu}_{\rm s}$), because any appreciable regions of the ISM that have neutral fractions $\gtrsim 10^{-3}$ or plasma $\beta_{\rm plasma} > 1$  produce a super-exponential suppression of the scattering term $S_{\rm et,\,\pm}$ in this model. So again, to give the best-possible chance to reproduce observations, we follow \citet{hopkins:cr.transport.constraints.from.galaxies} and ignore any ion-neutral or dust-damping and calculate the scattering and damping rates everywhere assuming $\beta_{\rm plasma}<1$ (regardless of the real value of ${\bf B}$). We use the full integral expressions from YL04 in the simulations, but for reference, this gives an approximate scattering rate $\bar{\nu}_{\rm s} \sim c/(3\,\ell_{A}\,f_{\rm turb})$ where $f_{\rm turb} \sim {\rm MIN}[0.04\,c_{s}/v_{A,\,{\rm ideal}},\,\mathcal{M}_{A}\,(\nu_{v}/v_{A,\,{\rm ideal}}\,\ell_{A})^{1/3}\,(k_{\|}\,\ell_{A})^{1/6}]$, $\nu_{v}$ is the kinematic viscosity, and $\mathcal{M}_{A}$ is the \Alf{ic} Mach number of the turbulence at the driving scale. This is independent of rigidity.

We note these two models are akin to the ``\Alf-Max'' and ``Fast-Max'' models studied in ``single-bin'' CR models in \citet{hopkins:cr.transport.constraints.from.galaxies}, where we extensively varied the normalization and damping terms to try and fit the observed grammage as accurately as possible for $\sim 1-10\,$GV protons (see also \S~\ref{sec:nuisance}). We recover similar conclusions here for those rigidities.

However, we see in Fig.~\ref{fig:ET.defaults} that even if we freely re-normalize the scattering and/or damping rates in these models to fit the proton spectra and secondary-to-primary ratios as best as possible at $\sim 1-10$\,GV, there is a much bigger problem: both models qualitatively fail to produce the observed dependence of B/C on rigidity, or the correct CR spectral shapes. This is because, as discussed in \S~\ref{sec:problems:et}, at a fundamental level, if we allow for anisotropy/critical balance in  \Alf{ic} models (even ignoring damping) {\rm or} allow for finite damping/dissipation scales in  magnetosonic models (even ignoring possible anisotropy and some of the more severe damping terms), this implies $\delta_{\rm s} \le 0$. In other words, the scattering rate cannot decrease as a function of CR rigidity as required by observations at $\gtrsim 0.1-1\,$GV.\footnote{At sufficiently low CR energies $\lesssim 100\,$MeV, it is notable in Figs.~\ref{fig:SC.defaults}-\ref{fig:ET.defaults} that even some models which produce qualitatively incorrect $\delta_{\rm s}$ and qualitatively incorrect behaviors at higher energies can reproduce the spectral shapes and secondary-to-primary ratios of some species. This is because, as shown explicitly in \paperone, at these very low energies the rapidly-increasing rate of Coulomb and ionization losses means that the residence time (at least in the disk midplane) can actually determined by the CR loss timescales, and thus becomes independent from the predicted scattering rates $\bar{\nu}_{\rm s}$.}

For completeness, we have also run simulations assuming an undamped, isotropic, $\mathcal{E}(k) \propto k^{-3/2}$ cascade (chosen to have roughly the correct $\delta_{\rm s}$)  across {\em all} energies and wavenumbers (i.e.\ ignoring all anisotropy terms, and all damping terms, and all SC terms, at all scales). This trivially gives $e_{A} \sim e_{\rm B}\,(k_{\|}\,\ell_{A})^{-1/2}$. But as noted in \S~\ref{sec:problems:et}, this is not only unphysical but gives CR scattering rates a factor $\sim 1000$ too large at all energies, vastly over-predicting e.g.\ secondary-to-primary ratios. We discuss models of this variety further below.

Given how widely we vary the amplitudes and damping rates and spectral indices of the ET models above, it should ultimately come as no surprise that subtleties such as the difference in the simulation-resolved properties of turbulence around the driving scale (e.g.\ the locally-varying values of $e_{\rm B}$ and $\ell_{A}$, or equivalently local $\mathcal{M}_{A}$) between different Milky Way like simulated galaxies, different resolution levels, different initial ${\bf B}$-field strengths, and other variations in \S~\ref{sec:nuisance} make no difference to our conclusions. Even if we ignoring any of the resolved turbulence structure and simply assume a spatially-universal $\mathcal{M}_{A}$, we obtain the same results (which again, is expected, given that our simple analytic toy model from \S~\ref{sec:problems:et} predicts the same discrepancies with observations).

\subsection{Alternative Damping Requires Discarding Other Damping Models}
\label{sec:results:damping}

We now consider the ``alternative damping'' model from \S~\ref{sec:rescue:damp} \&\ \ref{sec:important.variations}, with two examples illustrated in Fig.~\ref{fig:TurbdampEcr.bestvsbad}. First, we simply replace the ``standard'' linear damping mechanisms ($\Gamma_{\rm in}+\Gamma_{\rm dust}+\Gamma_{\rm turb/LL}+\Gamma_{\rm nll,\,\pm}$) with a new $\Gamma_{\rm new,\,damp,\,\pm} \propto e_{\rm cr}^{\prime}$. We use a best-fit normalization of the variants we have explored, which is  $\Gamma_{\rm new,\,damp,\,\pm} \sim (v_{A,\,{\rm ideal}}/\ell_{A})\,(k_{\|}\,\ell_{A})^{\xi_{k}}\,(e_{\rm cr}^{\prime}/e_{\rm B})^{\xi_{\rm cr}}$ with $0.1 \lesssim \xi_{k} \lesssim 0.4$ and $\xi_{\rm cr} = 1$. This  has the desired effect, discussed in  \S~\ref{sec:rescue:damp}, of cancelling the $e_{\rm cr}^{\prime}$ dependence in the SC driving term $S_{\rm sc,\,\pm}$, which is responsible for the instability/solution collapse problems (see Appendix~\ref{sec:sc.equilibrium.models}). Thus we can obtain a stable result in at least qualitative agreement with the observed behavior at all rigidities, and {\em independent} of the CR energy density in the ICs (i.e.\ we converge to the same answer for ``low'' and ``high'' start ICs). 

However, the challenge with this model is ensuring that $\Gamma_{\rm new,\,damp,\,\pm}$ dominates over other terms (specifically other damping terms) in the $D_{t} e_{\pm}$ equation (Eq.~\ref{eqn:ea}), the balance of which set the equilibrium value of $e_{\pm}$ in the volume-filling ISM. Among the other standard terms in Eq.~\ref{eqn:ea}, we can retain or remove the ``gradient terms'' (i.e.\ the ``advective'' term $\nabla \cdot (v_{A,\,\pm}\,e_{\pm}\,\bhat)$ and ``PdV'' term $(e_{\pm}/2)\,\nabla \cdot {\bf u}_{\rm gas}$), and/or the non-linear Landau damping term ($\Gamma_{\rm nll,\,\pm}$), and/or the ``default'' (theoretically-favored, but weak) ET driving term $S_{\rm et,\,\pm}$, without qualitatively changing the behavior seen in the top panels of Fig.~\ref{fig:TurbdampEcr.bestvsbad}. But unless we artificially remove or suppress the standard turbulent/linear Landau ($\Gamma_{\rm turb/LL}$), dust ($\Gamma_{\rm dust}$), and ion-neutral damping ($\Gamma_{\rm in}$) terms, they tend to dominate $\Gamma_{\pm}$ (e.g.\ $\Gamma_{\rm turb/LL} \gg \Gamma_{\rm new,\,damp}$). This causes the total damping $\Gamma_{\pm}$ to  once again be dominated by terms that are independent of $e_{\rm cr}^{\prime}$, and the ``solution collapse'' problem returns. Similarly, we cannot simply increase $\Gamma_{\rm new,\,damp}$ until it dominates over all the other damping mechanisms at all CR energies: even though this will cure the instability, it will necessarily over-damp the scattering modes, producing too-low CR spectra and secondary abundances. One example of this failure is shown in Fig.~\ref{fig:TurbdampEcr.bestvsbad}. This illustrates that the discrepancy is not small -- it would require more than order-of-magnitude changes in the expected strengths of turbulent, dust, and ion-neutral damping for typical Milky Way conditions.

In summary, while a version of this model that can reproduce  observations does exist, it requires a radical revision to our understanding of damping mechanisms. Not only must one introduce a novel damping mechanism with the desired $e_{\rm cr}^{\prime}$ dependence, but one must also argue that the standard turbulent/linear Landau, dust, and ion-neutral (in diffuse but partially-ionized phases) damping mechanisms are much weaker than currently understood, in order for this new damping mechanism to dominate with the correct normalization at all relevant rigidities.

\begin{figure*}
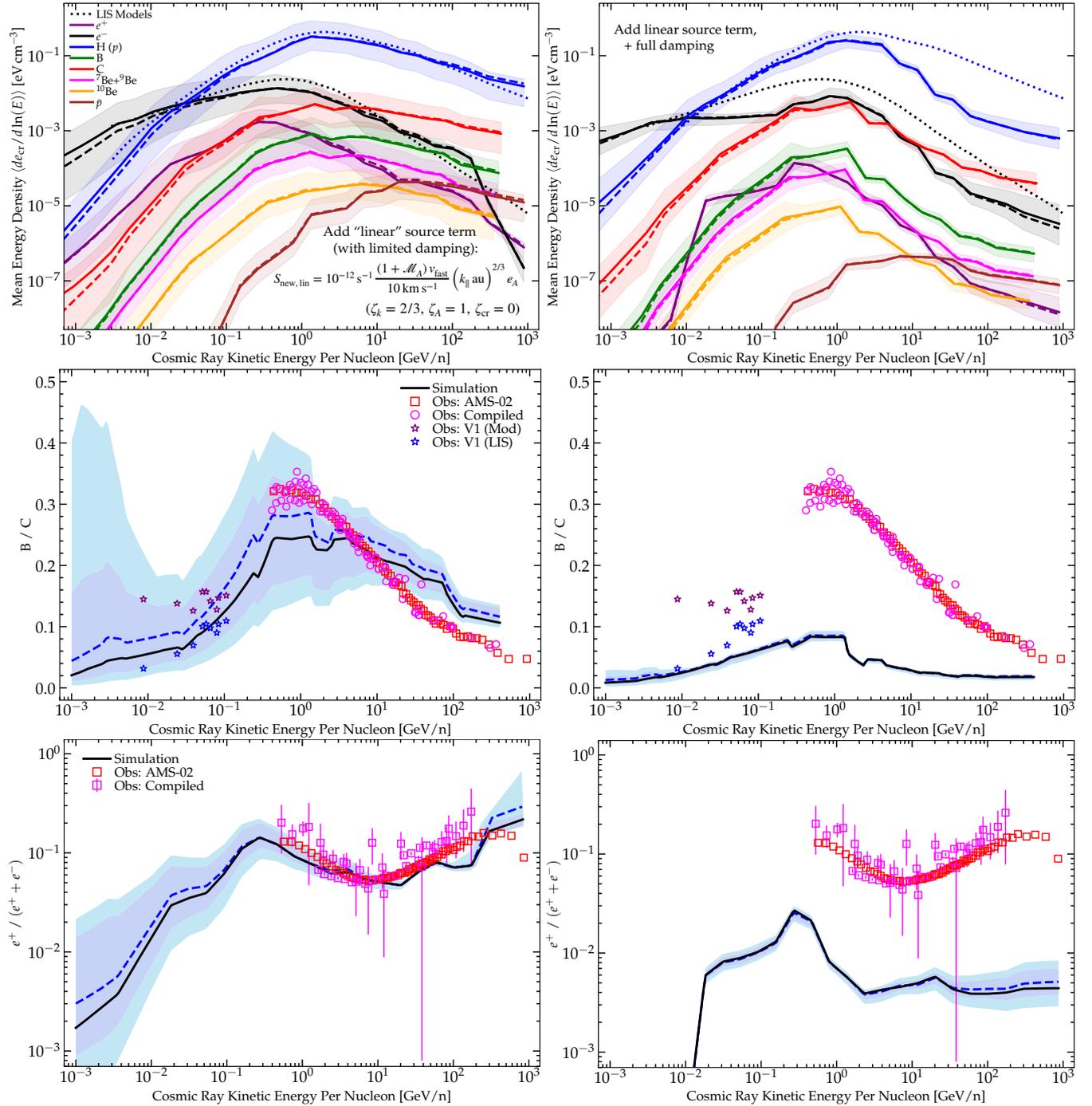

	\figrowlimv{output_02_21_21_newrsol_m10k_SC_model0_coeff1_anisoclosure_loinit_Sextlin0pt66_GammaLinINwPdV_someSextern_revFcas_slope4pt3}{output_02_21_21_newrsol_m10k_SC_model0_coeff1_anisoclosure_loinit_Sextlin0pt66_GammaLinINwDust_moreSextern_revFcas_slope4pt3}
	\vspace{-0.1cm}
	\caption{As Fig.~\ref{fig:SC.hilow}, for models where we consider alternative linear scattering-mode driving/source terms ($S_{\rm new,\,lin} \propto e_{A}$, i.e.\ $\zeta_{A}=1$; \S~\ref{sec:results:sources:linear}).
	{\em Left:} A model where we add linear source term $S_{\rm new,\,lin} \sim \Psi_{\rm new}\,e_{A}$ with $\Psi_{\rm new} \sim  10^{-12}\,{\rm s^{-1}}\,(k_{\|}\,{\rm au})^{2/3}\,(1+\mathcal{M}_{A})\,(v_{\rm fast}/10\,{\rm km\,s^{-1}})$, removing specifically the turbulent/linear Landau and dust damping terms (keeping all other damping), and reducing the standard SC driving term by a factor $\sim 0.01$. Note the agreement with observed $^{10}$Be/$^{9}$Be and $\bar{p}/p$ (not shown) is also good.
	We also obtain broadly similar results for a simpler model with $\Psi_{\rm new} 10^{-12}\,{\rm s^{-1}}\,(k_{\|}\,{\rm au})^{2/3}$, but the fit is not quite as good (this leads to flatter B/C at high-energies). 
{\em Right:} As {\em left} but re-introducing the dust and turbulent/linear Landau damping terms, which are usually larger than the linear growth term, so the behavior reverts to be closer to the ``default'' SC model and collapses at intermediate and high-energies to the unconfined solutions.
	\label{fig:Slin.best.worst}\vspace{-0.3cm}}
\end{figure*}

\subsection{Alternative Sources}
\label{sec:results:sources}

Figures~\ref{fig:Slin.best.worst} \&\ \ref{fig:Sext.best} now consider the ``alternative driving'' or ``alternative sources'' models discussed in \S~\ref{sec:rescue:drive} \&\ \ref{sec:important.variations}.

\subsubsection{Local/Linear Source Terms}
\label{sec:results:sources:linear}

First, in Fig.~\ref{fig:Slin.best.worst}, we consider adding an alternative linear driving/source ($\zeta_{A}=1$ or $S_{\rm new} \propto e_{A}$) term. We take the form $S_{\rm new,\,\pm} = S_{\rm new,\,lin} \equiv \Psi_{\rm new}\,e_{A}$ with $\Psi_{\rm new} \equiv \Psi_{0}\,(k_{\|}/k_{0})^{\zeta_{k}}$ (where we set $k_{0} \equiv {\rm au^{-1}}$ for convenience without loss of generality). Because this is a linear driving term, the ``net'' linear driving+damping $S_{\rm new,\,lin} - Q_{\pm} = (\Psi_{\rm new} - \Gamma_{\pm})\,e_{A}$ is only weakly influenced by $S_{\rm new,\,lin}$ if $\Psi_{\rm new} \lesssim \Gamma_{\pm}$. So, for an initial experiment we ignore the turbulent/linear Landau and dust damping mechanisms. 
In Fig.~\ref{fig:Slin.best.worst}, we show that this form can give a plausible fit to the observed spectra and ratios for $0.6 \lesssim \zeta_{k} \lesssim 0.9$. The required normalization is  modest, e.g.\ $\Psi_{0} \sim 10^{-12}\,{\rm s^{-1}}\,(1+\mathcal{M}_{A})\,(v_{\rm fast}/10\,{\rm km\,s^{-1}}) \sim \delta v_{\rm turb}/{\rm pc}$ (or even $\Psi_{0} \sim 10^{-12}\,{\rm s^{-1}}\sim\,$constant), or similarly $S_{\rm new,\,lin} \sim 10^{-5}\,k_{\|}\,v_{\rm fast}\,(k_{\|}\,{\rm au})^{-1/3}$. In other words, the driving/growth rate favored can be as little as $\sim 10^{-5}$ of the fast mode crossing rate. 

The obvious challenge here, akin to the  alternative damping discussed above (\S~\ref{sec:results:damping}), is ensuring $\Psi_{\rm new} \gtrsim \Gamma_{\pm}$. Going through all terms in Eq.~\ref{eqn:ea}, the effects of $S_{\rm new,\,lin}$ are robust to retaining or removing the ``gradient terms,'' or the other default source terms ($S_{\rm sc,\,\pm}$, $S_{\rm et,\,\pm}$), or the ion-neutral damping term ($\Gamma_{\rm in}$),\footnote{Unlike in the ``modified damping''  case ($\Gamma_{\rm new,\,damp,\,\pm}$; \S~\ref{sec:results:damping}), it appears that while $\Gamma_{\rm in} \gtrsim \Psi_{\rm new}$ in dense neutral ISM phases (CNM, molecular), which have a low volume-filling fraction and therefore do not strongly alter CR spectra in diffuse gas, we generally have $\Gamma_{\rm in} \lesssim \Psi_{\rm new}$ in warmer and/or more diffuse phases, even if they are partially-neutral. This is especially true if we adopt a version of $\Psi_{\rm new}$ that scales with $\mathcal{M}_{A}$ or $\delta v_{\rm turb}$, which is larger in warm or cool phases.} as well as retaining or modifying/expanding the non-linear damping terms ($\Gamma_{\rm nll,\,\pm}$). However, if we do include the standard turbulent/linear Landau ($\Gamma_{\rm turb/LL}$) or dust ($\Gamma_{\rm dust}$) damping terms in their ``default'' forms, these cause $\Gamma_{\pm} \gtrsim \Psi_{\rm new}$ in the volume-filling ISM, negating the effects of our added source term $S_{\rm new,\,lin}$. 

Thus while not totally implausible, this model does have theoretical challenges in dealing with the turbulent/linear-Landau and dust damping terms, akin to the modified damping scenario (\S~\ref{sec:results:damping}). As discussed in \S~\ref{sec:rescue:drive}, the advantage of this scenario is that it is quite easy to imagine linear instabilities operating on these scales with roughly the correct growth rate and $k_{\|}$-dependence. For example, multi-fluid instabilities like the Kelvin-Helmholtz instability would have growth rates $\sim (\delta\rho/\rho)\,k\,\delta v$, so would only require $(\delta\rho/\rho)\,(\delta v/v_{\rm fast}) \sim 10^{-5}$ on these scales to grow at roughly the correct rate. RDIs in the ``mid-$k$'' range, which may be applicable at these scales, and Rayleigh-Taylor instabilities (RTIs) similarly have growth rates of $\sim \sqrt{a\,k}$ where $a$ is some differential acceleration between e.g.\ dust and gas (for RDIs) or a fluid interface (for the RTI). Given the low normalizations, even a very small differential acceleration $a \sim 10^{-4}\,v_{A}^{2}/\ell_{A}$ could be sufficient to drive the required growth rates. Of course, for ``interface'' instabilities one must ask what the interface would be, while for  RDIs, the ``high-$k$'' modes often have a less-favored scaling $\propto k^{1/3}$, and more extreme conditions and/or certain modes (e.g.\ the ``cosmic ray like'' RDI modes) could produce over-confinement (see \citealt{squire:2021.dust.cr.confinement.damping} and Ji et al., in prep.). It is even conceivable that $S_{\rm new,\,lin}$ could arise at CR energies $\gg$\,GV from the action of the \citet{bell.2004.cosmic.rays} instability sourced by the dominant $\sim$\,GV CRs (i.e.\ the long-wavelength regime of the instability sourced by lower-energy CRs); however, this could introduce some (but not all) of the ``instability'' problems from SC (\S~\ref{sec:problems:sc}).

In these regimes, it is also not implausible to assume that the usual $\Gamma_{\rm turb/LL}$ and $\Gamma_{\rm dust}$ terms would be strongly modified. The expressions for ``turbulent/linear Landau'' damping reviewed in \S~\ref{sec:damping} are derived specifically assuming that the modes are sheared out by {\em external} turbulence from a standard GS95 cascade from larger scales, where the dominant driving of any modes that are {\em not} exactly parallel modes driven by SC comes from that ET cascade (see \citealt{yan.lazarian.02,farmer.goldreich.04,zweibel:cr.feedback.review}). But if $S_{\rm new,\,lin}$ dominates over $S_{\rm sc,\,\pm}$ and $S_{\rm et,\,\pm}$, then some of these assumptions will not apply -- so there is not necessarily any reason to think the modes would be sheared out in this manner. Similarly, the dust damping rate $\Gamma_{\rm dust}$ is derived specifically under the assumption that the RDIs are negligibly weak/inactive (indeed the damping and instability arise from similar physical effects) -- if there is sufficient dust drift to cause an RDI, the dust switches from being a ``damping'' to a ``driving'' term (see \citealt{squire:2021.dust.cr.confinement.damping}).

\subsubsection{External/Fixed-Rate Source Terms}
\label{sec:results:sources:external}

\begin{figure*}
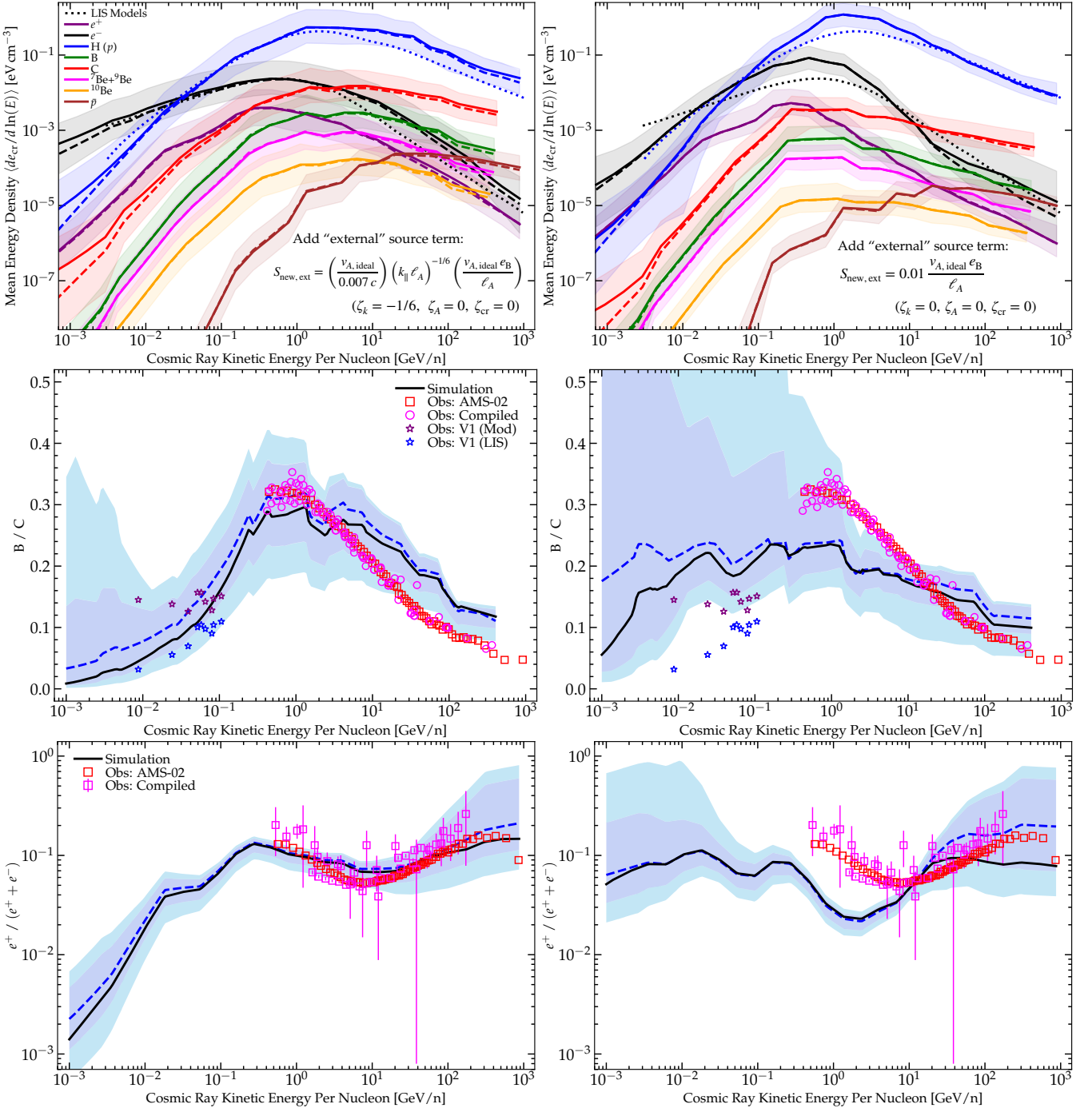

	\figrowlimv{output_01_22_21_newrsol_m10k_SC_model0_coeff1_anisoclosure_loinit_ExtW0pt66CoeffLoNorm}{output_01_26_21_newrsol_m10k_SC_model0_coeff1_anisoclosure_hiinit_ExtConstant_normaldamp}
	\vspace{-0.1cm}
	\caption{As Fig.~\ref{fig:SC.hilow}, for models where we consider alternative fixed or ``external'' ($e_{A}$-independent, $\zeta_{A}=0$) scattering-mode driving terms ($S_{\rm new,\,ext}$; \S~\ref{sec:results:sources:external}). Here, we {\em do not} disable any of the standard damping or other driving terms, we simply add this additional source term.
	{\em Left:} A model with $S_{\rm new,\,ext} = (v_{A,\,{\rm ideal}}/0.007\,c)\,(v_{A,\,{\rm ideal}}/\ell_{A})\,(k_{\|}\,\ell_{A})^{-1/6}\,e_{\rm B}$ 
	(we also obtain similar results for $S_{\rm new,\,ext} \sim 0.005\,(v_{A,\,{\rm ideal}}/\ell_{A})\,e_{\rm B}\,(k_{\|}\,\ell_{A})^{-1/6}$). Adding a weakly scale-dependent driving term of this form ($\zeta_{k}\approx-1/6$, $\zeta_{A}=\zeta_{\rm cr}=0$), with amplitude comparable to $\sim 1\%$ of turbulent or \Alf\ dissipation rates can produce reasonable behaviors, without having to strongly modify known damping or other driving terms. Note agreement with $^{10}$Be/$^{9}$Be and $\bar{p}/p$ is good as well.  
	{\em Right:} A model with $S_{\rm new,\,ext} = 0.01\,(v_{A,\,{\rm ideal}}/\ell_{A})\,e_{\rm B}$ (roughly $\sim 0.01\,\rho\,\delta v_{\rm turb}^{3} / \ell_{\rm turb}$ on the driving/resolved ISM scales). This has similar amplitude and behavior but slightly different wavelength-dependence ($\zeta_{k}=0$). While it does  not fail catastrophically, the agreement with observations is notably worse, demonstrating that the favored range of $-0.25 \lesssim \zeta_{k} \lesssim -0.1$ is relatively constrained.
	\label{fig:Sext.best}\vspace{-0.3cm}}
\end{figure*}

We now consider adding a ``constant'' or ``external'' alternative driving/source term ($\zeta_{A}=0$ or $S_{\rm new} \propto e_{A}^{0}$), in Fig.~\ref{fig:Sext.best}. This is the most straightforward successful model variant we consider. Keeping everything else in our ``reference'' model fixed, we can simply add a source term $S_{\rm new,\,ext} \propto k_{\|}^{\zeta_{k}}$ (where we find best-fits with $-0.25 \lesssim \zeta_{k} \lesssim -0.1$), and normalization e.g.\ $S_{\rm new,\,ext} \sim 0.005\,(v_{A,\,{\rm ideal}}/\ell_{A})\,(k_{\|}\,\ell_{A})^{-1/6}\,e_{\rm B}$ or $\sim 0.01\,\delta v_{\rm turb}^{3}/\ell_{A}$, i.e.\ $\sim 1\%$ of the typical turbulent dissipation rates. This produces remarkably good behavior across all diagnostics we consider. 
 
While a detailed exploration is outside the scope of this work, we have also applied the analysis pipeline from \paperone\ to this model to explore its predictions for the spatially-resoled $\gamma$-ray emissivities and spectra and CR ionization rates in the Galaxy (since any successful model must reproduce these, as well) and we find agreement there as well, broadly similar to the favored phenomenological model from \paperone\ which is shown in Fig.~\ref{fig:demo.cr.spectra.fiducial}.

Unlike the alternate linear-damping or driving models, we do not have to ``remove'' or re-tune any of the known terms (SC or ET driving or different damping mechanisms) to see good behavior here. In other words, this model works with no other unwarranted modifications to the wave-damping or source physics.  The one caveat is the SC driving in ``high-start'' IC cases. Adding $S_{\rm new,\,ext}$ with $\zeta_{A}=0$ prevents the SC instability from collapsing to the ``free-streaming'' solution, because $S_{\rm new,\,ext}$ sets a minimum driving even if $e_{\rm cr}^{\prime}$, and hence $S_{\rm sc,\,\pm}$, is low. But if $e_{\rm cr}^{\prime}$ is sufficiently high and we still include our standard $S_{\rm sc,\,\pm}$ in $S_{\pm}$, we can have $S_{\rm sc,\,\pm} \gtrsim S_{\rm new,\,ext}$, with $S_{\rm sc,\,\pm}$ large enough to push the system into the ``infinite-confinement'' branch of solution collapse (so that this causes $e_{\rm cr}^{\prime}$ and $S_{\rm sc,\,\pm}$ to continue to rise). This scenario cannot be halted by the added $S_{\rm new,\,ext}$ term, and indeed does still occur if we just add $S_{\rm new,\,ext}$ in  ``high start'' IC simulations. While it is plausible that such collapse could occur physically in extreme regions -- e.g.\ galactic nuclei, or starburst galaxies, which are observed to be at the proton calorimetric limit in $\gamma$-ray emission; see \citealt{lacki:2011.cosmic.ray.sub.calorimetric,tang:2014.ngc.2146.proton.calorimeter,griffin:2016.arp220.detection.gammarays,wjac:2017.4945.gamma.rays,wang:2018.starbursts.are.proton.calorimeters}) -- it obviously does not occur for typical MW conditions.  So to ensure it does not occur, we find that our results are most stable if we reduce $S_{\rm sc,\,\pm}$ by a factor $\sim 10-100$ from its ``reference'' value. 
But as discussed above in \S~\ref{sec:problems:sc} and in e.g.\ \S~5.3.4 of \citet{hopkins:cr.transport.constraints.from.galaxies}, such a renormalization of $S_{\rm sc,\,\pm}$ is plausible, based on corrections to $S_{\rm sc}$ from more careful detailed PIC modeling of pitch-angle dependence, helicity, non-linear, and non-gyro-resonant effects \citep[e.g.][]{bai:2019.cr.pic.streaming,holcolmb.spitkovsky:saturation.gri.sims}.
Of course, these would need to be revisited in more realistic situations with some $S_{\rm new,\,ext}$ term present.

It is also noteworthy that this functional dependence of $S_{\rm new,\,ext}$ on $k$, $e_{A}$, and $e_{\rm cr}$ (e.g.\ $\zeta_{A}=0$, $\zeta_{\rm cr}=0$, and $|\zeta_{k}|$ small) is superficially similar to what one would obtain in the simplest ``classical'' isotropic, undamped, inertial-range K41-like turbulent cascade, where $S_{\rm et,\,\pm}\sim\,$constant is the turbulent dissipation rate. This, plus the fact that the dimensional dependence of $\Gamma_{\rm turb/LL}$ and $\Gamma_{\rm dust}$ on $k_{\|}$ are similar to the turbulent cascade rate, is indeed why, as many have noted previously, the observed $\delta_{\rm s}$ is not so different from what one would naively obtain from a ``traditional'' isotropic undamped ET model with a spectrum similar to $\mathcal{E}(k) \propto k^{-3/2}$ (neglecting dissipation, anisotropy, and finite dynamic-range effects; \citealp[see e.g.\ discussion in][]{blasi:cr.propagation.constraints,vladimirov:cr.highegy.diff,gaggero:2015.cr.diffusion.coefficient,2016ApJ...819...54G,2016ApJ...824...16J,cummings:2016.voyager.1.cr.spectra,2016PhRvD..94l3019K,evoli:dragon2.cr.prop,2018AdSpR..62.2731A}). 
But there are fundamental physical differences here. Most importantly, as argued above and in Appendix~\ref{sec:turb.review} in detail, this $S_{\rm new,\,ext}$ cannot stem from a traditional undamped \Alf{ic} or magnetosonic cascade from large ISM scales. All of the effects reviewed therein would prevent $S_{\rm new,\,ext}$ from having the form assumed. It is possible that some sort of ``mini-cascade'' could occur over a small range of scales, with smaller-scale driving, provided it could avoid the anisotropy and damping problems we have outlined. But as justified formally in Appendix~\ref{sec:turb.review}, we easily avoid all of these conceptual difficulties if we simply assume $S_{\rm new,\,ext}$ represents driving of \Alf{ic} modes competing directly with damping {\em at each scale} separately -- we are simply arguing for a driving mechanism whose power is only weakly scale-dependent.
Such an effect could possibly arise, for example, if reconnection played an important role in MHD turbulence at small scales. Such a scenario would require that flux ropes formed by reconnection  between sheets in the perpendicular plane \citep{schekochihin:2020.mhd.turb.review} subsequently broke up in the parallel direction with the right spectrum, which is plausible but highly speculative. However, it is worth emphasizing that since  the required power in $S_{\rm new,\,ext}$ is two or three orders of magnitude smaller than the power in the turbulent cascade, these fluctuations should be strongly subdominant and would be very difficult to observe in simulations. 
Finally, we note that the true best-fit driving favors a modest scale-dependence $-0.25 \lesssim \zeta_{k} \lesssim -0.1$ (c.f. left and right panels of Fig.~\ref{fig:Sext.best}); this is not steep, but is distinctly different from the predictions of any turbulence models in the literature.

\subsubsection{Summary of Requirements}
\label{sec:results:summary}

We can summarize the required scaling for a viable driving/source term $S_{\rm new}$ for ``linear'' $S_{\rm new,\,lin}$ (\S~\ref{sec:results:sources:linear}) and ``external'' $S_{\rm new,\,ext}$ (\S~\ref{sec:results:sources:external}) cases as follows:
\begin{align}
\label{eqn:S.needed}
S_{\rm new,\,lin} &\sim 10^{-12}\,{\rm s^{-1}}\,e_{A} \left( \frac{k_{\|}}{{\rm au^{-1}}} \right)^{\zeta_{k}}\,f_{\rm S}(...) & \hfill (0.6 \lesssim \zeta_{k} \lesssim 0.9) \\
\nonumber S_{\rm new,\,ext} &\sim 0.01\,\frac{v_{A,\,{\rm ideal}}}{\ell_{A}}\,e_{\rm B}\,\left( \frac{k_{\|}}{{\rm au^{-1}}} \right)^{\zeta_{k}} f_{\rm S}(...) & \hfill (-0.25 \lesssim \zeta_{k} \lesssim -0.1)
\end{align}
where $f_{\rm S}(...)$ is some function of ISM/plasma properties. Any viable driving mechanism must therefore satisfy the following conditions: 
{\bf (1)} It must drive modes of interest, i.e.\ \Alf{ic} modes with $k_{\|}$ in the relevant range and that are not too-extreme in their anisotropy.\footnote{As shown in Appendix~\ref{sec:turb.review}, the modes do not have to be specifically parallel or isotropic, but should at least obey $| k_{\|} | \gg (|\delta{\bf B}(k_{\|})|/|{\bf B}|)\,| k_{\bot} | \sim 0.0003\,R_{\rm GV}^{0.2}\,| k_{\bot} |$.} We stress that Eq.~\ref{eqn:S.needed} refers to the driving rate of these modes, specifically, not to other (e.g.\ nearly-perpendicular) modes, which are generically less efficient scatterers and would require a larger $S_{\rm new}$.
{\bf (2)} It must be able to drive modes across the wavelength scales of interest. For rigidities $R \sim 0.001-1000\,$GV studied here, this is $1/k_{\|} \sim r_{g,{\rm cr}} \sim 3\,B_{\rm \mu G}^{-1} \times 10^{9-15}\,{\rm cm}$. However, it is possible that  very low-energy CRs ($\lesssim 100\,$MeV) have residence times that are primarily regulated by ionization/Coulomb losses (as argued empirically in \citealt{hopkins:cr.multibin.mw.comparison} and found in some of our experiments), which would increase the lower limit to $\sim 10^{11.5}\,B_{\rm \mu G}^{-1}\,{\rm cm}$. Similarly, it is plausible that  gyro-radii approach/exceed the dissipation scales of fast magnetosonic turbulence (so ``traditional'' ET theory becomes viable) above the scales relevant to few-hundred GV CRs \citep[e.g.][]{fornieri:2021.comparing.et.models.data.et.only.few.hundred.gv}, in which case the upper limit could decrease to $\sim (0.3-1)\times10^{15}\,B_{\rm \mu G}^{-1}\,{\rm cm}$.
{\bf (3)} It must have one of the forms above in Eq.~\ref{eqn:S.needed}, with the range of $\zeta_{k}$ corresponding to the extrinsic or linear driving ($\zeta_{A}=1$ or $=0$), with $f_{\rm S}(...)$ parameterizing all the dependence on the ISM plasma physics. 
{\bf (4)} In order to match the normalization in Eq.~\ref{eqn:S.needed}, the appropriate volume or scattering-rate weighted average $\langle f_{\rm S}(...) \rangle$ (parameterized in the same way) must be $\sim 1$ integrated from CR sources to the Solar LISM in a MW-like galaxy through most of the volume-filling ISM. 
{\bf (5)} By definition, $f_{\rm S}$ must depend weakly or not at all on CR properties (e.g.\ the distribution function $f$, number density $n_{\rm cr}$, energy density $e_{\rm cr}$, streaming speed $v_{\rm st}$, etc.); weakly or not at all on $k_{\|}$ (such that $S_{\rm new}$ has the correct $k_{\|}$ dependence parameterized  by the range of $\zeta_{k}$); and weakly or not at all on the local mode amplitude $e_{A}$ or $\delta{\bf B}(k_{\|})$ (i.e.\ the driver has $\zeta_{A}\approx 0$ or $\approx1$, appropriately). 

Briefly, it is worth noting that the favored ranges of $\zeta_{k}$ for these driving mechanisms (or the alternative damping in \S~\ref{sec:results:damping}) in our simulations are slightly different from that analytically anticipated from our simple steady-state back-of-the-envelope calculations in \S~\ref{sec:rescue:damp}-\ref{sec:rescue:drive}. This is not surprising: in our simple model we neglected losses, adiabatic terms, contributions to transport from \Alf{ic} streaming, the interplay of multiple damping/source mechanisms, and finite source/scattering halo distributions, all of which contribute some additional rigidity-dependence to the final behavior, in a way that only our full simulations can accurately capture. But crucially, the qualitative behaviors and conclusions are identical, with only modest quantitative corrections. This suggests that the general physical principles are robust.

\subsection{Can Different Galaxy Properties Rescue SC or ET Models?}
\label{sec:galaxy.props.rescue}

It is natural to ask whether there might be some different galaxy properties (perhaps some difference between the real Milky Way and our models or assumptions here) that could resolve the discrepancies with observations, without invoking new driving or damping mechanisms. We have attempted to explore this with both our general analytic arguments and, to the extent possible in our simulations, with the parameter variations discussed in \S~\ref{sec:nuisance}. Specifically, for the general SC and ET models in Figs.~\ref{fig:SC.defaults}-\ref{fig:ET.defaults}, we have run simulations using three different cosmologically-selected Milky-Way mass galaxies, which -- while all selected to have properties that are broadly similar to the Milky Way -- differ in detail (e.g.\ different sizes, gas density and star formation rate distributions, presence or absence of bars and arms, etc.). We have also arbitrarily  re-normalized the initial magnetic fields and CR energy densities in the simulations by large factors as discussed above. And for all of our simulations, we have a large number of independent snapshots sampling several galaxy dynamical times -- we have checked to confirm that the results in our Figures are robust (approximately steady-state) in time, and to see whether there could be even a transient phase where the SC and ET models produce good simultaneous agreement with different observables. As relates to all these differences (variations in time,  between different Milky-Way mass galaxies, or between modified initial conditions), our key conclusions are robust. Indeed,  the differences between galaxies or different times are much smaller than the differences between models (see \paperone\ for more detailed comparisons). 

However, it is not possible in computationally-expensive simulations like ours to survey all possible galaxy properties. So one might ask whether there still exists some hypothetical combination of plasma parameters that would allow the SC and/or ET models to reproduce observations. This is essentially the question explored in \citet{kempski:2021.reconciling.sc.et.models.obs}, of which we became aware during the writing of this manuscript. While our experiments in this paper might be described as ``constraining which CR scattering models can reproduce observations, given a set of galaxy models,'' \citet{kempski:2021.reconciling.sc.et.models.obs} effectively consider the complementary question ``given a fixed CR scattering-rate model, what galaxy model could reproduce observations?''. Specifically, they consider analytically parameterized models of a stratified disk+CGM and show, in agreement with our conclusions, that neither SC nor ET models can possibly reproduce the observations alone.\footnote{See also \citet{fornieri:2021.comparing.et.models.data.et.only.few.hundred.gv}, who similarly concluded that ET models alone (\Alf{ic} or akin to our ``Fast-Max'' $S_{\rm et,\,\pm}$) could not reproduce observations below a few hundred GV, even allowing for arbitrary freely-fit galaxy/ISM properties in a parameterized analytic model.} However,  they do argue that the combination SC+ET allows a match to observations, in principle, if the stratified disk+CGM follows a specific particular model. However, as \citet{kempski:2021.reconciling.sc.et.models.obs} caution, this requires a very specific and fine-tuned set of assumptions: their model requires that the profile of the \Alf\ speed, turbulence strength, ionization fraction, and $e_{\rm cr}^{\prime}$ follows a specific profile as a function of scale-height. This allows ET driving with a scaling close to our modified ``Fast-Max'' model in Fig.~\ref{fig:ET.defaults} to dominate within the thick disk (with a certain strength), while SC driving with non-linear-Landau damping and the  ``collapsed'' \Alf{ic} streaming solution only dominates outside the disk in the CGM (with that following a specific vertical \Alf-speed profile). Essentially, in their model, the profiles of relevant plasma properties (like $v_{A}$), which appear in the scalings of $\bar{\nu}_{\rm s}$ for the SC and ET models,  are chosen such that they ``cancel out'' the fundamental problematic scalings of SC or ET alone.

We have attempted to explore a model akin to this best fit of \citet{kempski:2021.reconciling.sc.et.models.obs}, by (1) replacing all our driving and damping terms with just the combination of SC driving plus the ``Fast-Max'' ET driving model (the same as the scalings adopted in \citealt{kempski:2021.reconciling.sc.et.models.obs}), with just non-linear Landau damping, while also (2) re-normalizing ${\bf B}$ and $e_{\rm cr}^{\prime}$ in our ICs to match the vertical profile of $e_{\rm cr}^{\prime}$ and $v_{A}$ assumed therein. But we find this experiment quickly undergoes the same ``solution collapse'' akin to our ``normal'' or ``high'' start ICs in Figs.~\ref{fig:SC.defaults}-\ref{fig:SC.hilow}. The difference may be that it is simply not possible to exactly reproduce all of the assumptions of the analytic model in our initial conditions; e.g.\ because the galaxy density profile cannot be freely renormalized in our simulations, and/or because we include SC+ET terms together, while \citet{kempski:2021.reconciling.sc.et.models.obs} consider a model where only one or the other contributes meaningfully at any given scale-height.  But the bigger challenge may be that our simulations dynamically evolve quantities like ${\bf B}$ and $e_{\rm cr}^{\prime}$, and these will quickly deviate from their initial values as e.g.\ super-bubbles and clustered SNe explode. These will then push the system away from equilibrium and into one of the solution-collapse regimes. This suggests, at least, that this fine-tuning is not trivial to achieve in practice, and is unlikely to be the case in the Milky Way (as, indeed, is concluded by \citealt{kempski:2021.reconciling.sc.et.models.obs} also).

\subsection{What About the ``Meso-Scale''?}
\label{sec:mesoscale}

In thinking about our conclusions, it is helpful to separate the enormous hierarchy of scales into three groups: ``macro,'' ``micro,'' and ``meso'' scales. For our purposes, we can think of the ``macro-scale'' structures as those which are at least semi-resolved by our simulations (larger than a few thousand Solar masses). This includes e.g.\ the multi-phase structures of the ISM, and clumping of gas (e.g.\ the existence of GMCs); global galactic structure (the nucleus, disk, and bulge, bars, and spiral arms); the scale heights of the cold, star-forming disk (and young stellar disk) and the warm/thick gas and stellar disks, and the associated driving scales of ISM turbulence; clumping of star formation and SNe (in space and time), and associated super-bubbles and galactic chimneys; galactic fountains and the ISM-CGM interface; the existence of a turbulent CGM and galactic outflows, and the interaction with satellite galaxy ISM/CGM structure. All of these, it is worth noting, have been extensively studied and compared to observations with simulations identical to those here (modulo the assumed CR scattering rate scalings; see references in \S~\ref{sec:intro} \&\ \S~\ref{sec:methods} and \citealt{chan:2018.cosmicray.fire.gammaray,hafen:2018.cgm.fire.origins,emami:2019.bursty.dynamics.similar.to.fire.in.obs.but.fire.too.rapid.in.intermediate,ji:fire.cr.cgm,benincasa:2020.gmc.lifetimes.fire,gurvich:2020.fire.vertical.support.balance,chan:2021.cosmic.ray.vertical.balance,ponnada:fire.magnetic.fields.vs.obs,kim:2022.hot.gas.cgm,trapp:2022.gas.radial.transport.fire}). The point of our numerical simulations, fundamentally, was to see if non-linear effects from structure (e.g.\ varying values of terms which go into estimating scattering rates, such as $|{\bf B}|$ or $n$, as we discussed immediately above in \S~\ref{sec:galaxy.props.rescue}) could somehow introduce qualitatively different behaviors from those predicted by the simple analytic arguments in \S~\ref{sec:problems}, and so somehow ``rescue'' traditional SC/ET models from the problems we anticipated. We also wanted to explore whether macroscopic ``back-reaction'' or cosmic ray ``feedback'' effects which should be resolve-able would somehow lead to a kind of feedback loop which could alter our analytic conclusions. This includes e.g.\ the effects of CRs changing galactic wind/outflow dynamics, driving instabilities such as the ``staircase,'' or altering the phase structure of the CGM, or exerting ``pressure'' to change the vertical balance or turbulent structure of the ISM, or altering the global ionization structure of cold clouds -- all of these can (and as noted above, many do to some extent) occur in our simulations, which include all of the required physics and coupling terms.

Of course, the simulations have finite resolution and as we clearly noted from the beginning of this study, they cannot even approach resolving the ``micro-scale'' by which we refer to gyro-resonant scales for the CRs of interest ($\lesssim 100\,$AU). These are the scales of actual ``scattering'' physics, where PIC-type methods are needed to treat the CR dynamics. In the simulations, CR scattering is therefore explicitly ``sub-grid.'' Another way of saying this is that we cannot predict CR scattering rates from first principles, but instead are here testing different models for how the ``micro-scale'' CR scattering rates depend on ``macro-scale'' parameters. This allows us to show that some key physics or assumptions must be missing from these models.

But it is also worth mentioning that, given this scale separation, there could be interesting dynamics in the ``meso-scale'' as well, by which we mean scales much larger than the gyro-resonant scales, but much smaller than the resolved simulation scales or driving/coherence scales of the volume-filling warm ISM/CGM components (and with small volume-filling factors). Consider, for example, stellar magnetospheres: we know from the Heliosphere that these represent regions with order-one changes in the magnetic field on scales $\sim 100\,$AU, vastly smaller than the \Alf/coherence scale of magnetic fields in the volume-filling warm ionized ISM ($\sim 100-200\,$pc), and that this can (and does) strongly scatter/deflect the pitch angles of CRs with energies $\ll$\,TeV. In a sense, we can think of this as a tiny patch of the ISM interior to which the local \Alf\ scale $\ell_{A}$ decreases from $\sim 100\,$pc to $\sim 100\,$AU. These are obviously un-resolved in our simulations. But it is, at least in principle, possible to imagine models in which such ``meso-scale'' structures dominate CR scattering, and introduce effects like those we sought to explore on the ``macro-scale'' in our simulations, and so could strongly modify the ``effective'' scattering rates and residence times of CRs. We stress that any such model would still represent a radical departure from traditional CR transport theory: in traditional models such as those explored here, scattering is dominated via the sum of many small-angle/weak scattering events, and the CR residence times (hence ``effective'' scattering rates) are dominated by the statistically homogeneous, relatively smooth, volume-filling phases of the ISM (e.g.\ the WIM and warm inner CGM, for CRs observed in the LISM; see \paperone\ and references therein). And any such model would still have to solve the problems we present here: it would have to predict a physical means by which such structures could introduce any (let alone the {\em correct}) energy dependence to the ``effective'' CR scattering rates over the required energy range. In this sense, one can think of such models as a mechanism by which something like our alternative damping or driving rates (required on macro-scales) could be achieved, just via an intermediate scale effect. But in addition, such a model would necessarily have to show that meso-scale structures actually dominate CR scattering between their initial acceleration and observation in the LISM. For the example of stellar magnetospheres given above, this appears impossible: the mean-free-path between magnetospheres in the ISM (given a stellar density of $\sim 1\,{\rm pc^{-3}}$ and radius $\sim 100\,$AU) is $\sim$\,Mpc, while the observationally-inferred mean-free-path for deflection/scattering of $\sim$\,GeV CRs is $\sim 10\,$pc ($10^{5}$ times smaller). And implicit in the above, it would still be necessary in such a model to explain how the diffuse ISM outside of such structures does not undergo solution collapse or SC runaway to over-confinement. Still, it is very much worth keeping such models in mind, as there is a diverse ensemble of phenomology in the ISM on scales not captured in either the simple analytic scalings or numerical galaxy-scale simulations explored here (for examples, see the scattering processes examined in \citealt{bai:2022.streaming.instab.sims,beattie:vA.scattering.turb}),  and we will in future work (Butsky et al., in prep.) try to map out in more detail some of the requirements of such meso-scale models.

\section{Conclusions}
\label{sec:discussion}

We have combined analytic models and a suite of detailed numerical simulations of CR transport in fully-dynamical Galactic environments to explore the physics of CR scattering at CR energies $\sim $\,MeV-TeV. From all of these, we show that standard SC and ET models cannot even qualitatively reproduce basic features of the observed CR spectra and secondary-to-primary or radioactive species ratios. The model failures are not superficial and, across our extensive survey, we find no ``tweaking'' that acts as a remedy; this is expected, because we argue that the problems arise due to fundamental and indeed {\em defining} assumptions of SC and ET models. Specifically, for SC models, the fact that the term driving the growth of the CR scattering rate itself depends on the CR flux or energy density causes the SC ``instability'' or ``solution collapse'' problem, wherein, regardless of any details of the functional form of SC scattering rates or damping mechanisms, CRs quickly converge to either the trapped/infinite-scattering limit or the free-streaming/escape-at-$c$ limit. For ET models, the assumption that the scattering modes arise from an MHD ``cascade,'' or other transfer between scales over a large dynamic range, forces the scattering modes to obey the qualitatively incorrect scaling as a function of rigidity at scales below the \Alf\ and/or dissipation scale of turbulence (which includes all CRs in the ISM below a few hundred GeV). 

We therefore phenomenologically approach the problem and ask ``what would be needed'' -- in terms of either the driving or damping of CR scattering modes -- to resolve all of these issues and reproduce  CR observations. While previous studies have empirically quantified this in terms of an ``effective diffusivity'' or ``mean scattering rate'' that best fits  observations  (e.g.\ fitting some constant in space and time or a simply parameterized function for the diffusion coefficient as a function of CR rigidity), we go further and actually solve the dynamical equations for the CR scattering rates, incorporating what is known about driving and damping rates of parallel magnetic fluctuations. For the first time, we constrain ``what is needed'' directly in terms of the {\em local} driving rate $S_{\pm}$ or damping rate $\Gamma_{\pm}$ of CR scattering modes, on scales of order the gyro-resonant  wavelengths. These are the quantities that can actually be predicted by detailed theoretical calculations and PIC simulations of CR scattering physics. We identify three classes of model that could, at least qualitatively, reproduce the CR observations, and quantify what is needed for each.

\begin{enumerate}

\item{Alternative Damping:} All the key problems introduced by the dominant SC term at $\sim$\,MeV-TeV energies can be resolved if the linear damping rate scales with the CR energy density (at some rigidity) $\Gamma_{\rm new,\,damp,\,\pm} \propto e_{\rm cr}^{\prime} \propto de_{\rm cr}/d\ln{R_{\rm cr}}$, e.g.\ $\Gamma_{\rm new,\,damp,\,\pm} \sim (v_{A,\,{\rm ideal}}/\ell_{A})\,(k_{\|}\,\ell_{A})^{\xi_{k}}\,(e_{\rm cr}^{\prime}/e_{\rm B})$ with $0.1 \lesssim \xi_{k} \lesssim 0.4$. However, there are two key issues: (1) it is not obvious what could physically produce such a scaling, and (2) this damping must dominate over all other linear damping mechanisms in the volume-filling ISM, which effectively requires discarding or drastically reducing the normalization of standard linear damping mechanisms such as ion-neutral, dust, turbulent/linear Landau, and non-linear Landau damping.

\item{Alternative Linear Driving/Sources:} Alternatively, if the CR scattering waves with energy $e_{A} \sim  |\delta{\bf B}(k_{\|})|^{2}$ are driven by a linear source term, $S_{\rm new,\,lin} \propto e_{A}$, where $S_{\rm new,\,lin}$ does {\em not} depend on CR energy, this can avoid the problems of SC models and reproduce observations. A form such as $S_{\rm new,\,lin} \sim 10^{-12}\,{\rm s^{-1}}\,e_{A}\,(k_{\|}\,{\rm au})^{\zeta_{k}}$ with $0.6 \lesssim \zeta_{k} \lesssim 0.9$ provides reasonable results. Note that SC models are intrinsically based on such a ``linear'' source term (from CR gyro-resonant instabilities), but the problem is that their dependence on the CR energy density  introduces the instability/solution-collapse problems, and the $k$ dependence scales incorrectly to reproduce observations. But a wide variety of other known linear instabilities -- e.g.\ a host of multi-fluid instabilities that are known to operate on the relevant scales -- could potentially explain this scaling, and only very modest power is needed in the relevant modes. The problem with this solution is that the linear source must compete with linear damping and SC driving, so reproducing observations with this model class requires somewhat weaker linear damping. This problem is not as severe as for ``alternative damping'' above, but in particular the standard turbulent/linear-Landau and dust damping scalings are too strong and would need to be revised.

\item{Alternative External Driving/Sources:} Instead, an alternative source term that is independent of $e_{A}$ and $e_{\rm cr}^{\prime}$, and only weakly dependent on $k$ -- for example, $S_{\rm new,\,ext} \sim d E(k_{\|})/d\ln{k_{\|}}\,d t\,d\,{\rm Volume} \sim 0.01\,(v_{A,\,{\rm ideal}}/\ell_{A})\,e_{\rm B}\,(k_{\|}\,{\rm au})^{\zeta_{k}}$ with $-0.25 \lesssim \zeta_{k} \lesssim -0.1$ -- can resolve the key problems of SC and ET models and reproduce observations. This version requires remarkably little revision to other known damping or driving terms. While this ``external'' scaling is qualitatively similar to ET models in that $S_{\rm new,\,ext}$ is independent of $e_{\rm cr}^{\prime}$ and $e_{A}$, it cannot derive from a standard turbulent cascade from large scales without introducing the anisotropy and damping problems, but is better thought of as a driver acting over a wide range of scales (or some modification of standard MHD turbulence paradigms). The total power needed is modest ($\sim 1\%$ of the dissipation rate in ISM turbulence) and it is plausible to imagine a variety of physical  mechanisms that could act in this way: the challenge may be to ensure such a mechanism can act over the entire relevant dynamic range of $\sim 10^{3}-10^{6}$ in $k_{\|}$.

\end{enumerate}

It is important to note that, although we demonstrate the conclusions above over a wide range of CR energies $\sim\,$MeV-TeV, it may be possible to somewhat reduce the dynamic range of CR energies (and therefore wavenumbers $k_{\|}$) over which alternative physics must play a key role. For example, as argued in \paperone\ and seen in some (but not all) of the models here, the residence time of very low-energy CRs at $\lesssim 10-100\,$MeV could be regulated by Coulomb/ionization losses (making predictions consistent with observations and nearly independent of scattering rates), so long as the scattering rates at these energies are sufficiently high so that the diffusion/escape time is longer than loss times. And depending on  detailed ISM properties, at some energy $\gtrsim 0.1-1\,$TeV, CR gyro radii will eventually become comparable to the dissipation scales of turbulence, so the ``classical'' ET scenario of scattering from an undamped extrinsic turbulent cascade becomes a reasonable approximation (provided there is an isotropic fast-magnetosonic inertial-range cascade with approximately $k\,\mathcal{E}(k) \propto k^{-3/2}$ and the correct normalization). It is at intermediate energies, where most of the CR energy density resides, that the problems described here are most acute. 

We stress that we are not here advocating for any one specific physical process as the explanation for CR scattering. Instead our goal is to identify and further investigate generic problems  with current SC and ET models; some of these problems were already known, some  have been first identified here. We further identify  classes of scalings for either driving or damping of CR scattering modes that could, in principle, explain the observations, discussing  several possible physical mechanisms above.
In future work, we hope to explore some of these candidate processes in more detail to assess if any can actually produce the correct scaling and normalizations needed to explain  observations. It may be that the quantitative details will differ by a modest amount, as there are a variety of effects that lead to e.g.\ exact deviations from simple power-law behavior, but we expect that the key qualitative requirements identified herein are robust.
If mechanisms can be identified that meet these criteria, it will be important to also test them in microphysical MHD-PIC-like simulations, then use the results as input to galactic simulations such as those explored here. This will allow a quantitative comparison to CR observations, providing further valuable constraints on the important processes at play.

\acknowledgments{Support for PFH was provided by NSF Research Grants 1911233 \&\ 20009234, NSF CAREER grant 1455342, NASA grants 80NSSC18K0562, HST-AR-15800.001-A. Support for JS  was
provided by Rutherford Discovery Fellowship RDF-U001804 and Marsden Fund grant UOO1727, which are managed through the Royal Society Te Ap\=arangi. Numerical calculations were run on the Caltech compute cluster ``Wheeler,'' allocations FTA-Hopkins/AST20016 supported by the NSF and TACC, and NASA HEC SMD-16-7592.}

\datastatement{The data supporting this article are available on reasonable request to the corresponding author.} 

\bibliography{ms_extracted}

\clearpage

\begin{appendix}

\section{Equilibrium Self-Confinement Models}
\label{sec:sc.equilibrium.models}

\subsection{Basic Equations and Setup}

Here we consider the behavior of SC models in steady-state. Given that (as shown in \S~\ref{sec:local.steady.state.detailed.solns} below) the CR flux $F_{e,\,{\rm cr}}^{\prime}$ and $e_{\pm}$ equations converge to local steady state on a timescale much shorter than the CR energy $e_{\rm cr}^{\prime}$ equation, we can safely assume their steady-state values (in \S~\ref{sec:local.steady.state.detailed.solns}) in evaluating the CR energy equation. We will assume only SC driving (take $S_{\rm ext,\,\pm}=S_{\rm new,\,\pm}=0$),  giving (from Eq.~\ref{eqn:dtecr.eqm}):
\begin{align}
\label{eqn:Dtecr.sc} D_{t}e_{\rm cr}^{\prime} &\rightarrow -\nabla\cdot ( F_{e,\,{\rm cr}}^{\prime}\,\bhat ) + \mathcal{S}_{\rm eff}^{\prime} 
\end{align}
where $F_{e,\,{\rm cr}}^{\prime} \equiv -({v_{\rm cr}^{2}}/{3\,\bar{\nu}_{\rm s}})\,\bhat\cdot\nabla e_{\rm cr}^{\prime} + \bar{v}_{A}\,e_{\rm cr}^{\prime}$ includes the ``diffusive'' ($\propto \bar{\nu}_{\rm s}^{-1}$) and ``streaming'' ($\propto \bar{v}_{A}$, with $\bar{v}_{A} \rightarrow -v_{A,\,{\rm eff}}\,{\rm sign}[\bhat\cdot\nabla e_{\rm cr}^{\prime}]$) terms, and $S_{\rm eff}^{\prime} \equiv -P_{\rm cr}^{\prime}\,\nabla \cdot ({\bf u}_{\rm gas} + \bar{v}_{A}\,\bhat) + \mathcal{S}_{\rm other,\,cr}^{\prime}$ includes the ``adiabatic'' and ``streaming loss'' terms ($\propto \nabla \cdot [{\bf u}_{\rm gas} + \bar{v}_{A}\,\bhat]$) and all injection and radiative/catastrophic losses in $\mathcal{S}_{\rm other,\,cr}^{\prime}$. 

In steady state ($D_{t}e^{\prime}_{\rm cr} \rightarrow 0$), integrating Eq.~\ref{eqn:Dtecr.sc} over some volume $V$ with surface $\partial V$ immediately gives: 
\begin{align}
\label{eqn:flux.eqm.global} \langle F_{e,\,{\rm cr}}^{\prime} \rangle\,A_{\rm eff} &\equiv \oint_{\partial V} {F}_{e,\,{\rm cr}}^{\prime} \bhat \cdot {\rm d}{\bf A} = \int_{V} {\rm d}^{3}{\bf x} \, \mathcal{S}_{\rm eff}^{\prime} \equiv \dot{E}_{\rm inj,\,eff}. 
\end{align}
Here $\langle F_{e,\,{\rm cr}}^{\prime} \rangle$ is the weighted-mean scalar flux from the integral over $F_{e,\,{\rm cr}}^{\prime}$, $A_{\rm eff} \equiv \oint_{\partial V} |{\rm d}{\bf A}|$, and $\dot{E}_{\rm inj,\,eff}$ is the net CR energy production inside $\partial V$. For simplicity, we will consider CR primary species at rigidities $\gtrsim\,$GV where $v_{\rm cr} \sim c$ (so $P_{\rm cr}^{\prime} \approx e_{\rm cr}^{\prime}/3$) and empirical constraints (see text and \paperone) indicate losses are negligible, so $\dot{E}_{\rm inj,\,eff} \approx \int_{V}\,{\rm d}^{3}{\bf x} j_{{\rm inj},\,e}(R_{\rm cr}) = \dot{E}_{\rm inj}$ is approximately the total injection rate.

\subsection{Behavior of Phenomenological or ET Models}

First, consider a typical phenomenological model, where $\bar{\nu}_{\rm s}$ is taken to be constant with $\bar{\nu}_{\rm s} \sim 10^{-9}\,{\rm s^{-1}}\,R_{\rm GV}^{-0.5}$ as in Fig.~\ref{fig:demo.cr.spectra.fiducial}. With $\bar{\nu}_{\rm s}=$\,constant, Eq.~\ref{eqn:Dtecr.sc} indeed behaves like a diffusion equation, with the diffusive term much larger than streaming terms on scales of interest, and if we assume tangled magnetic fields the effective isotropic flux is just $\langle F_{e,\,{\rm cr}}^{\prime} \rangle \approx \kappa_{\rm iso}\,\langle | \nabla e_{\rm cr}^{\prime} | \rangle$ with $\kappa_{\rm iso} \equiv (c^{2}/9\,\bar{\nu}_{\rm s})$. For the Galaxy, take $\dot{E}_{\rm inj,\,eff} \approx \dot{E}_{\rm inj} \approx 0.1\,\dot{E}_{\rm SNe}\,f_{\rm inj}(R_{\rm cr}) \sim 3\,\dot{N}_{{\rm SNe},\,100}\,R_{\rm GV}^{-0.2}\,\times10^{40}\,{\rm erg\,s^{-1}}$ where $\dot{N}_{{\rm SNe},\,100}$ is the SNe rate inside $\partial V$ in units of $\sim 1/(100\,{\rm yr})$ and $f_{\rm inj}(R_{\rm cr}) \equiv (1/\dot{E}_{\rm inj}^{\rm total})\,{\rm d}\dot{E}_{\rm inj}^{\rm total}/{\rm d}\ln{R_{\rm cr}} \sim R_{\rm GV}^{-0.2}$ is the fraction injected at the given $R_{\rm cr}$ (according to our assumed standard injection slope in the text).  Assuming e.g.\ spherical symmetry or a vertically-stratified model, the steady-state $e_{\rm cr}^{\prime}$ profile is then trivially solved by $\nabla e_{\rm cr}^{\prime}  = - \dot{E}_{\rm in} / (\kappa_{\rm iso}\,A_{\rm eff})$. If we assume approximate spherical symmetry at large Galacto-centric radii $\langle F_{e,\,{\rm cr}}^{\prime} \rangle = \dot{E}_{\rm inj}/(4\pi\,r^{2})$ we obtain $e_{\rm cr}^{\prime} \sim 0.6\,{\rm eV\,cm^{-3}}\,\dot{N}_{{\rm SNe},\,100}\,R_{\rm GV}^{-0.7}$ at the Solar circle ($r\sim 8.3\,{\rm kpc}$), in excellent agreement with observations (by construction, of course, since $\bar{\nu}_{\rm s}$ was originally fit to the data).

In standard ET models, $\bar{\nu}_{s} \rightarrow \bar{\nu}_{s}(k_{\|},\,{\bf B},\,\ell_{A},\,f_{\rm ion},\,...)$ can be some arbitrary function of ISM properties, but (crucially) is -- like in the phenomenological model above -- {\em independent} of $e_{\rm cr}^{\prime}$ ($\zeta_{\rm cr} = \xi_{\rm cr} = 0$). This means that, again, solutions always exist for a steady-state $e_{\rm cr}^{\prime}$ profile, given by solving Eq.~\ref{eqn:flux.eqm.global}: $\nabla e_{\rm cr}^{\prime} \sim -\dot{E}_{\rm in} / (\kappa_{\rm iso}\,A_{\rm eff}) \sim -9\,\dot{E}_{\rm in}\,\bar{\nu}_{\rm s} / (c^{2}\,A_{\rm eff}) = \mathcal{F}(r,\,k_{\|},\,{\bf B},\,\ell_{A},\,f_{\rm ion},\,...)$. Whether or not these solutions have the correct observed behavior (as a function of e.g.\ CR rigidity) is what we investigate in the main text.

\subsection{Behavior of SC Models}

But now consider SC models, with $\bar{\nu}_{\rm s} \approx (3\pi\,\Omega_{\rm cr}/16)\,(e_{A}/e_{\rm B})$ where $e_{A}$ is set (see \S~\ref{sec:local.steady.state.detailed.solns}) by the competition between damping ($\Gamma_{\pm}$) and driving with $S_{\rm sc} \rightarrow |{v}_{A,\,{\rm eff}}\,\bhat \cdot \nabla P_{\rm cr}^{\prime}|$, giving $e_{A}/e_{\rm B} \rightarrow (\Gamma_{\rm lin}/2\,\Gamma_{\rm nll}^{0})\,(-1 + [1 + 4\,S_{\rm sc}\, \Gamma_{\rm nl}^{0}/\Gamma_{\rm lin}^{2}\,e_{\rm B} ]^{1/2})$, where $\Gamma_{\rm lin} \equiv \Gamma_{\rm in} + \Gamma_{\rm turb/LL} + \Gamma_{\rm dust} + \Gamma_{\rm new,\,damp} + ...$ collects all linear damping terms and $\Gamma_{\rm nl}^{0}$ collects the prefactors of any non-linear terms (e.g.\ $\Gamma_{\rm nl}^{0} = \Gamma_{\rm nll}^{0}=(\sqrt{\pi}/8)\,c_{s}\,k_{\|}$ for non-linear Landau damping). The dependence of $e_{A}$ on $e^{\prime}_{\rm cr}$ introduces fundamentally distinct behavior. 

\subsubsection{Linear Damping}

First assume linear damping dominates\footnote{So long as $\Gamma_{\rm lin} \gtrsim \Gamma_{\rm nl}$, explicitly including small-but-finite $\Gamma_{\rm nl}$ changes none of our conclusions above. The limit $\Gamma_{\rm lin} \lesssim \Gamma_{\rm nl}$ is discussed below.} ($\Gamma_{\rm lin} \gtrsim \Gamma_{\rm nl} = \Gamma_{\rm nl}^{0}\,(e_{A}/e_{\rm B})$). Then $e_{A} \rightarrow S_{\rm sc}/\Gamma_{\rm lin}$, giving the ``diffusive'' flux $F_{e,\,{\rm cr}}^{\prime} \rightarrow (16/3\pi)\,(c\,e_{\rm B}\,r_{g,{\rm cr}}\,\Gamma_{\rm lin})/v_{A,\,{\rm eff}} = (4c/3\pi^{3/2})\,R_{\rm cr}\,\rho_{\rm ion}^{1/2}\,\Gamma_{\rm lin}$. If $D_{t}e_{\rm cr}^{\prime}$ is dominated by the diffusive term,  then since $\Gamma_{\rm lin}$ for any known damping mechanism depends on properties extrinsic to the CRs (e.g.\ turbulent velocities, $e_{\rm B}$, etc.) and so it and therefore $F_{e,\,{\rm cr}}^{\prime}$ are independent of the CR energy density, this means there exist {\em no steady state solutions} for $e_{\rm cr}^{\prime}$. It does not seem possible to construct an $e_{\rm cr}^{\prime}$ profile that ensures $\langle F_{e,\,{\rm cr}}^{\prime} \rangle = \dot{E}_{\rm inj,\,eff}/A_{\rm eff}$.\footnote{One might imagine a (contrived) special case where the  properties (e.g.\ ${\bf B}$, $c_{s}$) which enter $\Gamma_{\rm lin}$ scale exactly as required with both $\dot{E}_{\rm inj}$ and position ${\bf x}$ such that $\langle F_{e,\,{\rm cr}}^{\prime} \rangle = \dot{E}_{\rm inj}/A_{\rm eff}$. However, not only does this require exceptional fine-tuning, but (1) because $\dot{E}_{\rm inj}$ and $F_{e,\,{\rm cr}}^{\prime}$ scale {\em differently} with $R_{\rm GV}$, it is impossible to satisfy this at any two CR energies simultaneously, and (2) such a solution cannot ``respond'' to adjust to any perturbations to the source rate $\dot{E}_{\rm inj}$ or to the gas quantities which enter $F_{e,\,{\rm cr}}^{\prime}$.} 

In practice what this means is that there are only two real equilibrium solutions: if $\langle F_{e,\,{\rm cr}}^{\prime} \rangle < \dot{E}_{\rm inj}/A_{\rm eff}$, since the diffusive flux is independent of $e_{\rm cr}^{\prime}$, the CR energy density will continue to build up (increasing $\bar{\nu}_{\rm s} \propto e_{\rm cr}^{\prime}$ and lowering the effective diffusivity or streaming speed) until the streaming term $\propto v_{A,\,{\rm eff}}\,e_{\rm cr}^{\prime}$ dominates $F_{e,\,{\rm cr}}^{\prime}$ or catastrophic loss terms (also $\propto e_{\rm cr}^{\prime}$) dominate $D_{t} e_{\rm cr}^{\prime}$. Thus CRs collapse to the \Alf{ic} streaming and/or calorimetric limit, with maximal isotropically-averaged streaming speed $\approx v_{A,\,{\rm eff}}/2$. This is problematic for two reasons: first, the implied residence time (neglecting losses) to escape the Galaxy and CR scattering halo ($\sim 10\,$kpc) is $\sim 10\,{\rm Gyr}\,n_{1}^{1/2}\,B_{\mu {\rm G}}^{-1}$, far longer than observationally allowed. Second, even if we arbitrarily re-normalized the \Alf\ speed and/or Galaxy+halo size, the streaming/escape/residence time would (by definition) be {\em independent} of CR energy (i.e.\ $\delta_{\rm s}=0$), also ruled out. Alternatively, if $\langle F_{e,\,{\rm cr}}^{\prime} \rangle > \dot{E}_{\rm inj}/A_{\rm eff}$, then $e_{\rm cr}^{\prime}$ will deplete until $\bar{\nu}_{\rm s}$ is so low\footnote{As noted in the main text, if extrinsic turbulence is present, then at some sufficiently-low $\bar{\nu}_{\rm s}$, $S_{\rm et,\,\pm}$ will dominate so $\bar{\nu}_{\rm s}$ will not vanish entirely, but in this case the system is in the entirely-ET dominated limit.} that the CRs free-stream and escape at $\sim c$, vastly faster than observed (residence times $\lesssim 10^{4.5}\,$yr), with again $\delta_{\rm s}=0$. 

An alternative way to see this is to simply insert the full expression for the SC-predicted $\bar{\nu}_{\rm s}$ directly into Eq.~\ref{eqn:Dtecr.sc}. As noted by many going back to \citet{skilling:1971.cr.diffusion} and \citet{1971BAAS....3..480C}, the ``diffusive'' part of the flux-gradient term then formally takes the form of a source or sink term: 
\begin{align}
\label{eqn:dtecr.source.term} D_{t}e_{\rm cr}^{\prime} =& \pm \left( \frac{4\,c\,R_{\rm cr}}{3\,\pi^{3/2}}\right)\,\nabla \cdot (\rho_{\rm ion}^{1/2}\,\Gamma_{\rm lin}\,\bhat) \\
\nonumber \sim& \pm \frac{\rm eV\,cm^{-3}}{\rm Myr}\,B_{\mu {\rm G}}^{3/2}\,R_{\rm GV}^{1/2}\,\left(\frac{10\,{\rm pc}}{\ell_{\nabla,\Gamma\rho^{1/2}}}\right)
\end{align}
where the sign is determined by the gradient in $e_{\rm cr}^{\prime}$; ${\ell_{\nabla,\Gamma\rho^{1/2}}} \equiv \rho_{\rm ion}^{1/2}\,\Gamma_{\rm lin}/|\nabla \cdot (\rho_{\rm ion}^{1/2}\,\Gamma_{\rm lin}\,\bhat)|$ is the gradient scale length of $f_{\rm ion}^{1/2}\,\rho^{1/2}\,\Gamma_{\rm lin}$ which can vary on $\lesssim $\,pc scales \citep{hopkins:cr.transport.constraints.from.galaxies}; and in the second equality we inserted the scalings for $\Gamma_{\rm lin}$ for turbulent/linear Landau damping (to give a typical value). From Eq.~\ref{eqn:dtecr.source.term}, it is clear that within a timescale $\sim$\,Myr, the CR energy $e_{\rm cr}^{\prime}$ will either (1) be driven to negligible values if $D_{t}e_{\rm cr}^{\prime}$ is negative (making all other terms in $D_{t}e_{\rm cr}^{\prime}$ smaller, until the ``free escape'' limit is reached), or (2) be driven to increase if $D_{t}e_{\rm cr}^{\prime}$ is positive, until the other terms in $D_{t}e_{\rm cr}^{\prime}$ such as the streaming term $\propto \bar{v}_{A,\,{\rm eff}}\,e_{\rm cr}^{\prime}$ dominate (the ``over-confined'' limit).

\subsubsection{Non-Linear Damping}

Now instead assume non-linear Landau (NLL) damping dominates. Let us first ask when this might occur: for NLL damping to set $e_{A}$ (see \S~\ref{sec:local.steady.state.detailed.solns}) requires the dimensionless $\psi_{\rm nl} \equiv |4\,S_{\rm sc}\,\Gamma_{\rm nl}^{0}/\Gamma_{\rm lin}^{2}\,e_{\rm B}|^{1/2} \gg 1$. 
Taking the standard linear damping scalings from \S~\ref{sec:damping}, $\psi_{\rm nl} \gg 1$ requires the gas is highly ionized ($f_{\rm neutral} \lesssim 10^{-3}$, so $\Gamma_{\rm in}$ is small), has a low dust-to-gas ratio ($f_{\rm dg} \lesssim 10^{-3}$, so $\Gamma_{\rm dust}$ is small), and is weakly turbulent with a high CR energy density at the given $R_{\rm cr}$ ($\mathcal{M}_{A}^{2} \lesssim e_{\rm cr}^{\prime} / {\rm eV\,cm^{-3}}$, so $\Gamma_{\rm turb/LL} \ll \Gamma_{\rm nll}$). 
While specific, this is not impossible around $\sim 1\,$GV where $e_{\rm cr}^{\prime}$ peaks, in the diffuse warm/hot ISM/CGM. 

Assuming $\psi_{\rm nl} \gg 1$ with NLL damping dominating, we have  $e_{\rm A} \rightarrow (S_{\rm sc}\,e_{\rm B}/\Gamma_{\rm nl}^{0})^{1/2}$, so the ``diffusive'' $F_{e,\,{\rm cr}}^{\prime} \rightarrow (2^{5/2}/3^{3/2}\,\pi^{3/4})\,c\,(|\bhat \cdot \nabla e^{\prime}_{\rm cr}|\,c_{s}\,e_{\rm B}\,r_{g,{\rm cr}}/v_{A})^{1/2}$. This does formally have a steady-state solution given by $\langle |\bhat\cdot \nabla e_{\rm cr}^{\prime}| \rangle \approx (27\pi^{3/2}/32)\,(v_{A,\,{\rm eff}}/c^{2}\,c_{s}\,e_{\rm B}\,r_{g,{\rm cr}})\,(\dot{E}_{\rm inj}/A_{\rm eff})^{2}$ with $\langle \bar{\nu}_{\rm s} \rangle \rightarrow (9\pi^{3/2}/32)\,(v_{A,\,{\rm eff}}\,\dot{E}_{\rm inj}/c_{s}\,e_{\rm B}\,r_{g,{\rm cr}}\,A_{\rm eff})$. But this solution has some very strange features: using $A_{\rm eff} \sim 4\pi\,r^{2}$ as above and evaluating it at the Solar circle we obtain: 
$e_{\rm cr}^{\prime} \approx 2000\,{\rm eV\,cm^{-3}}\,R_{\rm GV}^{-1.4}\,N_{\rm SNe,\,100}^{2}\,(n_{1}\,T_{4})^{-1/2}$ (with $T_{4} = T/10^{4}\,$K) and $\bar{\nu}_{\rm s} \rightarrow 10^{-5}\,{\rm s^{-1}}\,R_{\rm GV}^{-1.2}\,N_{\rm SNe,\,100}\,(n_{1}\,T_{4})^{-1/2}$. These are enormously unphysically high CR energy densities and scattering rates, which also exhibit a clearly ruled-out dependence on $R_{\rm GV}$. In practice this means that, beginning from any physically realistic (much smaller) $e^{\prime}_{\rm cr}$, $F_{e,\,{\rm cr}}^{\prime} \ll \dot{E}_{\rm cr}/A_{\rm eff}$ will drive $D_{t} e^{\prime}_{\rm cr} > 0$ so $e^{\prime}_{\rm cr}$ increases until either the streaming or loss terms (which scale $\propto e_{\rm cr}^{\prime}$, while the diffusive term scales $\propto \sqrt{e_{\rm cr}^{\prime}}$) dominate $D_{t}e_{\rm cr}^{\prime}$ (e.g.\ the streaming flux will dominate $F_{e,\,{\rm cr}}^{\prime}$ once $e_{\rm cr}^{\prime} \gtrsim 0.5\,{\rm eV\,cm^{-3}}\,R_{\rm GV}\,(n_{1}^{3/2}\,T_{4}^{1/2}\,B_{\mu {\rm G}}^{-2}\,{\rm kpc}/\ell_{\nabla,{\rm cr}})$).\footnote{From the scalings above, in this limit the flux should be dominated by the \Alf{ic} streaming component at all galacto-centric radii interior to $\lesssim {\rm Mpc}\,B_{\rm \mu G}\,N_{\rm SNe,\,100}\,n_{1}^{-1}\,T_{4}^{-1/2}\,R_{\rm GV}^{-1.2}$} 

So again we see immediate ``solution collapse,'' but the conditions where non-linear damping dominates, which require higher $e_{A}$ and therefore higher $e_{\rm cr}^{\prime}$, are such that they always drive the collapse to the over-confined/streaming solution.

\subsubsection{Summary}

These behaviors above are what we refer to in the text as the SC models being globally ``not stable.'' This is not necessarily a linear-stability analysis (in fact the ``collapsed'' free streaming or over-confined spherical equilibrium solutions above, if we perturb just $e_{\rm cr}^{\prime}$ infinitesimally and ignore all interactions with the gas, are formally linearly stable). But it is common, when SC is discussed, to refer to ``super-\Alf{ic} streaming,'' i.e.\ flux in excess of $v_{A,\,{\rm eff}}\,e_{\rm cr}^{\prime}$ with an effective contribution to $F_{e,\,{\rm cr}}^{\prime}$ from the $\bar{\nu}_{\rm s}$ term as defined above (i.e.\ a finite-but-not-infinite CR transport speed in excess of $v_{A,\,{\rm eff}}$). This can arise trivially, in any infinitesimal local patch, if one defines $\bar{\nu}_{\rm s}$ for a {\em given} $e_{\rm cr}^{\prime}$ (e.g.\ choosing a fixed $e_{\rm cr}^{\prime}$ similar to the Solar circle value) -- in fact we show this  below in \S~\ref{sec:local.steady.state.detailed.solns}, where we derive the values of $e_{\pm}$ given by assuming local steady-state of the CR flux equation. But generically these solutions will {\em not} give a self-consistent steady-state for the CR {\em energy} equation: converging to such a ``local equilibrium'' value of $\bar{\nu}_{\rm s}$ for a given $e_{\rm cr}^{\prime}$ (as determined by the CR flux and $e_{\pm}$ equations, which evolve on very short timescales)  will mean necessarily that the energy equation is out-of-steady-state. This then forces $e_{\rm cr}^{\prime}$, and correspondingly $\bar{\nu}_{\rm s}$, to evolve either towards ``bottleneck'' and the infinite-strong-scattering \Alf{ic} streaming regime, or towards ``escape,'' de-confinement, and the negligible-scattering streaming-at-$c$ regime. Any initial condition rapidly collapses (over $\sim$\,Myr) towards one of these two states for all spatial and CR energy scales of interest if $S_{\rm sc}$ is the dominant driving term.

\section{Local Steady-State Solutions for Scattering Rates In Detail}
\label{sec:local.steady.state.detailed.solns}

\subsection{Relevant Equations and Limits}

Consider the CR flux and $e_{\pm}$ equations in more detail. From the general versions of Eqs.~\ref{eqn:specific.cr.energy}, \ref{eqn:ea}, and \ref{eqn:scr}, in the text we can write:
\begin{align}
\label{eqn:f.app} D_{t} F^{\prime}_{e,\,{\rm cr}} + c^{2}\,\bhat\cdot \left( \nabla \cdot \mathbb{P}_{\rm cr}^{\prime} \right) &= -\frac{\omega}{e_{\rm B}}\,(e_{+}+e_{-})\,F^{\prime}_{e,\,{\rm cr}} \\
\nonumber & \ \ \ + \frac{\omega}{e_{\rm B}}\,(e_{+}-e_{-})\,H^{\prime}_{\rm cr} \\ 
\label{eqn:ea.app} D_{t} e_{\pm} + \nabla \cdot (v_{A,\,\pm}\,e_{\pm}\,\bhat) &= -\frac{e_{\pm}}{2}\nabla \cdot {\bf u}_{\rm gas} -\Gamma_{\rm L}\,e_{\pm} - \Gamma_{\rm NL}\,\frac{e_{\pm}}{e_{\rm B}}\,e_{\pm}  \\
\nonumber &\ \ \  
+ \frac{\omega\,v_{A,\,\pm}\,e_{\pm}}{c^{2}\,e_{\rm B}}\,\left(F^{\prime}_{e,\,{\rm cr}} -H^{\prime}_{\rm cr} \right)
+ S_{\rm et,\,\pm}
\end{align}
where $H^{\prime}_{\rm cr} \equiv 3\,\chi\,v_{A,\,\pm}\,(e_{\rm cr}^{\prime} + P_{\rm cr}^{\prime})$, $\omega \equiv (\pi\,\hat{\nu}_{\rm s}/4)\,\Omega_{\rm cr}$, $v_{A,\,\pm} \equiv \pm v_{A,\,{\rm eff}}$, and we have expanded the damping rates in terms of the various linear ($\xi_{A}=0$, $\Gamma_{\rm L}$) and non-linear ($\xi_{A}=1$, $\Gamma_{\rm NL}$) damping terms as $Q_{\pm} = \Gamma_{\rm L}\,e_{\pm} + \Gamma_{\rm NL}\,(e_{\pm}/e_{\rm B})\,e_{\pm}$ (so for e.g.\ non-linear Landau damping, $\Gamma_{\rm NL} = (\sqrt{\pi}/8)\,c_{s}\,k_{\|}$). 

As noted in the text, these equations evolve towards local steady-state on a timescale $\sim \bar{\nu}_{\rm s}^{-1} \sim 30\,{\rm yr}\,R_{\rm GV}^{0.5}$ (if we take empirically-estimated $\bar{\nu}_{\rm s}$ values), much faster than the timescales for the CR energy equation or bulk ISM fluid motion timescales on the scales of interest. Similarly, the ``gradient terms'' in Eq.~\ref{eqn:ea.app} (the $\nabla \cdot (v_{A,\,\pm}\,e_{\pm}\,\bhat)$ and $\nabla \cdot {\bf u}_{\rm gas}$) involve timescales of order those same ISM timescales and are much smaller than the other terms in Eq.~\ref{eqn:ea.app}. We will justify these assumptions more formally below. Let us therefore assume these equations reach {\em local} steady-state ($D_{t} \rightarrow 0$, or $|D_{t}|\ll \bar{\nu}_{s}$) in the comoving \Alf\ frame (neglecting the ``gradient terms'' in Eq.~\ref{eqn:ea.app}) -- although we stress this does not mean the CR {\em energy} equation is in  steady state. In this case, we can re-write them in the dimensionless form:
\begin{align}
\tilde{g} &= -(x_{+}+x_{-})\,\tilde{f} + (x_{+}-x_{-})\,\tilde{h}, \\ 
\pm \tilde{f}\,x_{\pm}   &= \tilde{h}\,x_{\pm} +  \gamma_{\rm L}\,x_{\pm} + \gamma_{\rm NL}\,x_{\pm}^{2} - s_{\rm et} ,
\end{align}
where $x_{\pm} \equiv e_{\pm}/e_{\rm B}$, $\tilde{f} \equiv (v_{A,\,{\rm eff}}/c)^{2}\,(F^{\prime}_{e,\,{\rm cr}} / e_{\rm B}\,v_{A,\,{\rm eff}})$, $\tilde{h} \equiv (v_{A,\,{\rm eff}}/c)^{2}\, (H^{\prime}_{\rm cr} / e_{\rm B}\,v_{A,\,{\rm eff}} )$, $\tilde{g} \equiv (v_{A,\,{\rm eff}}/\omega\,e_{\rm B})\,\bhat\cdot (\nabla \cdot \mathbb{P}^{\prime}_{\rm cr})$, $\gamma_{\rm L} \equiv \Gamma_{\rm L}/\omega$, $\gamma_{\rm NL} \equiv \Gamma_{\rm NL}/\omega$, and $s_{\rm et} \equiv (S_{\rm et,\,+}+S_{\rm et,\,-}) / (2\,\omega\,e_{\rm B})$. This has solutions
\begin{align}
\label{eq: 5th order poly soln}
4\,s_{\rm et} &= \bar{x}\,[2\,(\gamma_{\rm L}+\tilde{h})+\gamma_{\rm NL}\,\bar{x}] - \frac{\tilde{g}^{2}\,[2\,(\gamma_{\rm L}+\tilde{h}) + \gamma_{\rm NL}\,\bar{x}]}{\bar{x}\,(\gamma_{\rm L}+\gamma_{\rm NL}\,\bar{x})^{2}} 
\end{align}
where $\bar{x} \equiv x_{+}+x_{-}$, with $x_{-} = x_{+} + \tilde{g}/(\gamma_{\rm L} + \gamma_{\rm NL}\,\bar{x})$ and $\tilde{f} \equiv [(x_{+}-x_{-})\,\tilde{h} - \tilde{g}]/\bar{x}$ following immediately.

\subsection{Local Steady-State Behavior in ET and SC Limits}

Unfortunately Eq.~\ref{eq: 5th order poly soln} is still a fifth-order polynomial for $\bar{x}$, whose general solutions are neither closed-form analytic nor particularly instructive. The solutions do, however, become simple in various limits. First consider the case where linear damping dominates over non-linear ($\gamma_{\rm NL}$ can be neglected). Then\footnote{The solution here assumes $\Gamma_{\rm L}>0$, i.e.\ linear damping. If instead there were net linear {\rm driving} from a driving source not considered here, so $\Gamma_{\rm L}<0$ in our language, then the physical solution branch for $\gamma_{\rm NL}$ small and $\gamma_{\rm L} < -\tilde{h}$ becomes $\bar{x} \rightarrow s_{\rm et}\,(1-\sqrt{1+\Phi^{2}})/(\gamma_{\rm L} + \tilde{h})$.} $\bar{x} \rightarrow s_{\rm et}\,(1+\sqrt{1+\Phi^{2}})/(\gamma_{\rm L} + \tilde{h})$ with 
\begin{align}
|\Phi| &\equiv \frac{|\tilde{g}|}{s_{\rm et}}\,\left[1 + \frac{\tilde{h}}{\gamma_{\rm L}} \right] 
\sim \frac{v_{A,\,{\rm eff}}\,|\nabla P^{\prime}_{\rm cr}|}{S_{\rm et,\,\pm}}\,\left[1 + \frac{\pi}{2}\frac{\Omega_{\rm cr}\,e_{\rm cr}^{\prime}}{\Gamma_{\rm L}\,\rho_{\rm ion}\,c^{2}} \right] 
\end{align}
Small $|\Phi|\ll 1$ here corresponds to the ET limit, large $|\Phi| \gg 1$ to the SC limit. 

Now consider each of those limits (SC and ET-dominated) in turn, but retain the non-linear term $\gamma_{\rm NL}$.

In the ET limit ($|\Phi| \ll 1$): the dimensionless ``streaming speed'' $\bar{v}_{A}/v_{A,\,{\rm eff}} = (x_{+}-x_{-})/(x_{+} + x_{-}) \rightarrow \Phi/2 \ll 1$ is small and $\bar{x} \rightarrow \gamma_{\rm NL}^{-1}\,(\gamma_{\rm L} + \tilde{h})\,(-1 + [1 + 4\,\gamma_{\rm NL}\,s_{\rm et}/(\gamma_{\rm L}+\tilde{h})]^{1/2})$, which corresponds to\footnote{The $\phi$ term here accounts for the fact that diffusive re-acceleration produces a net transfer of energy from the scattering modes ($e_{A}$) to the CRs ($e_{\rm cr}^{\prime}$) when $e_{+} \approx e_{-}$ (in the ET limit), so acts like an additional linear damping term even when $\Gamma_{\rm L} \rightarrow 0$.} $e_{+} \approx e_{-} \sim S_{\rm et} / (\Gamma_{\rm L}\,[1+\phi])$ (with $\phi \equiv \tilde{h}/\gamma_{\rm L} \sim (v_{A,\,{\rm eff}}/c)^{2}\,(\pi\,\hat{\nu}_{\rm s}/4)\,(\Omega_{\rm cr}/\Gamma_{\rm L})\,(e_{\rm cr}^{\prime} + P_{\rm cr}^{\prime})/e_{\rm B}$) when linear damping dominates and $e_{+} \approx e_{-} \sim (S_{\rm et}\,e_{\rm B}/\Gamma_{\rm NL})^{1/2}$ when non-linear damping dominates (which occurs when $4\,\gamma_{\rm NL}\,s_{\rm et} \gtrsim (\gamma_{\rm L} + \tilde{h})^{2}$). 

In the SC limit ($|\Phi|\gg 1$): the streaming speed $\bar{v}_{A}/v_{A,\,{\rm eff}} = (x_{+}-x_{-})/(x_{+} + x_{-}) \rightarrow -{\rm sign}(\tilde{g})$, so $\bar{v}_{A}$ is just the effective \Alf\ speed ($\pm v_{A,\,{\rm eff}}$) directed down the CR pressure gradient. Only the $x_{\pm}$ aligned in this direction is large (the other vanishes), with the relevant $x \approx \bar{x} \sim (\gamma_{\rm L}/2\,\gamma_{\rm NL})\,(-1 + [1 + 4\,|\tilde{g}|\,\gamma_{\rm NL}/\gamma_{\rm L}^{2}]^{1/2})$ which corresponds to $e_{A} \sim S_{\rm sc}^{0}/\Gamma_{\rm L}$ (with $S_{\rm sc}^{0} \equiv |{\bf v}_{A} \cdot \nabla \cdot \mathbb{P}_{\rm cr}^{\prime}|$) when linear damping dominates, and $e_{A} \sim (S_{\rm sc}^{0}\,e_{\rm B}/\Gamma_{\rm NL})^{1/2}$ when non-linear damping dominates (which occurs when $4\,\gamma_{\rm NL}\,|\tilde{g}| \gtrsim \gamma_{\rm L}^{2}$).

\subsection{Justification of Approximations}

This allows us to formally justify some of the approximations used in the text to estimate scalings: if we write $S_{\pm} \sim S_{\rm et} + S_{\rm sc}^{0}$ as the ``total'' driving and set $S_{\pm}$ equal to $Q_{\pm}\sim(\Gamma_{\rm L} + \Gamma_{\rm NL}\,e_{A}/e_{\rm B})\,e_{A}$ to solve for $e_{A}$ (this was done in the text to justify our more approximate scalings), we obtain the correct qualitative behaviors of $e_{A}/e_{\rm B}$ in all the relevant limits above. The transition between ET and SC limits here, $S_{\rm sc}^{0} \gtrsim S_{\rm et}$, corresponds to $|\tilde{g}|/s_{\rm et}$, which usually determines $\Phi$ (though there can, in greater detail, be non-negligible corrections from the $\tilde{h}/\Gamma$ term in determining which limit is most relevant). 

One interesting limit where this allows us to resolve some ambiguities is the case in highly-neutral gas ($f_{\rm ion} \ll 1$), with gyro-resonant \Alf\ frequencies larger than the ion-neutral collision time, so in the expressions above $v_{A,\,{\rm eff}} \approx (|{\bf B}|^2 / 4\pi\,\rho_{\rm ion})^{1/2} = v_{A,\,{\rm ideal}}\,f_{\rm ion}^{-1/2} \gg v_{A,\,{\rm ideal}}$. In this case, $S_{\rm et}$ is suppressed by strong ion-neutral damping (which usually leads to $\Gamma_{\rm L} \sim \Gamma_{\rm in} \gg \Gamma_{\rm NL}$), while $\tilde{g}^{2} \propto 1/f_{\rm ion}$ and $\tilde{h} \propto 1/f_{\rm ion}^{1/2}$ are enhanced, so for conditions of relevance in e.g.\ GMCs this means $|\Phi| \gg 1$ and the system rapidly converges to the SC regime with streaming at $v_{A,\,{\rm eff}} \propto 1/f_{\rm ion}^{1/2}$, essentially independent of the strength of $\nabla e_{\rm cr}^{\prime}$ or $S_{\rm et}$ on larger scales.

We can also return to the approximations regarding timescales made above. In Eq.~\ref{eqn:f.app}, we see from our steady-state solutions that the $c^{2}\,\nabla \cdot \mathbb{P}_{\rm cr}^{\prime}$ term is never negligible (it acts as a source term), while the relative importance of the $F_{e,\,{\rm cr}}^{\prime}$ and $H_{\rm cr}^{\prime}$ terms depends on whether the flux is super or sub-\Alf{ic}. In any case, noting that the $F_{e,\,{\rm cr}}^{\prime}$ term can be written $D_{t} F_{e,\,{\rm cr}}^{\prime} \sim \bar{\nu}_{\rm s}\,F_{e,\,{\rm cr}}^{\prime} + ...$ we immediately confirm that the equation is driven towards steady-state on the very short scattering timescale $\sim  \bar{\nu}_{\rm s}^{-1} \sim 30\,{\rm yr}\,R_{\rm GV}^{1/2}$ (for empirically-favored $\bar{\nu}_{\rm s}$ values). In Eq.~\ref{eqn:ea.app}, in steady state the dominant driving ($S_{\rm et,\,\pm}$ or SC $F_{e,\,{\rm cr}}^{\prime}-H_{\rm cr}^{\prime}$ term) terms have magnitude of order the damping terms $\Gamma_{\pm}\,e_{\pm}$, so the equation $D_{t}e_{\pm} = -\Gamma_{\rm L}\,e_{\pm} + ...$ is driven to steady-state on the local damping time $\sim \Gamma_{\pm}^{-1} \sim (30-300)\,{\rm yr}\,(R_{\rm GV}/B_{\mu {\rm G}})^{1/2}\,(10\,{\rm km\,s^{-1}}/v_{A,\,{\rm eff}})$ (using the scalings from \S~\ref{sec:damping} for $\Gamma_{\rm turb/LL}$ and $\Gamma_{\rm dust}$, assuming typical LISM properties; if other damping terms are also important, then $\Gamma_{\pm}^{-1}$ will be even smaller). This is similar to the scattering time. The ``gradient terms'' $\mathcal{O}(\nabla [u_{\rm gas},\,v_{A,\,{\rm eff}}]\,e_{\pm})$ are smaller than the other terms in Eq.~\ref{eqn:ea.app} by a factor $\mathcal{O}(|\nabla [u_{\rm gas},\,v_{A,\,{\rm eff}}]| / \Gamma_{\pm}) \sim 10^{-4}\,(R_{\rm GV}/B_{\mu {\rm G}})^{1/2}\,\ell_{\nabla,{\rm ISM},10}^{-1}$ where $\ell_{\nabla,{\rm ISM},10} = \ell_{\nabla,{\rm ISM}}/10\,{\rm pc}$ with $\ell_{\nabla,{\rm ISM}}$ the gradient scale-length of the bulk ISM properties (${\bf u}_{\rm gas}$ or $v_{A,\,{\rm eff}}$), justifying their neglect above. These timescales $\mathcal{O}(1/\nabla [u_{\rm gas},\,v_{A,\,{\rm eff}}])$ are of course also the same as the characteristic timescales for bulk ISM properties to evolve (e.g.\ $v_{A,\,{\rm eff}}$, ${\bf B}$, ${\bf u}_{\rm gas}$, $\rho$ from the usual MHD equations). Thus, this justifies our assumption that these MHD properties can be taken as approximately constant over the timescale for Eqs.~\ref{eqn:f.app}-\ref{eqn:ea.app} to reach local steady-state.

Now consider the CR energy equation $D_{t}e_{\rm cr}^{\prime}=...$ assuming the CR flux $F_{e,\,{\rm cr}}^{\prime}$ and $e_{\pm}$ equations have reached local steady-state (Eq.~\ref{eqn:dtecr.eqm} in the text). The source/sink term $\mathcal{S}_{\rm other,\,cr}^{\prime}$ is small compared to other terms at rigidities $\gtrsim $\,GV (except in special environments, e.g.\ at sources). Examination shows that the ``diffusive reacceleration'' term $\propto (v_{A,\,{\rm eff}}^{2} - \bar{v}_{A}^{2})/c^{2}$ is always small: it vanishes identically in the SC limit, but even in the ET limit it is suppressed by $\mathcal{O}(v_{A,\,{\rm eff}}^{2}/c^{2})$ for any plausible $\bar{\nu}_{\rm s}$ \citep[see][]{hopkins:cr.multibin.mw.comparison}. The ``streaming'' and ``adiabatic'' terms ($\sim \nabla \cdot (\bar{v}_{A}\,e_{\rm cr}^{\prime}\,\bhat)$ and $\sim P_{\rm cr}^{\prime}\,\nabla \cdot({\bf u}_{\rm gas} + \bar{v}_{A}\,\bhat)$) involve the same ``gradient'' or ISM bulk-property timescales $\mathcal{O}(1/\nabla [u_{\rm gas},\,v_{A,\,{\rm eff}}])$ as defined above. The term in the $D_{t}e_{\rm cr}^{\prime}$ equation which can evolve most rapidly is the ``diffusive'' term $\sim \nabla \cdot[\kappa_{\|}\,\bhat\bhat\cdot\nabla e_{\rm cr}^{\prime}]$ (with $\kappa_{\|}\equiv v_{\rm cr}^{2}/3\,\bar{\nu}_{\rm s}$), which drives $e_{\rm cr}^{\prime}$ to equilibrium on the diffusive timescale $\sim \ell_{\nabla,{\rm cr}}^{2}/\kappa_{\|}$ with $\ell_{\nabla,{\rm cr}} \sim e_{\rm cr}^{\prime} / |\nabla e_{\rm cr}^{\prime}|$ the CR energy gradient scale length. This is larger than the CR scattering time $\bar{\nu}_{\rm s}^{-1}$ for $\ell_{\nabla,{\rm cr}} \gtrsim c/(3\,\bar{\nu}_{\rm s}) \sim 3\,{\rm pc}\,R_{\rm GV}^{1/2}$ (i.e.\ $\ell_{\nabla,{\rm cr}}$ larger than the CR scattering mean free path). Since we observe and expect $\ell_{\nabla,{\rm cr}} \gtrsim$\,kpc, we confirm that $e_{\rm cr}^{\prime}$ will generically converge to steady-state on much longer timescales than $F_{e,\,{\rm cr}}^{\prime}$ or $e_{\pm}$.

\section{Extrinsic Turbulence: Basic Scalings \&\ Results}
\label{sec:turb.review}

It is helpful to briefly review some properties of extrinsic turbulence. In a turbulent cascade with velocity and magnetic field fluctuations $\delta{\bf v}$, $\delta{\bf B}$ on a scale $\lambda\sim 1/k$, most of the energy ($|\delta{\bf v}^{2}(k)| \sim k\,\mathcal{E}(k)$) is concentrated around the driving scale $\lambda \sim \ell_{\rm drive}$ ($\gtrsim 0.1-1\,$kpc in the ISM/CGM), with \Alf\ Mach number $\mathcal{M}_{A} \sim \langle |\delta{\bf v}^{2}(\lambda \sim \ell_{\rm drive})|\rangle^{1/2}/v_{A,\,{\rm ideal}} \gtrsim 1$. On the largest super/trans-sonic/\Alf{ic} scales this can give rise to a compressible and weakly-pressurized  \citet{burgers1973turbulence}-like power-spectrum ($|\delta{\bf v}^{2}(\lambda)| \propto \lambda$; \citealt{schmidt:2009.isothermal.turb,konstandin:2012.lagrangian.structfn.turb,hopkins:2012.intermittent.turb.density.pdfs,squire.hopkins:turb.density.pdf}). Below the \Alf\ scale $\ell_{A}$, where $|\delta{\bf v}^{2}(\lambda \sim \ell_{A})|\rangle^{1/2} \sim v_{A,\,{\rm ideal}}$ the fluctuations are sub-\Alf{ic}, by definition. In the Galactic ISM, typically $\ell_{A} \sim 10-100\,{\rm pc}$ \citep{elmegreen:sf.review,mac-low:2004.turb.sf.review}. The defining feature of a traditional strong inertial-range  cascade is the energy condition, $S_{\rm turb} \sim e_{\rm  turb} / \tau_{\rm cas} \propto |\delta{\bf v}(\lambda)|^{2} / \tau_{\rm cas}(\lambda) \sim $\,constant, where $\tau_{\rm cas}$ is the cascade or energy-transfer or decoherence time, which can be parameterized over the inertial range as $\tau_{\rm cas} \sim (\ell_{A}/v_{A,\,{\rm ideal}})\,(\lambda/\ell_{A})^{\alpha}$ with some $\alpha>0$ (with $\alpha\le1$ almost always required).\footnote{From the energy condition, this immediately gives $e_{\rm turb}(k) \propto k^{-\alpha}$, or the one-dimensional $\mathcal{E}(k) \propto k^{-(1+\alpha)}$. So e.g.\ the commonly cited ``K41-like'' ($\mathcal{E}(k) \propto k^{-5/3}$), ``IK-like'' ($\mathcal{E}(k) \propto k^{-3/2}$), and \citet{burgers1973turbulence}-like ($\mathcal{E}(k) \propto k^{-2}$) isotropic power spectrum scalings correspond to $\alpha=(2/3,\,1/2,\,1)$, respectively.} For now, we neglect the difference between the parallel (to $\bhat$) $k_{\|}$ and perpendicular $k_{\bot}$ components of $k$, but recall that what we need to calculate CR scattering rates is $e_{\rm turb}(k_{\|})$, since it is the parallel component $k_{\|}$ which controls the scattering terms (e.g.\ \citealt{1975RvGSP..13..547V}; and for gyro-resonance, $k_{\|} \sim 1/r_{g,{\rm cr}}$). If there is strong damping/dissipation, then the dissipation/Kolmogorov scale of the turbulence $k_{\rm diss} \sim 1/\lambda_{\rm diss}$ occurs when some dissipation/damping rate $\sim \Gamma(k,\,...)\,|\delta{\bf v}|^{2}$ becomes larger than the turbulent dissipation/cascade/transfer rate $\sim |\delta{\bf v}|^{2}/\tau_{\rm cas}$, i.e.\ $\Gamma \gtrsim 1/\tau_{\rm cas}$. For example, for some kinematic viscosity $\Gamma_{\rm visc} \sim \nu_{\rm visc}\,k^{2}$, we have $\lambda_{\rm diss} \sim \ell_{A}\,(\ell_{A}\,v_{A,\,{\rm ideal}} / \nu_{\rm visc})^{-1/(2-\alpha)}$ (i.e.\ $\ell_{A}\,v_{A,\,{\rm ideal}} / \nu_{\rm visc}$ is the Reynolds number). 

The gyro-resonant scale $\lambda_{g} \sim 1/k_{g} \sim r_{g,{\rm cr}} \sim 10^{-6}\,{\rm pc}\,R_{\rm GV}/B_{\rm \mu G}$, so $\lambda_{g} \ll \ell_{A}$ at energies of interest, hence $|\delta{\bf v}(\lambda_{g})| \ll v_{A,\,{\rm ideal}}$ and $|\delta{\bf B}(\lambda_{g})| \ll |{\bf B}|$, but we also know this empirically, since the observationally-constrained CR scattering rates require $|\delta{\bf B}|/|{\bf B}| \sim 3\times10^{-4}\,R_{\rm GV}^{0.2}$. This means we can at least approximately decompose the fluctuations into a linear superposition of \Alf, slow, and fast magnetosonic modes, with $|\delta{\bf B}|/|{\bf B}| \sim |\delta{\bf v}|/v_{A,\,{\rm ideal}} \ll 1$, and we can treat the gas as weakly compressible and smooth (gradient length scales of bulk ISM properties are much larger than $\lambda_{g}$). 

\subsection{Anisotropy: \Alf{ic} and Slow Cascades}
\label{sec:turb.review.alf}

First consider the \Alf{ic} case. \Alf\ waves are generally weakly damped by collisionless processes in ionized gas down to the ion gyro scale, so assume we can temporarily neglect damping. Since these are incompressible modes, we can rewrite the MHD equations in terms of the convenient Elsasser variables:
\begin{align}
\label{eqn:incompressible.mhd}
\partial_{t} {\bf Z}^{+}  - ({\bf v}_{A}\cdot\nabla){\bf Z}^{+} + ({\bf Z}^{-}\cdot\nabla)\,{\bf Z}^{+} = -\nabla p/\rho \\
\nonumber \partial_{t} {\bf Z}^{-} + ({\bf v}_{A} \cdot \nabla){\bf Z}^{-} + ({\bf Z}^{+}\cdot \nabla)\,{\bf Z}^{-} = -\nabla p/\rho
\end{align}
where ${\bf v}_{A} = v_{A}\,\hat{\bf b}$, ${\bf Z}^{+} \equiv \delta{\bf v} + \delta{\bf B}/(4\pi\rho)^{1/2}$, ${\bf Z}^{-} \equiv \delta {\bf v} -  \delta{\bf B}/(4\pi\rho)^{1/2}$. 

There are  two possible limits to Eq.~\ref{eqn:incompressible.mhd}. In limit {\bf (1)} the non-linear term ($({\bf Z}^{-}\cdot\nabla)\,{\bf Z}^{+}$ or $({\bf Z}^{+}\cdot\nabla)\,{\bf Z}^{-}$) is small. If this term is negligible, then we trivially recover the equations for \Alf\ wave packets without any interactions: i.e.\ the equations do not feature any ``cascade'' per se, but simply admit whatever spectrum of \Alf\ waves we wish to externally impose, by introducing some other source term (e.g.\ SC driving or our proposed novel driving mechanisms). If the non-linear term is not completely ignored but still small (e.g.\ $\langle |({\bf Z}^{-}\cdot\nabla)\,{\bf Z}^{+}|^{2} \rangle \ll \langle | ({\bf v}_{A}\cdot\nabla){\bf Z}^{+}|^{2} \rangle$), then we obtain the classic assumptions of ``weak'' \Alf{ic} turbulence. The conditions where this might occur in practice (in the absence of some other small-scale driving) are restrictive \citep[see][]{lazarian:2016.cr.wave.damping}, but there is a bigger problem. As shown elegantly in \citet{1994ApJ...432..612S} (see also \citealt{schekochihin:2020.mhd.turb.review} for a pedagogical presentation), an {\em isotropic} IK-type weak ``cascade'' as envisioned by e.g.\ \citet{kraichnan:1965.ik.aniso.turb} cannot exist (it is neither physically nor mathematically self-consistent): instead the weak cascade occurs purely along $k_{\bot}$, conserving $k_{\|}$, so there is again {\em no} cascade to define $e_{\rm turb}(k_{\|})$ nor is there any connection between $e_{\rm turb}(k_{\|})$ for different $k_{\|}$ (weak \Alf{ic} turbulence simply re-distributes this energy to different $k_{\bot}$ at the same $k_{\|}$, which has no effect to leading order on CR scattering). In other words, we once again simply recover whatever \Alf\ spectrum $e_{\rm turb}(k_{\|})$ we choose to impose by introducing some {\em other}, non-ET source term. 

In limit {\bf (2)},  the non-linear term is not negligible (e.g.\ $\langle |({\bf Z}^{-}\cdot\nabla)\,{\bf Z}^{+}|^{2} \rangle \gtrsim \langle | ({\bf v}_{A}\cdot\nabla){\bf Z}^{+}|^{2} \rangle$). In this limit, a cascade linking $\mathcal{E}$ at different $k_{\|}$ is possible, and making additional assumptions leads to, for example, the specific ``strong'' turbulence cascade of \citet{GS95.turbulence} (GS95, or variations proposed in \citealt{boldyrev:2005.dynamic.alignment} or others reviewed in \citealt{schekochihin:2020.mhd.turb.review}), all of which give $\mathcal{E} \propto k_{\|}^{-1}$, i.e.\ $\delta_{\rm s}=0$, as noted in the main text. 
More generally, a simple argument that $\delta_{\rm s}$ must be $\le 0$ in this regime goes as follows. Define the parallel scale of a mode as $l_{\|} \sim 1/k_{\|}$, such that $\mathcal{O}[({\bf v}_{A}\cdot\nabla){\bf Z}^{+} ] \sim v_{A}\,Z_{\lambda} / l_{\|}$, and note that since $|\delta {\bf v}| \ll v_{A}$ this limit {\bf (2)} requires $l_{\|} \gtrsim l_{\bot}\,(v_{A}/|\delta {\bf v}|) \gg l_{\bot}$ (or else the non-linear term would again be negligible, putting us in limit {\bf (1)}). This means $k \approx k_{\bot}$, so $\lambda \sim 1/k = l_{\bot}$. Without loss of generality, define $l_{\|} \propto v_{A}\,\tau_{\rm cas}\,(\ell_{A} / l_{\|})^{\alpha_{\|}}$ over some dynamic range, such that $\mathcal{O}[ ({\bf v}_{A}\cdot\nabla){\bf Z}^{+}] \sim \mathcal{O}[ (l_{\|}/\ell_{A})^{\alpha_{\|}}\,Z_{\lambda}/\tau_{\rm cas} ]$. Trivially, $\alpha_{\|} \ge 0$ is required so that the linear term is equal to or smaller than the non-linear \&\ ``cascade'' terms (otherwise, if $\alpha_{\|} < 0$, for $l_{\|} \sim \lambda_{g} \ll \ell_{A}$ we would immediately arrive back in limit {\bf (1)}). Note that the critical balance assumption corresponds specifically to $\alpha_{\|} = 0$. Now, we can also allow for some arbitrary losses from the cascade across scales by defining the cascade rate $S \sim e_{\rm turb}/\tau_{\rm cas} \sim S_{0}\,(l_{\|}/\ell_{A})^{\alpha_{S}}$, where again any physical cascade requires $\alpha_{S} \ge 0$ (an un-damped cascade corresponds to $\alpha_{S}=0$, but non-zero dissipation or  losses can decrease the energy on smaller scales). Now if we recall $\nu \propto \Omega\,|\delta{\bf B}(k_{\|} \sim 1/r_{g,{\rm cr}})|^{2}/|{\bf B}|^{2} \sim k_{\|}\,e_{\rm turb}(k_{\|})|_{k_{\|}\sim 1/r_{g,{\rm cr}}} \sim S_{0}\,(l_{\|}/\ell_{A})^{\alpha_{S}}\,\tau_{\rm cas}/ l_{\|} \propto l_{\|}^{\alpha_{\|} + \alpha_{S}}$, we have $\delta_{\rm s} = -(\alpha_{\|} + \alpha_{S}) \le 0$. 

In short, it is not possible to construct an internally-consistent \Alf{ic} {\em cascade} with $\delta_{\rm s} > 0$. Anisotropy in the form of critical balance with an un-damped \Alf{ic} cascade gives $\delta_{\rm s} = 0$. Adding losses/dissipation at scales between gyro-resonant and driving only further decreases $\delta_{\rm s}$.
Violating the critical balance-type assumptions (by e.g.\ introducing a non-zero $\alpha_{\|}$, in our notation above) leads to one of two outcomes. {\bf (1)} There is no ``cascade'' or any interaction between modes with different parallel wavenumbers (if $\alpha_{\|}<0$), and the power $|\delta{\bf B}(k_{\|})|^{2}$ must be set not by ET but by some other source term driving modes independently on each scale. Or {\bf (2)} the cascade produces $\delta_{\rm s}<0$ if $\alpha_{\|} > 0$, i.e.\ if the anisotropy is even larger than required for critical balance.\footnote{Of course, as many have pointed out, any \Alf{ic} cascade which does not obey critical balance will be pushed (by a weak cascade or causality/de-correlation) towards a state of critical balance. We simply wish to stress that even transient violations of this condition fail to produce an \Alf{ic} cascade with $\delta_{\rm s}>0$.}

For more rigorous discussion, we refer interested readers to \citet{schekochihin:2020.mhd.turb.review} for a review of more detailed \Alf{ic} turbulence models, demonstrating that even models which are imbalanced, intermittent, decaying, damped, or otherwise strongly modified all obey $\delta_{\rm s}\le 0$ in our language. 

Note that, as many others have pointed out, slow modes are subject to a similar anisotropy constraint to \Alf\ waves as described above (which again leads to $\delta_{\rm s} \le 0$), {\em and} are subject to additional magnetosonic damping terms, which further constrain $\delta_{\rm s} \le 0$ as we discuss for fast modes below \citep[see e.g.][and  references therein]{cho.lazarian:2003.mhd.turb.sims,yan.lazarian.04:cr.scattering.fast.modes,Schekochihin2009}. Thus we do not discuss them further.

\subsection{Fast Modes} 
\label{sec:turb.review.fast}

Now consider instead a fast magnetosonic cascade. It is at least theoretically possible, in principle, that {\em within the inertial-range} these could produce an isotropic cascade ($k_{\|} \sim k_{\bot} \sim k$) with the desired scaling of $e_{\rm turb}(k) \propto k^{-\alpha}$: if e.g.\ $\delta_{\rm s} \sim 0.6$ is observationally required, this would imply an inertial-range $\tau_{\rm cas} \sim \lambda^{0.4}$ or $|\delta {\bf v}| \propto \lambda^{0.2}$ ($\alpha \sim 0.4$). But it is important to stress that even the inertial-range behavior on small scales ($\ll \ell_{A}$) is not theoretically clear: while e.g.\ \citet{cho.lazarian:2003.mhd.turb.sims,ferrand:2020.acoustic.turb.sims.high.dynrange} argue for a spectrum with a \citet{1970SPhD...15..439Z}-type weak cascade $e_{\rm turb}(k) \propto k^{-1/2}$ below the sonic/\Alf\ scale (which would give $\delta_{\rm s} \sim 0.5$ in the inertial range, within the observationally-allowed range), others have argued from both analytic theoretical grounds \citep{1973SPhD...18..115K,1976ZPhyB..23...89E,1992PhLA..166..243S,2000JPlPh..63..447G,2011PhyA..390.1534S,2008PhPl...15f2305K,2011PhRvL.107m4501G,2016MPLB...3050297S} and numerical simulations \citep{1974PhLA...47..419E,1990ThCFD...2...73E,2006MNRAS.370..415M,2010PhRvE..81a6318L,2010ApJ...720..742K,2020PhRvX..10c1021M} that the spectrum should be closer to $e_{\rm turb}(k) \propto k^{-1}$ (giving $\delta_{\rm s} = 0$). And the classic \citet{kolmogorov:turbulence} (K41) type scaling $e_{\rm turb}(k) \propto k^{-2/3}$ ($\delta_{\rm s}=1/3$ in the inertial range), though often cited in older ``leaky box'' models for CR transport which did not include a scattering halo, actually provides a poor fit to the observations in modern models that include any extended scattering halo \citep[see e.g.][]{blasi:cr.propagation.constraints,vladimirov:cr.highegy.diff,gaggero:2015.cr.diffusion.coefficient,2016ApJ...819...54G,2016ApJ...824...16J,cummings:2016.voyager.1.cr.spectra,2016PhRvD..94l3019K,evoli:dragon2.cr.prop,2018AdSpR..62.2731A,hopkins:cr.multibin.mw.comparison,delaTorre:2021.dragon2.methods.new.model.comparison}. 

It is also not clear that isotropy is a good assumption on small scales even for fast modes \citep[see e.g.][and references therein]{2008PhPl...15f2305K,2010PhRvE..81a6318L,2011RPPh...74d6901B}. If there is significant anisotropy, for reasons similar to those above it will generically tend to decrease $\delta_{\rm s}$. 

But as discussed in the text, an entirely un-ambiguous problem is that isotropic fast modes at gyro-resonant scales are very strongly damped. Even in a fully-ionized medium, collisionless damping of fast modes is orders-of-magnitude more efficient than for \Alf\ modes on these scales.\footnote{The dominant fast-mode damping terms (in addition to the weaker terms in the main text which also apply to \Alf\ waves) are: viscous damping $\Gamma_{\rm fast,\,visc}$ (both by neutrals and \citealt{braginskii:viscosity} viscosity from ions)  and collisionless/Landau damping $\Gamma_{\rm fast,\,L}$: 
\begin{align}
\Gamma_{\rm fast,\,visc} &\equiv k^{2}\,\nu_{\rm visc,\,eff} \\ 
\Gamma_{\rm fast,\,L} &\equiv \frac{\sin^{2}(\theta)}{\cos{\theta}}\,k\,v_{\rm fast}\,f_{\rm fast,\,L} \\
\nu_{\rm visc,\,eff} &\equiv \nu_{\rm visc,\,ion,\,0}\,f_{\rm v,\,ion}(\theta)\,f_{\rm ion} + \nu_{\rm visc,\,neutral}\,f_{\rm neutral} \\ 
\nu_{\rm visc,\,ion,\,0} &\sim 0.6\times10^{18}\,{\rm cm^{2}\,s^{-1}}\,T_{4}^{5/2}\,n_{1}^{-1} \\ 
\nu_{\rm visc,\,neutral} &\sim 3\times10^{20}\,{\rm cm^{2}\,s^{-1}}\,T_{4}^{1/2}\,n_{1}^{-1} \\
f_{\rm v,\,ion}(\theta) &\approx 
\begin{cases}
\sin^{2}{(\theta)} \hfill & \ ( \beta_{\rm plasma} \ll 1) \\ 
|1-3\,\cos^{2}(\theta)|^{2} \hfill & \ (\beta_{\rm plasma} \gg 1) 
\end{cases}
\end{align}
Where for small $\theta$, at $\beta_{\rm plasma}$ not too large $f_{\rm fast,\,L} \equiv (\omega_{\rm fast}/k\,v_{\rm fast})\,(\sqrt{\pi\,\beta_{\rm plasma}}/4)\,\sqrt{m_{e}/m_{p}}\,\exp{[-m_{e}/m_{p}\,\beta_{\rm plasma}\,\cos^{2}(\theta)]}$ \citep{1961PThPS..20....1G}, while for very large $\beta_{\rm plasma}$, $f_{\rm fast,\,L} = (2/\cos^{2}(\theta))\,(\omega_{\rm fast}/k\,v_{\rm fast})\,(\omega_{\rm fast}/\omega_{\rm c,\,i})$ (with $\omega_{\rm fast}$ the fast-mode frequency at wavenumber $k$, and $\omega_{c,\,i}$ the ion cyclotron frequency; \citealt{foote.kulsrud:1979.damping}). Note that viscous damping in ionized gas with $\beta_{\rm plasma} \gtrsim 1$ acts similar to isotropic (neutral) damping, in that it strongly damps parallel fast waves. In evaluating the full CR scattering rate expressions, this has the same practical effect of strongly truncating the gyro-resonant scattering term (giving $\delta_{\rm s} \ll 1$).}
Depending on the assumptions of ISM/CGM properties, mode angles, and the cascade timescale, if we define the ``damping scale'' $k_{\rm diss} \sim 1/\lambda_{\rm diss}$ as that where for {\em some} mode angle $\cos{\theta} \equiv k_{\|}/k$, the most-rapid fast-mode damping rate $\Gamma_{\rm fast}$ is larger than the  cascade rate $1/\tau_{\rm cas}$, we would obtain $\lambda_{\rm diss} \sim 10^{-4} - 10\,$pc (see e.g.\ Fig.~1 and Table~1 in \citealt{yan.lazarian.04:cr.scattering.fast.modes}). More importantly, accounting for the combination of viscous, collisionless, and neutral damping (with realistic ISM/CGM scalings), it is essentially impossible to make $\lambda_{\rm diss}$ smaller than $r_{g,{\rm cr}}$ at rigidities $< 100-1000\,$GV.\footnote{Given the most optimistic possible assumptions for reducing $\lambda_{\rm diss}$, it may be possible in some phases of the ISM, such as the WIM, to make $r_{g,{\rm cr}} > \lambda_{\rm diss}$ at $>100$\,GV, while more typical assumptions for the WIM and even the most optimistic assumptions for the WNM and CNM or GMCs require $\gtrsim 1000\,$GV. In hot gas in the Galactic coronae, HIM, and CGM/halo, $\lambda_{\rm diss}$ becomes much larger, and it is plausible that $\lambda_{\rm diss} \gtrsim r_{g,{\rm cr}}$ up to $>10^{6}$\,GV (i.e.\ up to PeV CR energies).} As a result, we argued in the text that $\delta_{\rm s} \le 0$. Detailed numerical calculations showing $\delta_{\rm s} \lesssim 0$ is always the case for all $R_{\rm cr}\lesssim 100-1000\,$GV for fast-mode ET, accounting in detail for exact expressions of the scattering rates and their detailed dependence on pitch angle, mode angle, and wavelength, along with the full range of angle-dependent damping rates from different processes (following more exact integral expressions for CR scattering physics), have been extensively presented, including in YL04 (their Fig.~2), \citet{yan.lazarian.02,yan.lazarian.04:cr.scattering.fast.modes,yan.lazarian.2008:cr.propagation.with.streaming}, and \citet{kempski:2021.reconciling.sc.et.models.obs}. And of course our fast-mode ET model in the main text (Fig.~\ref{fig:ET.defaults}; {\em right}) is one such calculation as well. So here we only seek to justify this heuristically (see \citealt{kempski:2021.reconciling.sc.et.models.obs} as well for a similar discussion). 

First, consider the effects of damping on gyro-resonant CR scattering. If the damping is isotropic (as with e.g.\ neutral viscosity or ion-neutral damping per \citealt{spitzer:1978.book}, in regions with neutral fraction $f_{\rm neutral} \gtrsim 0.001-0.01$; see text and \citealt{hopkins:cr.transport.constraints.from.galaxies}), the spectrum is, by definition, truncated at $k_{\|} \gtrsim k_{\rm diss}$ exponentially or  super-exponentially,\footnote{Even if we assumed the mathematically ``weakest possible'' cutoff for the spectrum below the damping scale, i.e. we assume $S \propto S_{0}\,(k_{\rm diss}/k)^{\alpha_{\rm S}}$ continues for $k > k_{\rm diss}$, we must have $\alpha_{\rm S}>0$. Equating this driving with the (by definition dominant) neutral damping $Q_{\rm visc,\,fast} = \Gamma_{\rm visc,\,fast}\,e_{A} \sim \nu_{\rm visc,\,n}\,k^{2}\,e_{A}$, we have $e_{A} \propto k^{-(2+\alpha_{\rm S})}$, i.e.\ $\delta_{\rm s} = -1 - \alpha_{\rm s} < -1$.} equivalent to $\delta_{\rm s} \ll 0$. But even in a fully-ionized medium with $\beta_{\rm plasma} \ll 1$ assuming the dominant damping is e.g.\ from Braginskii viscosity or collisionless damping, which are anisotropic and do not damp parallel modes, we have a damping rate of the form: $\Gamma_{\rm fast} \propto k^{1+\alpha_{k}}\, \sin^{2}(\theta)$, where $0 \le \alpha_{k} \le 1$ depends on e.g.\ whether collisionless or viscous damping dominates (and $\beta_{\rm plasma}$). At scales $\lambda < \lambda_{\rm diss}$, modes with $\theta > \theta_{c}$ will have $\Gamma_{\rm fast} \gg 1/\tau_{\rm cas}$ and be truncated, so if we make the most optimistic assumption that the remaining modes simply continue their cascade uninterrupted, the surviving modes are confined to a narrow bicone with $|\theta| < \theta_{c}\ll1$, where $\theta_{c}$ becomes smaller with increasing $k$. Equating $\Gamma_{\rm fast}$ and $\tau_{\rm cas}$ gives $\sin^{2}(\theta_{c}) \sim \theta_{c}^{2} \sim \tau_{\rm cas}^{-1}(k)\,k^{-(1+\alpha_{k})}$. So if $S\sim S_{0}\sim$\,constant on large scales is the total cascade power and begins (by assumption) isotropic, the power on smaller scales is necessarily reduced by a factor proportional to the solid angle of the undamped cone $\propto \theta_{c}^{2}$. Thus the energy of scattering modes with a given $k_{\|}$ must scale as: $e_{A} \propto S\,\tau_{\rm cas}\,\theta_{c}^{2} \propto S_{0}\,k^{-(1+\alpha_{k})}$, i.e.\ $\delta_{\rm s} = -\alpha_{k} \le 0$. Note further that if there is any ``leakage,'' i.e.\ transfer of energy between the weakly-damped cone and broader mode angles which are rapidly damped, then $S$ must decrease further even along the ``surviving'' directions, so we take $S\rightarrow S_{0}\,(\lambda\,k_{\rm diss})^{\alpha_{\rm S}}$ with $\alpha_{\rm S}>0$, giving $\delta_{\rm s} = -(\alpha_{k} + \alpha_{\rm s}) < 0$, and further reducing $\delta_{\rm s}$.

As pointed out in YL04 and others, if a spectrum is strongly suppressed or truncated at scales $\lambda_{\rm diss} \gg r_{g,{\rm cr}}$,  then transit-time damping (TTD) from the larger-scale modes near $\lambda_{\rm diss}$ could still produce CR scattering which dominates over the gyro-resonant term. But for TTD, we must replace our gyro-resonant expression from the main text ($\nu_{\rm s} \sim \Omega\,|\delta{\bf B}(k_{\|})|^{2}/|{\bf B}|^{2} \sim (v_{\rm cr}\,k_{\|})\,(k_{\|}\,\mathcal{E}(k_{\|})) / e_{\rm B}$) with $\nu_{\rm s} \rightarrow \int (\Omega_{\rm cr}^{2}/|{\bf B}|^{2})\mathcal{E}(k)\,dk\, |J^{\prime}[k_{\bot}\,v_{\rm cr,\,\bot}/\Omega_{\rm cr})]|^{2}\, 1/(k_{\|}\,v_{\|})\mathcal{R}(k_{\|}\,v_{\|} - \omega \, | \, k,\,k_{g},\,...) \sim e_{\rm B}^{-1} \int_{0}^{k_{\rm diss}} dk \mathcal{E}(k)\, k\,v_{\rm cr}\,\mathcal{R} \sim (k_{\rm diss}\,v_{\rm cr})\,|\delta{\bf B}(k_{\rm diss})|^{2}/|{\bf B}|^{2}\,\mathcal{R}(k_{g},\,k_{\rm diss},\,...) \sim ({\rm constant}) \times \mathcal{R}$ from e.g.\ \citet{1975RvGSP..13..547V}. Here $\mathcal{R}$ is some appropriate dimensionless ``response'' or resonance function. Heuristically, this is just the statement that a CR is scattered in pitch angle by a random amplitude $|\Delta \mu| \sim |\delta {\bf B}(k)| / |{\bf B}|$ as it crosses a mode in time $\Delta t \sim \lambda/v_{\rm cr} \sim 1/k\,v_{\rm cr}$, so will random-walk to an order-unity change in pitch angle after $N \sim |{\bf B}|^{2}/|\delta{\bf B}(k)|^{2}$ events, implying a scattering time $\nu_{\rm s}^{-1} \sim N^{2}\,\Delta t \sim [(k\,v_{\rm cr})\,|\delta{\bf B}(k)|^{2}/|{\bf B}|^{2}]^{-1}$. But if this is dominated by the integral over larger-scale modes, then it is by definition independent of $R_{\rm cr}$, so $\delta_{\rm s}=0$. Moreover, if we account for any non-trivial response function  $\mathcal{R}$ (describing how efficiently a mode of scale $k$ can deflect the pitch-angle of a CR with gyro-radius $r_{g,{\rm cr}} \sim 1/k_{g}$), it must be the case that $\mathcal{R}$ is a decreasing function of $k_{g}/k$ for $k_{g} \gg k$, hence we must have $\delta_{\rm s} < 0$. 

Finally, note that because of how the actual scaling of the spectrum $S$ factors out in the above, and that these fast-mode damping mechanisms act on all scales of interest, the above conclusion that fast-mode damping requires $\delta_{\rm s} \le 0$ applies not just to fast modes sourced by a larger-scale cascade, but {\em any} isotropically-driven population of fast modes, even if they were driven or sourced around the gyro-resonant scales.

\subsection{Summary}
\label{sec:turb.review.summary}

In summary, {\em it is not possible to construct a self-consistent ``cascade'' model which produces $\delta_{\rm s} > 0$} for CRs, as required (observed $\delta_{\rm s} \sim 0.5-0.7$). We stress that the arguments above are quite generic: this is not a statement specific to one particular model of turbulence or to various controversial or uncertain assumptions. Rather, they arise from fundamental features of the MHD equations themselves (which require that any \Alf{ic} or slow ``cascade'' linking different parallel $k_{\|}$ have $\delta_{\rm s} \le 0$) or the fundamental nature of magnetosonic damping (which means any magnetosonic cascade where the most-rapidly-damped-modes begin to be appreciably damped on a spatial scale $\lambda_{\rm diss}$ larger than the gyro-radius $r_{g,{\rm cr}}$ at some $R_{\rm cr}$ must have $\delta_{\rm s} \le 0$ at all smaller $R_{\rm cr}$). 

There is, however, one rather straightforward way to provide the desired scattering: as we show in \S~\ref{sec:turb.review.alf}, if there exists some {\em other} source/driving term of \Alf\ waves (other than a cascade related to any mode-coupling from larger or smaller scales) at parallel $k_{\|}$, and those waves are not extremely-anisotropic (i.e.\ have typical $k_{\|} \gtrsim (|\delta {\bf v}(k_{\|})| / v_{A})\,k_{\bot} \gg 0.0003\,k_{\bot}$, so limit {\bf (1)} in \S~\ref{sec:turb.review.alf} applies), then it is perfectly allowed to construct an arbitrary spectrum $\mathcal{E}(k_{\|}) \propto k_{\|}^{\delta_{\rm s}-2}$ with the desired $\delta_{\rm s}$. These can be weakly-damped (by e.g.\ the mechanisms in the text), and undergo a weak cascade mixing the perpendicular wavenumbers $k_{\bot}$ but leaving $\mathcal{E}(k_{\|})$ unmodified, and satisfy all consistency constraints we discuss above. As noted in the text, standard SC theory would be one example of precisely this case, except that (for entirely different reasons) the form of the driving term $S_{\rm sc}$ produces the incorrect spectrum $\mathcal{E}(k_{\|})$ (unless one also modifies the damping terms as we discuss). The other driving terms we propose: $S_{\rm new,\,lin}$ and $S_{\rm new,\,ext}$ also function in this manner. 

At large CR rigidities, $\gtrsim 0.1-1\,$TV, it becomes possible (at least in some ISM phases) to have $r_{g,{\rm cr}} \gtrsim \lambda_{\rm diss}$, so a ``traditional'' ET-type theory can apply and at least in principle one could obtain a reasonable CR scattering rate from turbulence with an isotropic inertial-range spectrum with $k_{\|}\,\mathcal{E}(k_{\|}) \propto k_{\|}^{-1/2}$ \citep[see e.g.][]{fornieri:2021.comparing.et.models.data.et.only.few.hundred.gv}. But of course it is also possible that additional source terms like those we propose could still be important on these scales as well.

\end{appendix}

\end{document}